\documentclass[aps,prd,floatfix,showpacs,twocolumn,superscriptaddress,nofootinbib]{revtex4-2}
\usepackage{dsfont}
\usepackage{amsmath}
\usepackage{amsfonts}
\usepackage{amssymb}
\usepackage{graphicx}%
\usepackage{subfig}
\usepackage{braket}
\usepackage{float}
\usepackage{pslatex}
\usepackage{slashed}
\usepackage{hyperref}
\usepackage{color}
\usepackage{lipsum}
\usepackage{tensor}
\usepackage{physics}
\usepackage{tikz-feynman}

\definecolor{darkgreen}{rgb}{0,0.35,0}

\newcommand{\MSbar}{\overline{\mbox{MS}}}
\newcommand{\MSbartiny}{\overline{\mbox{\tiny{MS}}}}
\newcommand{\DIS}{\mbox{\tiny{DIS}}}

\newcommand{\p}{\partial}

\newcommand{\omu}{\overline{\mu}}
\newcommand{\lms}{\Lambda_{\overline{\mbox{\tiny{MS}}}}}

\newcommand{\ddd}{\ensuremath{\mathrm{d}}}

\newcommand{\be}{\begin{equation}}
\newcommand{\ee}{\end{equation}}
\newcommand{\bea}{\begin{eqnarray}}
\newcommand{\eea}{\end{eqnarray}}

\newcommand{\unict}{University of Catania, Department of Physics and Astronomy, Via Santa Sofia 64, I-95123 Catania, Italy}
\newcommand{\infnct}{INFN Sezione di Catania, Via Santa Sofia 64, I-95123 Catania, Italy}
\newcommand{\kulak}{KU Leuven Campus Kulak Kortrijk, Department of Physics, Etienne Sabbelaan 53 bus 7657, 8500 Kortrijk, Belgium}
\newcommand{\ughent}{Ghent University, Department of Physics and Astronomy, Krijgslaan 281-S9, 9000 Gent, Belgium}
\newcommand{\uerj}{Departamento de F\'{\i }sica Te\'{o}rica, Instituto de F\'{\i }sica, UERJ - Universidade do Estado do Rio de Janeiro, Rua São Francisco Xavier 524, 20550-013, Maracanã, Rio de Janeiro, Brazil}

\begin{document}

\title{Dynamically massive linear covariant gauges: setup and first results}
\author{Giorgio Comitini}\email{giorgio.comitini@dfa.unict.it}\affiliation{\unict}\affiliation{\infnct}\affiliation{\kulak}
\author{Tim De Meerleer}\email{timdemeerleer07@gmail.com}\affiliation{\kulak}
\author{David Dudal}\email{david.dudal@kuleuven.be}\affiliation{\kulak}\affiliation{\ughent}
\author{Silvio Paolo Sorella}\email{sorella@uerj.br}\affiliation{\uerj}

\begin{abstract}
\noindent We discuss the possibility to obtain a massive Landau gauge, based on the local composite operator (LCO) effective action framework combined with the Zimmerman reduction of couplings prescription. As a way to deal with the gauge ambiguity, we check that the ghost propagator remains positive, a necessary condition for gluon field configurations beyond the Gribov region to be negligible. We pay attention to the BRST invariance of the construction, allowing for a future generalization to a class of massive linear covariant gauges. As a litmus test, we compare our predictions to the lattice data for the two-point functions in Landau gauge introducing the ``Dynamically Infrared-Safe'' renormalization scheme, including the renormalization group optimization of both the gap equation and the two-point functions. We also discuss the relation to and differences with the Curci-Ferrari model, the usefulness of which in providing an effective perturbative description of non-perturbative Yang-Mills theories became clear during recent years.
\end{abstract}

\maketitle

\section{Introduction}
Recently, the Curci-Ferrari model \cite{Curci:1976bt,Curci:1976kh} has witnessed a revived interest in works like \cite{Gracey:2002yt,Tissier:2010ts,Tissier:2011ey,Pelaez:2013cpa,Reinosa:2017qtf,Gracey:2019xom,DallOlio:2020xpu,Pelaez:2021tpq,Barrios:2021cks} thanks to its capability to describe rather well the $n$-point functions of gauge theories in the Landau gauge, next to allowing a perturbative sneak peek into the phase diagram, \cite{Reinosa:2014zta,Reinosa:2016iml,Maelger:2018vow}. As of now, however, the success in matching the model to lattice data relies on the fitting of both the gauge coupling $g$ and the Curci-Ferrari mass $m$ \cite{Tissier:2010ts,Tissier:2011ey,Gracey:2019xom}.

A possible route towards a first principle derivation of the model, including a determination of $m$ from the sole knowledge of $g$ at a given scale, was formulated in \cite{Serreau:2012cg} based on a weighing over the Gribov copies. It was implemented successfully in a class of non-linear gauges that contains the Landau gauge as a limiting case \cite{Tissier:2017fqf}. Unfortunately, the dynamical mass generation mechanism identified in this reference fails precisely in this limit. A more recent attempt was done in \cite{Reinosa:2020skx}, directly in the Landau gauge. It was found that the system exhibits two phases, one of which corresponds to a massive implementation of the Landau gauge bearing some resemblance with the Curci-Ferrari model, with however gapped ghost degrees of freedom. As discussed in \cite{Reinosa:2020skx}, the presence of massive ghosts is not incompatible with the lattice results for the latter are not a direct measurement of the ghost propagator but rather the averaging of the Faddeev-Popov operator $-\partial_\mu D_\mu$ which remains massless in both of the above mentioned phases.  A more serious difficulty is, however, that the gluon mass identified in \cite{Reinosa:2020skx} appears as a mere gauge-fixing parameter whose value is not determined in terms of $g$. Although it remains yet to be seen how much the correlators computed within this approach are sensitive to this gauge-fixing parameter, this could potentially compromise the comparison to lattice data.

On the other hand, it has been known for some time that dimension two condensates such as $\langle A_\mu^a A_\mu^a\rangle$ can be dynamically generated in the Landau gauge via the local composite operator (LCO) formalism \cite{Verschelde:2001ia} and the minimization of the vacuum energy. A natural question that emerges is then what connection do these condensates bear with the Curci-Ferrari model (or similar approaches), and to which extent do they allow one to reproduce the Landau gauge correlators evaluated on the lattice. The current note aims at investigating these questions. Since our goal is to eventually extend the approach to covariant gauges, we will pay particular attention to the BRST invariance of the construction.

Two dimensional condensates are of course not free of ambiguities due to their composite nature that requires extra renormalization. However, since the correlators of the primordial fields do not depend on this additional subtractions, at least not at an exact level, it is possible to exploit the reduction of coupling technique \cite{Zimmermann:1984sx,Heinemeyer:2019vbc} in order to fix this arbitrariness in some way, as we recall below. In fact, one could envisage using similar ideas to fix the arbitrariness in the choice of the gluon mass in the approach of \cite{Reinosa:2020skx}, at least for the evaluation of physical observables.

\section{The BRST invariant condensate}

\subsection{BRST invariant gauge field}
Our starting point is the Euclidean Yang-Mills action in $\smash{d=4-\epsilon}$ dimensions supplemented with a linear covariant gauge fixing,
\begin{equation}
S_{YM}^{(1)} \!=\!\!\int d^dx\bigg(\frac{1}{4}F^{a}_{\mu\nu}F^{a}_{\mu\nu}\!+
\frac{\alpha}{2}b^2\!+ib^{a}\partial_{\mu}A^{a}_{\mu}
+\bar{c}^{a}\partial_{\mu}D^{ab}_{\mu}c^{b}\bigg).
\label{S_FP}
\end{equation}
with $F_{\mu}^{a}=\partial_{\mu}A_{\nu}^{a}-\partial_{\nu}A_{\mu}^{a}+gf^{a}_{bc}A_{\mu}^{b}A_{\nu}^{c}$ the non-Abelian field-strength tensor, and $D_{\mu}^{ab}=\delta^{ab}\partial_{\mu}+gf^{acb}A_{\mu}^{c}$ the covariant derivative in the adjoint representation. Since our choice is to preserve BRST invariance at each step, we should only consider BRST invariant gluon mass operators, constructed out of a BRST invariant version of the gauge field. To this purpose, we insert into the corresponding path integral a unity $1=\mathcal{N}\int \mathcal{D}\xi\,\mathcal{D}\tau\,\mathcal{D}\bar\eta\,\mathcal{D}\eta\,e^{-S_1}\det(\Lambda(\xi))$, with
\begin{equation}\label{unity}
S_1=\int d^dx \Big(\tau^{a}\partial_{\mu}A^{h,a}_{\mu}+ {\bar \eta}^a\partial_{\mu}D^{ab}_{\mu}(A^h)\eta^{b} \Big)
\end{equation}
and
\begin{equation}\label{eq:lambdadef}
\Lambda_{ab}(\xi)=\frac{2i}{g}\,\text{Tr}\left\{t_{a}\frac{\partial h^{\dagger}}{\partial \xi^{b}}h\right\}\,,
\end{equation}
the appropriate normalization being collected in $\mathcal{N}$. Here, we introduced the local but non-polynomial composite field
\begin{equation}\label{local_Ah}
A^{h}_{\mu} \equiv A^{h,a}_{\mu}\,t^{a}=h^{\dagger}A_{\mu}h+\frac{i}{g}h^{\dagger}\partial_{\mu}h\,,
\end{equation}
with
\begin{equation}
h=e^{ig\xi}=e^{ig\xi^{a} t^{a}}\!,     \label{hxi}
\end{equation}
where the $t^a$ denote the generators of the $su(N)$ algebra and the $\xi^{a}$ are akin to Stueckelberg fields. The fields ${\bar \eta}^a$ and $\eta^a$ are additional (anti-commuting) ghosts that, together with the $\xi$-dependent determinant $\det(\Lambda(\xi))$, account for the Jacobian arising from the change of variables $\xi\to A^h$, which is itself needed in order to treat the functional distribution $\delta(\partial_\mu A^h_\mu)$ that appears after integration over $\tau$. As we will show in Appendix~\ref{app1}, so long as the theory is defined in dimensional regularization, the determinant gives no contribution to the partition function at any fixed order in perturbation theory. For this reason, in what follows we will drop $\det(\Lambda(\xi))$ and write the unity in the form $1=\mathcal{N}\int \mathcal{D}\xi\,\mathcal{D}\tau\,\mathcal{D}\bar\eta\,\mathcal{D}\eta\,e^{-S_1}$.

We stress that, when writing equations such as (\ref{S_FP}) or (\ref{unity}), we are disregarding the presence of Gribov copies. The justification is two-fold. First, in this work, we restrict to perturbation theory for the evaluation of both the correlation functions and the vacuum energy. Second, we will soon check that the dynamically generated condensate is such that the ghost propagator remains positive, a necessary condition for the functional integral to be dominated by configurations within the first Gribov region.\\

The action (\ref{unity}) is here used as a way to treat, within a local setup, a BRST-invariant quantity $A_\mu^h$ that becomes non-local on-shell. The non-local on-shell nature of $A_\mu^h$ becomes explicit after one solves the condition $\partial_\mu A^h_\mu = 0$ iteratively for $\xi$ using
\begin{equation}\label{Ah_expansion}
 (A^{h})^{a}_{\mu}=A^{a}_{\mu}- \partial_{\mu} \xi^a - gf^{abc} A^b_\mu \xi^c -\frac{g}{2}f^{abc}\xi^{b}\partial_\mu \xi^c
 + \ldots
 \end{equation}
Indeed, this leads to the infinite series of terms,
 \begin{eqnarray}
\xi &=&\frac{\partial A}{\partial ^{2}}+i\frac{g}{\partial ^{2}%
}\left[ \partial A,\frac{\partial A}{\partial ^{2}}\right]\nonumber\\
&& +\,i\frac{g}{%
\partial ^{2}}\left[ A_{\mu },\partial _{\mu }\frac{\partial A}{\partial ^{2}%
}\right] +\frac{i}{2}\frac{g}{\partial ^{2}}\left[ \frac{\partial A}{%
\partial ^{2}},\partial A\right] +\ldots  \label{phi0}
\end{eqnarray}
which eventually gives the transverse on-shell expression
 \begin{eqnarray}
A_{\mu }^{h}=\left( \delta _{\mu \nu }-\frac{\partial _{\mu }\partial
_{\nu }}{\partial ^{2}}\right) \phi _{\nu }\;,
\end{eqnarray}
with
\begin{eqnarray}
\phi _{\nu } =A_{\nu }-ig\left[ \frac{\partial A}{\partial ^{2}},A_{\nu
}\right]+\frac{ig}{2}\left[ \frac{\partial A}{\partial ^{2}},\partial
_{\nu }\frac{\partial A}{\partial ^{2}}\right] +\ldots  \label{min0}
\end{eqnarray}
It can be shown that the on-shell expression (\ref{min0}) is gauge/BRST invariant order per order. We will have nothing to say about large gauge transformations. At the level of the off-shell/local formulation (\ref{unity}), the BRST invariance is explicit if one supplements the usual BRST transformation
\begin{equation}
sA^{a}_{\mu}=-D^{ab}_{\mu}c^{b}\!,\, sc^{a}=\frac{g}{2}f^{abc}c^{b}c^{c}\!,\, s\bar{c}^{a}=ib^{a}\!,\, sb^{a}=0\,, \label{brst}
\end{equation}
with the following transformations for the remaining fields:
\begin{equation}
s \tau^a  = 0,\, s {\bar \eta}^a  =  s \eta^a = 0,\,s h^{ij} = -ig c^a (t^a)^{ik} h^{kj} \,.    \label{brst2}
\end{equation}
The BRST transformation remains nilpotent ($\smash{s^2=0}$) over the extended field space and it is immediate to check that
\begin{equation}
s (A^h)^a_\mu = 0\,,
\end{equation}
which is nothing but the infinitesimal version of $\smash{(A^U_\mu)^{U^{-1}h}=(A^{UU^{-1}}_\mu)^h=A^h_\mu}$.

For later purpose, we also mention that $A_\mu^h$ becomes $A_\mu$ in the Landau gauge (at least perturbatively) as it is easily checked from (\ref{min0}) using $\smash{\p_\mu A_\mu=0}$. We refer to e.g.~\cite{DellAntonio:1991mms,Lavelle:1995ty,Capri:2015ixa} for more details, including the connection of the representation \eqref{min0} with the (local) minimization of the $\ell_2$-norm of the gauge field.

\subsection{BRST invariant condensate}
Having gone through this preparatory phase, we are now ready to investigate the vacuum structure of the local action $\smash{S_{YM}^{(2)}\equiv S_{YM}^{(1)}+S_1}$, which is perturbatively equivalent to (\ref{S_FP}). In particular, we shall analyze the possible formation of a BRST-invariant condensate $\langle A^h_\mu A^h_\mu\rangle$. To this purpose, we couple the corresponding gluon mass operator to a source $J$:\footnote{The method presented here was first worked out in \cite{Verschelde:1995jj}, see also \cite{Knecht:2001cc,Verschelde:2001ia} and \cite{Dudal:2002pq}. We will however slightly adapt the discussion to point out some new, yet unnoticed, features, which also facilitate the interpretation.}
\begin{equation}\label{wj}
S^{(2)}_J\equiv S_{YM}^{(2)}+\int d^d x\left(Z_2 \frac{J}{2} A_\mu^h A_\mu^h-(\zeta+\delta \zeta)\mu^{-\epsilon}\frac{J^2}{2}\right).
\end{equation}
Here, we introduced the relevant renormalization factors and counterterms in the $J$-dependent piece of the action, as these will concern us most here,\footnote{The other corresponding factors in $S_{YM}^{(2)}$ are tacitly assumed but not spelled out.} next to the necessary powers of $\mu$ to ensure that $\smash{\dim \zeta=0}$ (for later use, we also note that $\smash{\dim J=2}$). More precisely, starting from bare fields, parameters and sources, which we denote by a subscript $b$, we write
\begin{equation}
J_b A_{\mu,b}^h A_{\mu,b}^h=J Z_J Z_{A_h} A_\mu^h A_\mu^h\equiv Z_2 J A_\mu^h A_\mu^h\,,
\end{equation}
from which we deduce that the renormalization of the source is
\begin{equation}\label{ren_source}
Z_J=\frac{Z_2}{Z_{A_h}}\,.
\end{equation}
Similarly,
\begin{equation}\label{eq:dzeta}
\zeta_b J_b^2 \equiv (\zeta+\delta\zeta)\mu^{-\epsilon} J^2\,.
\end{equation}
The (pure vacuum) counterterm $\propto\delta\zeta J^2$ is necessary to remove the vacuum divergences in the generating functional $W[J]$, with $W[J]=-\ln\int {\cal D}~\text{fields}~e^{-S_J}$, hence its appearance as an additive renormalization. Let us stress that these vacuum divergences do not affect the divergences appearing in correlation functions with at most one insertion of $A_hA_h$. Therefore, we can (and will) consider only renormalization schemes where all renormalization factors, including $Z_2$, are $\zeta$-independent.

In general, from \eqref{eq:dzeta} it follows that
\begin{eqnarray}\label{eq:dzetabis}
\mu\frac{\partial \zeta}{\partial\mu}&=& -2 \gamma_J \zeta + \delta\,,\nonumber\\
\delta&=& (\epsilon-2\gamma_J)\delta\zeta-\mu\frac{\partial \delta\zeta}{\partial\mu}
\end{eqnarray}
where $\mu\frac{\partial J}{\partial\mu}=\gamma_J J$. We immediately discarded terms that vanish in the $\epsilon\to 0^+$ limit. We will recall below that, in principle, it is possible to choose $\delta\zeta$ merely proportional to $\zeta$ such that $\zeta+\delta\zeta= Z_\zeta \zeta$, henceforth implementing multiplicative renormalization also for the vacuum divergences.

We note that from \eqref{eq:dzetabis}, or directly from \eqref{eq:dzeta}, it follows that a finite shift implies that \begin{equation}\label{shift}
\delta\zeta\to \delta\zeta +  \delta\zeta^{fin}  \Rightarrow \zeta\to \zeta-  \delta\zeta^{fin},
\end{equation}
making $\zeta+\delta\zeta$ invariant. The shift $\delta\zeta^{fin}$ can depend on all other variables. The invariance of $\zeta+\delta\zeta$ under such shifts was first discussed in \cite{Knecht:2001cc}.

 From the generating functional $W[J]$, one can introduce the field
 \begin{equation}\label{eq:sigma}
\sigma\equiv\frac{\delta W}{\delta J}=\frac{Z_2}{2}\langle A_\mu^hA_\mu^h\rangle_J-Z_\zeta \zeta \mu^{-\epsilon} J\,
 \end{equation}
 which becomes the argument of the effective action $\Gamma(\sigma)\equiv W(J)-\int d^dx J\sigma$, with $\delta\Gamma/\delta\sigma=-J$. As usual, the benefit of such a construct is that it allows one to access the zero source limit of $\sigma$, and therefore $\langle A_\mu^h A_\mu^h\rangle_{J\to 0}$, from the minimization of $\Gamma[\sigma]$. In particular, a non-trivial minimum is tantamount to the dynamical generation of a condensate $\langle A_\mu^h A_\mu^h\rangle_{J\to 0}\neq 0$. We mention here that the non-positivity of the integration measure associated to the action \eqref{S_FP} could potentially jeopardize the usual relation between the limit of zero sources and the minimization of $\Gamma[\sigma]$. However, we shall later verify that the dynamically generated condensate is such that the ghost propagator remains positive, suggesting that field configurations with negative measure have a subdominant effect and, therefore, that the minimization prescription can be used. Strictly speaking, a similar check should be done with the new ghost fields $\eta$ and $\overline{\eta}$.

 To actually compute the effective action, it is computationally simplest to rely on Jackiw's background field method \cite{Jackiw:1974cv,Peskin:1995ev}. In the present context, this is not easily done a priori due to the coupling of $J$ to a composite operator and to the presence of the quadratic term $\propto J^2$. We can however easily remedy this situation following \cite{Verschelde:1995jj,Verschelde:2001ia}. We insert a unity $1=\mathcal{N}\int\mathcal{D}\sigma~e^{-S_{\sigma}}$, with
\begin{eqnarray}\label{nieuweactie}
  S_\sigma &=&\frac{\mu^\epsilon}{2 Z_\zeta \zeta}\int d^dx\left(\sigma-\frac{Z_2}{2}A^h_\mu A^h_\mu+Z_\zeta \zeta\mu^{-\epsilon}J\right)^2\nonumber\\
  & = & \frac{\mu^\epsilon}{2 Z_\zeta \zeta}\int d^dx\left(\sigma-\frac{Z_2}{2}A^h_\mu A^h_\mu\right)^2\nonumber\\
  & + & \int d^dx\left(J\sigma-Z_2\frac{J}{2}A_\mu^hA_\mu^h+Z_\zeta \zeta \mu^{-\epsilon}\frac{J^2}{2}\right).
\end{eqnarray}
The last two terms of the third line cancel exactly the two $J$-dependent terms in (\ref{wj}) hereby defining a new sourced action
\begin{equation}
S_J^{(3)}=S_{YM}^{(3)}+\int d^dx \,J\sigma\,,
\end{equation}
with
$S_{YM}^{(3)}=S^{(1)}_{YM}+S_{1}+S_2$, and
\begin{equation}
S_2\!=\!\!\frac{\mu^\epsilon}{2 Z_\zeta \zeta}\int d^dx\left(\sigma^2-\!Z_2\,\sigma A^h_\mu A^h_\mu\!+\!\frac{Z_2^2}{4}(A^h_\mu A^h_\mu)^2\right).\label{eq:S2}
\end{equation}
The source now couples linearly to a primary field $\sigma$ (whose expectation value is of course (\ref{eq:sigma})) and the background field method can be implemented as usual. Of course, integrating exactly over $\sigma$, working with $S_{YM}^{(3)}$ will give completely equivalent result as with $S_{YM}^{(2)}$. The situation will only get interesting if the dynamics of the theory would prefer to assign a non-vanishing vacuum expectation value to $\sigma$. This is a possibility that we now investigate. Before doing so, we also notice that the bare action only depends on the combination\footnote{In the more general case where we would not have multiplicative renormalizability of the vacuum divergences, we simply have to replace $Z_\zeta\zeta$ by $\zeta+\delta\zeta$ in \eqref{eq:S2}.} $\zeta+\delta\zeta$, which we already argued to be independent of the finite parts in $\delta\zeta$. Without loss of generality, we can thus renormalize the vacuum divergences in a computationally efficient scheme as $\MSbar$.

To do so, we first notice that, given the BRST invariance of both the action and the mass operator $A_\mu^h A_\mu^h$ as well as the fact that the $\alpha$-dependent part of the action is BRST-exact in the limit of zero sources, the expectation value of $\sigma$ does not depend on $\alpha$ in this limit.\footnote{The argument goes as follows:$$\frac{d}{d\alpha}\langle\sigma\rangle_{J\to 0}\propto -i\int d^dx \langle s\big(A_\mu^{h,a} A_\mu^{h,a} \bar c^d\,b^d\big)\rangle=0\,.$$} Therefore, we can (and will) choose to work in the Landau gauge, $\alpha\to0$, in which case the $(\tau,\xi)$-integration can also be done exactly, leading to the on-shell identification $A^h\to A$ \cite{Capri:2018ijg}. At one-loop order, one obtains
\begin{eqnarray}\label{effpot0}
  V(\sigma) & = & \frac{\mu^{2\epsilon}\sigma^2}{2\zeta}\left(1-\frac{\delta\zeta}{\zeta}\right)\nonumber\\
 & + & (N^2-1)\frac{d-1}{2}\mu^\epsilon\int_q \ln\left(q^2-\frac{\mu^\epsilon\sigma}{\zeta}\right),
  \end{eqnarray}
where, as usual, we have divided the effective action by the space-time volume to compute the effective potential $V(\sigma)$ and we have included an extra factor $\mu^{\epsilon}$ to ensure that $\smash{\mbox{dim}\,V=4}$. We have also treated $\delta Z_2\equiv Z_2-1$ and $\delta Z_\zeta\equiv Z_\zeta-1=\delta\zeta/\zeta$ as higher loop corrections, neglecting them in the one-loop term, and expanding to first order in $\delta\zeta$ in the tree-level term. After explicitly computing the integral and absorbing the divergence in $\delta\zeta$ using the minimal subtraction scheme, one arrives at the expression
\begin{eqnarray}\label{effpot}
  V(m^2) & = & \zeta \frac{m^4}{2}-\frac{3(N^2-1)}{64\pi^2}m^4\left[\ln\frac{\bar\mu^2}{m^2}+\frac{5}{6}\right],
  \end{eqnarray}
where we have defined
\begin{eqnarray}\label{echtemassa}
m^2\equiv-\mu^\epsilon\sigma/\zeta.
\end{eqnarray}

 \subsection{Reduction of couplings}
 Various remarks are in order at this point. First, the trivial solution $\smash{m^2=0}$ is always a maximum of the potential since $\smash{V''(m^2\to 0)=-\infty}$. This  shows that, in the present approach and at the present order of evaluation, a condensate $\langle A_\mu^h A_\mu^h\rangle_{J\to 0}$ is dynamically generated (independently of the value of $\zeta$).

 Beyond this proof of existence, the next pressing question is the size of the condensate. Here, however, we face a serious problem: while the running of $\zeta$ with $\bar\mu$ is entirely fixed from the renormalization factor, its value at a chosen initial scale $\bar\mu_0$ is arbitrary and impacts directly on the size of the condensate. On the other hand, it is easily shown that $\delta W/\delta \zeta\propto \int J^2$. This means that, were we to work exactly, $\zeta$ should not influence any quantity with less than two $J$-derivatives, in the limit of zero sources, or, in other words, at the minimum of the effective action. This includes the correlation functions for the primary fields but also the condensate itself $\langle A_\mu^h A_\mu^h\rangle_{J\to 0}$. Of course, at a given order of approximation, a spurious dependence with respect to $\zeta$ is to be expected but the previous argument shows that, given a prescription for choosing $\zeta$, we can test a priori how the $\zeta$-independence (re)emerges as the approximation is improved. Let us now discuss two possible prescriptions that could be used.

A first possibility would be to impose the $\zeta$-independence of certain quantities such as the condensate or the value of the potential at the minimum. At the present order, this is not very useful because 1) the constraint does not fix any particular value of $\zeta$ and 2) it leads to $m^2=0$ which we have already argued to correspond to an unstable state. It would be interesting to see how this constraint is modified at the next order of approximation. We leave this interesting question for a future investigation.

Another possibility is to follow the prescription of \cite{Verschelde:1995jj,Verschelde:2001ia}. Indeed, despite the presence of two coupling constants, $g^2$ and $\zeta$, $g^2$ runs as usual, that is separately from $\zeta$. Moreover, at an exact level, the value of $\zeta$ does not affect the quantities that we are after, namely the correlation functions for the primary fields. This makes our situation suitable for the Zimmermann reduction of couplings program, \cite{Zimmermann:1984sx}, see also \cite{Heinemeyer:2019vbc} for a recent overview, in which case one coupling (here $\zeta$) is re-expressed as a series in the other (here $g^2$), so that the running of $\zeta$ controlled by $\gamma_\zeta(g^2)$ is then automatically satisfied. This selects one consistent coupling $\zeta(g^2)$ from a whole space of couplings $\zeta$, and it is also the one (unique) choice compatible with multiplicative renormalizability, $\zeta+\delta\zeta=Z_\zeta\delta\zeta$. This approach was also applied to the Gross-Neveu model in \cite{Verschelde:1997jx}, reporting very good agreement with the exactly known mass gap in this toy model.

In the $\MSbar$ scheme, one finds \cite{Verschelde:2001ia,Gracey:2002yt},
\begin{eqnarray}
\zeta(g^2)&=& \frac{N^2-1}{g^2N}\frac{9}{13}+\frac{161}{52}\frac{N^2-1}{16\pi^2}+\ldots\equiv \frac{\zeta_0}{g^2}+\zeta_1+\ldots\nonumber\\
\end{eqnarray}
which is a particular solution of
\begin{eqnarray}
\beta(g^2)\frac{\partial}{\partial g^2}\zeta(g^2)&=& -2\gamma_J(g^2)\zeta(g^2)+\delta(g^2)
\end{eqnarray}
which evidently draws from \eqref{eq:dzetabis}.

From our point of view, this choice is consistent with the perturbative approach followed in this work, where any quantity is assumed to admit an expansion (be it a Laurent expansion) in powers of $g^2$.

Noticing that
\begin{equation}
m^2=-\frac{g^2\mu^\epsilon\sigma}{\zeta_0+g^2\zeta_1}=m_0^2\left(1-g^2\frac{\zeta_1}{\zeta_0}\right),\label{eq:mexp}
\end{equation}
with $m_0^2\equiv -g^2\mu^\epsilon\sigma/\zeta_0$, and expanding to order ${\cal O}(g^0)$, the potential becomes
\begin{eqnarray}\label{eq:Vm0}
& & V(m_0^2)=\left(\frac{\zeta_0}{g^2}-\zeta_1\right)\frac{m^4_0}{2}-\frac{3(N^2-1)}{64\pi^2}m^4_0\left[\ln\frac{\bar\mu^2}{m^2_0}+\frac{5}{6}\right]\nonumber\\
& & \hspace{0.5cm}=\,\frac{9}{13}\frac{N^2-1}{N}\frac{m^4_0}{2g^2}-\frac{3(N^2-1)}{64\pi^2}m^4_0\left[\ln\frac{{\bar\mu}^2}{m^2_0}+\frac{113}{39}\right]\!.\nonumber\\
\end{eqnarray}
As already mentioned above, this potential admits a non-trivial minimum for $m^2_0>0$. Since $m^2_0=-g^2\sigma/\zeta_0$ with $\smash{\zeta_0>0}$, we should then expect $\smash{\sigma<0}$. The sign is compatible with (\ref{eq:sigma}) in the zero-source limit. Indeed, under the assumption that configurations beyond the Gribov horizon are negligible (which we check below), (\ref{eq:sigma}) implies $Z_2\sigma\geq 0$. Moreover, since $Z_2$ diverges negatively \cite{Verschelde:2001ia,Browne:2006uy}, it follows that $\sigma\leq 0$ (and in practice we find $\sigma<0$).

\section{Two-point functions}
Let us now study how the dynamically generated condensate affects the Landau gauge two-point correlation functions. Rewriting the field $\sigma$ in (\ref{eq:S2}) as it vacuum expectation value which we also denote $\sigma$ and a fluctuating part $\delta\sigma$, we find that the free gluon propagator is similar to the one in the Curci-Ferrari model
\begin{equation}
D_{\mu\nu}(p)=\frac{\delta^{ab}P^\perp_{\mu\nu}(p)}{p^2+m^2}\,, \mbox{ with } P^\perp_{\mu\nu}(p)=\delta_{\mu\nu}-\frac{p_\mu p_\nu}{p^2}\,.
\end{equation}
The ghost propagator is simply $G(p)=\delta^{ab}/p^2$, while the $\delta\sigma$ propagator is $\zeta/\mu^\epsilon$. Moreover, in addition to the usual Landau gauge vertices, we have a $AA\delta\sigma$ vertex
\begin{equation}
\frac{\mu^\epsilon}{2\zeta}\,\delta^{ab}\delta_{\mu\nu}\,,\label{eq:AAsigma}
\end{equation}
and a new $AAAA$ vertex
\begin{equation}
-\frac{\mu^\epsilon}{4!\zeta}\Big(\delta^{ab}\delta^{cd}\delta_{\mu\nu}\delta_{\rho\sigma}+\delta^{ac}\delta^{bd}\delta_{\mu\rho}\delta_{\nu\sigma}+\delta^{ad}\delta^{bc}\delta_{\mu\sigma}\delta_{\nu\rho}\Big)\,.\label{eq:AAAA}
\end{equation}
We mention that, strictly speaking, $\zeta^{-1}$, and therefore $m^2$, have a perturbative expansion in terms of $\zeta_0$, $\zeta_1$, $\dots$ which one needs to take into account in order to evaluate the correlation functions at a given order. One could define the Feynman rules in terms of this expanded parameters. However, it is more convenient, and equivalent, to consider the Feynman rules in terms of $\zeta^{-1}$ and $m^2$ and re-expand them only at the end of the calculation, when necessary.

\subsection{Ghost propagator and Gribov horizon}
Using the above derived Feynman rules, we find that the one-loop ghost propagator coincides with the one computed in the CF model. We can use this remark to elucidate the role of the dynamical condensate within the context of Gribov's construction  \cite{Gribov:1977wm} to avoid multiple solutions to the Landau gauge condition.

Following Gribov's original setup \cite{Gribov:1977wm}, the ambiguity related to the presence of Gribov copies is handled by restricting the domain of integration in the functional integral to the Gribov region $\Omega = \{A^a_\mu|  \;  \partial_\mu A^a_\mu =0,  \;    {\mathcal M}^{ab}(A) > 0\}$ where ${\mathcal{M}}^{ab}$ is the Faddeev-Popov operator, ${\mathcal{M}}^{ab}= - \partial_\mu D_{\mu}^{ab}$. As it is apparent from the definition of the region $\Omega$, the ghost propagator $\langle {\bar c}^a c^b \rangle_{p}$, i.e.~the inverse of  ${\mathcal{M}}^{ab}$, remains positive within $\Omega$.  The positiveness of the ghost propagator is thus a necessary condition for any approach to be compatible with a restriction to the Gribov region $\Omega$, and we can check to what extent our gluon mass scale is consistent with such condition.

We parameterize the ghost two-point vertex as
\begin{align}\label{p1}
\Gamma^{(2)}_{\bar cc} &= Z_{c}p^2+\Sigma_{gh}(p^2)=\\
\notag & = p^2\big(1-\sigma_{gh}(p^2)\big)  \;.
\end{align}
where $\Sigma_{gh}(p^2)=-p^{2}[\sigma_{gh}(p^2)+\delta Z_{c}]$, with $\delta Z_{c} = Z_{c}-1$, is the ghost self-energy and we have factored out the trivial color structure $\delta^{ab}$. At one-loop order, one finds~---modulo renormalization---
\begin{equation}
\sigma_{gh}(p^2) = g^2N \frac{p_\mu p_\nu}{p^2} \int \frac{d^dq}{(2\pi)^d} \frac{P^\perp_{\mu\nu}(q)}{(p-q)^2 (q^2+m^2)}  \;, \label{si1}
\end{equation}
where, to the present accuracy, $m^2$ can be expanded to leading order in $g^2$ that is replaced by $m_0^2$ (which we keep denoting $m^2$ for simplicity from here onwards). As $\sigma_{gh}(p^2)$ is a decreasing function of the momentum square $p^2$ \cite{Kroff:2018ncl}, the positivity of the ghost propagator will be ensured by demanding that
$\sigma_{gh}(0) < 1$. Using dimensional regularization in the $\MSbar$ scheme,\footnote{This can be checked against the more general one loop result for $\sigma(p^2)$ derived in \cite[Sect.IIb]{Cucchieri:2016jwg}, by setting there $a=\frac{1}{2}$, $v=m^2$, $w=0$, $b=0$.} one finds
\begin{eqnarray}\label{eq:ghostsigms}
\sigma_{gh}(p^2)&=&\frac{\lambda}{4}\bigg[-\frac{(p^{2}+m^{2})^{3}}{m^{2}p^{4}}\,\ln\left(\frac{p^{2}+m^{2}}{m^{2}}\right)+\\
\notag&&\quad\ \ +\frac{p^{2}}{m^{2}}\,\ln\frac{p^{2}}{m^{2}}+\frac{m^{2}}{p^{2}}-3\ln\frac{m^{2}}{\bar\mu^{2}}+5\bigg],
\end{eqnarray}
where we have reparameterized the coupling by defining
\begin{equation}
\lambda=\frac{Ng^{2}}{16\pi^{2}}\,.
\end{equation}
This leads to
\begin{equation}
\sigma_{gh}(0) = \frac{3\lambda}{4} \left[ \ln\left( \frac{{\bar \mu}^2}{m^2} \right)  +\frac{5}{6} \right], \label{s2}
\end{equation}
in the zero-momentum limit. The positivity condition translates then into
\begin{equation}
m^2 >  \bar\mu^2 \; e^{(\frac{5}{6} - \frac{4}{3 \lambda}   )} \;. \label{mcd}
\end{equation}
On the other hand, the minimum of the potential (\ref{effpot}) is located at
\begin{equation}
m^2 =\bar\mu^2 \; e^{(\frac{187}{78} - \frac{6}{13 \lambda}   )} \;.\label{eq:mofmu}
\end{equation}
It is easily checked that this value obeys the positivity bound (\ref{mcd}) for any value of $\lambda$, clarifying then the role the dynamical condensate plays in relationship to the issue of gauge copies.

\subsection{Gluon propagator}
Similarly, the gluon two-point vertex reads
\begin{eqnarray}
\Gamma^{(2)}_{A_\mu A_\nu} & = & Z_Ap^2P^\perp_{\mu\nu}+(1+\delta Z_2-\delta Z_\zeta)\,m^2\delta_{\mu\nu}+[\Pi^{m^2}_{\mbox{\tiny CF,1$\ell$}}]_{\mu\nu}\nonumber\\
& - & \frac{4\cdot 2}{2!}\left(\frac{\mu^\epsilon}{2\zeta}\right)^2\frac{\zeta}{\mu^\epsilon}\int_Q D_{\mu\nu}(Q)\nonumber\\
& + & \frac{4\cdot 3}{4!}\frac{\mu^\epsilon}{\zeta}\left[\delta_{\mu\nu}\int_Q D_{\rho\rho}^{cc}(Q)+2\int_Q D_{\mu\nu}(Q)\right],\label{eq:g2a}
\end{eqnarray}
where we have again factored out the trivial color structure $\delta^{ab}$ and it is understood that $m^2$ and $\zeta$ need to be expanded to the appropriate order, namely to leading order, except for the tree-level mass term $m^2$ that needs to be expanded to next-to-leading order according to (\ref{eq:mexp}). Up to the combination $Z_2-Z_\zeta$ whose relation to the mass counterterm in the CF model we have not worked out yet, the first line is the one-loop $\Gamma^{(2)}_{AA}$ as computed in the CF model with mass $m^2$. The two other contributions correspond to two new diagrams that arise from the vertices (\ref{eq:AAsigma}) and (\ref{eq:AAAA}).

Now, the terms involving $D_{\mu\nu}$ cancel between the last two lines of (\ref{eq:g2a}) and we are left with
\begin{eqnarray}
\Gamma^{(2)}_{A_\mu A_\nu} & = & Z_Ap^2P^\perp_{\mu\nu}+(1+\delta Z_2-\delta Z_\zeta)\,m^2\delta_{\mu\nu}+[\Pi^{m^2}_{\mbox{\tiny CF,1$\ell$}}]_{\mu\nu}\nonumber\\
& + & (N^2-1)\frac{d-1}{2}\frac{\mu^\epsilon}{\zeta}\delta_{\mu\nu}\int_q\frac{1}{q^2+m^2}\,.\label{eq:aux}
\end{eqnarray}
Writing (\ref{effpot0}) in terms of $m^2$ and setting $V'(m^2)=0$, we find
\begin{equation}
0=(1-\delta Z_\zeta)m^2+(N^2-1)\frac{d-1}{2}\frac{\mu^\epsilon}{\zeta}\int_q\frac{1}{q^2+m^2}\,.
\end{equation}
This implies that
\begin{eqnarray}\label{eq:final}
\Gamma^{(2)}_{A_\mu A_\nu}=Z_Ap^2P^\perp_{\mu\nu}+\delta Z_2\,m^2\delta_{\mu\nu}+[\Pi^{m^2}_{\mbox{\tiny CF,1$\ell$}}]_{\mu\nu}\,.
\end{eqnarray}
For the above equation to make sense, we see that $m^2\delta Z_2$ should correspond to the mass counterterm in the CF model $\smash{\delta m^2_{\mbox{\tiny CF}}=m^2(\delta Z_A+\delta Z^{\mbox{\tiny CF}}_{m^2})}$. We have checked that this is indeed the case using the value for $\delta Z_2$ given in \cite{Verschelde:2001ia}. This is of course no surprise since a constant source $J$ plays exactly the same role as the Curci-Ferrari mass and therefore $\smash{Z_J=Z_{m^2}^{\mbox{\tiny CF}}}$, or, owing to (\ref{ren_source}), $\smash{Z_2=Z_AZ_{m^2}^{\mbox{\tiny CF}}}$.

Interestingly, the constant $\zeta$ disappears from the gluon self-energy, see \eqref{eq:final}, once the latter is
computed on the shell of the gap equation. This could have been foreseen, given that such a
parameter was not present in the Faddeev-Popov action in the first place, see our earlier made comments. Nonetheless, an
implicit dependence on $\zeta$ still survives via the solutions of the mass gap equation.

From (\ref{eq:final}), it is evident that, although the condensate acts as a mass for the tree-level propagator, it disappears from the tree-level term at one-loop order.\footnote{Strictly speaking, the gluon propagator is obtained after considering $\Gamma_{A\delta\sigma}^{(2)}$, $\Gamma_{\delta\sigma A}^{(2)}$ and $\Gamma_{\delta\sigma\delta\sigma}^{(2)}$, and inverting
\begin{equation}\label{eq:inv}
\Gamma_{AA}^{(2)}-\Gamma_{A\delta\sigma}^{(2)}\big[\Gamma_{\delta\sigma\delta\sigma}^{(2)}\big]^{-1}\Gamma_{\delta\sigma A}^{(2)}\,.\nonumber
\end{equation}
Fortunately, using that $\zeta^{-1}\sim g^2$, it is easily shown, however, that $\Gamma_{A\delta\sigma}^{(2)}\sim \Gamma_{\delta\sigma A}^{(2)}\sim g^3$ while $\Gamma_{\delta\sigma\delta\sigma}^{(2)}\sim g^2$ which means that the extra term can be neglected at the present order of accuracy. We also note that $\Gamma_{A\delta\sigma}^{(2)}$ vanishes in the zero-momentum limit from symmetry considerations.} Thus the LCO approach does not lead to the CF model but rather to something more akin to the screened massive approach of \cite{Siringo:2015gia,Siringo:2018uho,Comitini:2020ozt} where the mass disappears from the loops as the order of approximation is increased. Of course, such a reorganization of the perturbative expansion does not necessarily lead to a trivial reformulation of the theory without condensates. In this respect, it would be interesting to push the present approach to the next order and identify quantities that remain constant and different from their corresponding predictions in the absence of condensate. We leave this interesting question for future work. Nonetheless, in Appendix~\ref{app2} we already offer a generic argument as to why the tree level mass indeed should cancel upon using the gap equation.

\subsection{The Dynamically Infrared-Safe (DIS) scheme and the Renormalization Group}
We would now like to compare our results to lattice data. To this purpose, we need to choose a renormalization scheme that is free of Landau poles and that yields a gluon propagator which displays the desired infrared behavior. The renormalization factors we need to fix are $Z_A$, $Z_c$, $Z_\lambda$ and $Z_2$. We will also redefine $Z_{\zeta}$ to better match our scheme, rather than using $\MSbar$ like we did in Sec.~IIB. Since $Z_{\zeta}$ ultimately only affects the solution to the gap equation, we will postpone this to the next section.

To begin with, since we are in the Landau gauge, in order to fix the coupling renormalization $Z_{\lambda}$ we can impose the Taylor condition
\begin{equation}
Z_\lambda Z_A Z_c^2=1\,.\label{eq:Taylor}
\end{equation}
In the Taylor scheme, the running of $\lambda$ is completely determined by the anomalous dimensions $\gamma_{A}$ and $\gamma_{c}$ of the gluon and ghost propagators,
\begin{equation}\label{eq:betalambda}
\beta_{\lambda}=\mu\frac{d\lambda}{d\mu}=-\mu\frac{d\ln Z_{\lambda}}{d\mu}\lambda=\lambda(\gamma_{A}+2\gamma_{c}),
\end{equation}
where
\begin{equation}\label{eq:gammadef}
\gamma_{A}=\frac{d\ln Z_{A}}{d\ln \mu}\ ,\qquad \gamma_{c}=\frac{d\ln Z_{c}}{d\ln \mu}\ .
\end{equation}
$Z_{A}$ and $Z_{c}$ themselves can be defined in the momentum subtraction (MOM) scheme, as is appropriate for a comparison with the lattice data. Namely, if we renormalize the gluon and ghost propagators at the scale $\mu$ so that
\begin{equation}\label{eq:PropCond}
\Gamma_{AA}^{(2)\perp}(p^{2}=\mu^{2})=\Gamma_{\overline{c}c}^{(2)}(p^{2}=\mu^{2})=\mu^{2}
\end{equation}
(here $\Gamma_{AA}^{(2)\perp}$ denotes the transverse component of the gluon two-point vertex), then, going back to \eqref{eq:final} and \eqref{p1},
\begin{align}
Z_{A}&=1-\delta Z_{2}\frac{m^{2}}{\mu^{2}}-\frac{\Pi_{\text{CF},1\ell}^{m^{2}\perp}(p^{2}=\mu^{2})}{\mu^{2}}\,,\label{eq:ZA0}\\
Z_{c}&=1-\frac{\Sigma_{gh}(p^{2}=\mu^{2})}{\mu^{2}}\,,\label{eq:Zc}
\end{align}
where $\Pi_{\text{CF},1\ell}^{m^{2}\perp}$ is the transverse component of the CF gluon polarization \cite{Tissier:2011ey},
\begin{widetext}
\begin{align}\label{eq:glupolcf}
\Pi_{\text{CF},1\ell}^{m^{2}\,\perp}(\mu^{2})&=-\frac{\lambda m^{2}}{6}\left(13t-\frac{9}{2}\right)\left(\frac{2}{\epsilon}+\ln\frac{\overline{\mu}^{2}}{m^{2}}\right)-\frac{\lambda m^{2}}{24t^{2}} \bigg[\frac{242}{3} t^{3}-126 t^{2}+2t+(t^{2}-2)t^{3}\ln t+\\
\notag&\qquad\quad  -2(t+1)^{3}(t^{2}-10t+1)\ln(t+1)-t^{3/2}(t+4)^{3/2}(t^{2}-20t+12)\ln\left(\frac{\sqrt{t+4}-\sqrt{t}}{\sqrt{t+4}+\sqrt{t}}\right)\bigg]\,,
\end{align}
\end{widetext}
with $t=\mu^{2}/m^{2}$. Observe that the first of the conditions in \eqref{eq:PropCond} is especially suitable to our model, which does not have a tree-level mass term in the inverse propagator. Indeed, due to the cancellations which occur in \eqref{eq:aux} as soon as the gap equation is enforced, \eqref{eq:PropCond} is favored over conditions such as $\Gamma_{AA}^{(2)\perp}(\mu^{2})=\mu^{2}+m^{2}$, which is often used for renormalizing the gluon propagator in the CF model \cite{Tissier:2011ey}.

At this stage, $\delta Z_{2}$ in \eqref{eq:ZA0} remains yet undetermined. Before discussing its definition, let us explore how $Z_{2}$ is relevant to renormalization, starting from the gluon mass $m^{2}$. In the present model, we do not really have a mass renormalization factor since $m^2=-g^2\sigma/\zeta_0$ is defined only at the renormalized level. This is at variance with the CF model, where it is found \cite{Tissier:2010ts,Tissier:2011ey,Dudal:2002pq} that $\beta_{m^2}=m^2(\gamma_A+\gamma_c)$. However, we can rewrite $m^2$ in terms of $\mu$-independent bare quantities modulo $\mu$-dependent renormalization factors. Explicitly, we have
\begin{equation}\
m^2=-\frac{g^2\overline{\mu}^\epsilon}{2\zeta_0}Z_2\langle A^2\rangle=-\frac{g^2_b\overline{\mu}^{\epsilon}}{2\zeta_0}\frac{Z_2}{Z_\lambda Z_A}\langle A^2_b\rangle\,.
\end{equation}
It follows that the $\mu$-dependence of $m^2$ is encoded in the prefactor
\begin{equation}\label{eq:massZs}
\frac{Z_2}{Z_\lambda Z_A}=Z_{2}Z_{c}^{2}\,,
\end{equation}
where we have used the Taylor condition, \eqref{eq:Taylor}. In particular, the $\mu$-dependence of $m^2$ is entirely governed by $Z_{2}$ and $Z_{c}$, with the square mass beta function $\beta_{m^{2}}$ reading
\begin{equation}
\beta_{m^2}=\mu\frac{dm^{2}}{d\mu}=m^2 (\gamma_{2}+2\gamma_c)\,.\label{eq:bm}
\end{equation}

Furthermore, as we have seen in Sec.~IIB, $Z_2$ plays the same role as $Z_AZ_{m^2}^{\mbox{\tiny CF}}$ in the CF model. Since the latter enjoys a non-nilpotent BRST symmetry which implies that $Z_{m^2}^{\mbox{\tiny CF}}Z_AZ_c$ is UV finite, we deduce that, in the present model, $Z_2Z_c$ should also be UV finite.\footnote{This can also be seen either by an explicit calculation starting from \eqref{eq:ZA0} and \eqref{eq:Zc}, or from the fact that a constant source $J$ plays the role of the square mass in the Curci-Ferrari model and therefore $Z_J=Z_2/Z_A=Z_{m^2}^{\mbox{\tiny CF}}$.} The most obvious condition to impose for defining $Z_{2}$ would then be $Z_2Z_c=1$ \cite{Dudal:2002pq}. However, there are good reasons to use a slightly different definition.

To see why an alternative condition to $Z_{2}Z_{c}=1$ is appropriate in the present renormalization scheme, it is instructive to compute the low-energy limit of the beta function $\beta_{\lambda}$ while keeping $\delta Z_{2}$ arbitrary. As $\mu^{2}\to 0$, we have---see \eqref{eq:ghostsigms}, \eqref{s2} and \eqref{eq:glupolcf}---
\begin{align}
\Pi_{\text{CF},1\ell}^{m^{2}\perp}(\mu^{2})&=\frac{3\lambda m^{2}}{4}\left(\frac{2}{\epsilon}+\frac{5}{6}+\ln\frac{\overline{\mu}^{2}}{m^{2}}\right)+O(\mu^{2})\,,\\
\Sigma_{gh}(\mu^{2})&=-\frac{3\lambda \mu^{2}}{4}\left(\frac{2}{\epsilon}+\frac{5}{6}+\ln\frac{\overline{\mu}^{2}}{m^{2}}\right)+O(\mu^{4})\,,\label{eq:sigmaghmu0}
\end{align}
where the $O(\mu^{2})$ term in $\Pi_{\text{CF},1\ell}^{m^{2}\perp}$ contains a logarithm of $\mu^{2}$. Plugging these into \eqref{eq:ZA0} and \eqref{eq:Zc}, and using \eqref{eq:gammadef} while neglecting any higher-order term in the renormalization scale and in the coupling, we find that the gluon and ghost anomalous dimensions have the following asymptotic behavior:
\begin{align}
\gamma_{A}&=\left[2\delta Z_{2}+\frac{3\lambda}{2}\left(\frac{2}{\epsilon}+\frac{5}{6}+\ln\frac{\overline{\mu}^{2}}{m^{2}}\right)\right]\frac{m^{2}}{\mu^{2}}+O(\mu^{0})\,,\\
\gamma_{c}&=O(\mu^{2})\,.\label{eq:gammacmu0}
\end{align}
By \eqref{eq:betalambda}, these yield
\begin{equation}
\beta_{\lambda}=\left[2\delta Z_{2}+\frac{3\lambda}{2}\left(\frac{2}{\epsilon}+\frac{5}{6}+\ln\frac{\overline{\mu}^{2}}{m^{2}}\right)\right]\frac{\lambda m^{2}}{\mu^{2}}+O(\mu^{0})\,.
\end{equation}
For most values of $\delta Z_{2}$, subject to the sole condition that $\gamma_{A}$ (or, equivalently, $\beta_{\lambda}$) do not contain divergences, we find that
\begin{equation}
\mu\frac{d\lambda}{d\mu}\propto \frac{\lambda^{2}m^{2}}{\mu^{2}}\qquad(\mu\to0)\,,
\end{equation}
which is equivalent to\footnote{Here we are assuming that $m^{2}$ does not vanish as $\mu\to 0$. As we will see later on, this is indeed the case.}
\begin{equation}
\lambda(\mu)\propto \frac{\mu^{2}}{m^{2}}\to 0\qquad(\mu\to0)\,.
\end{equation}
This behavior is consistent with the one found for the Taylor coupling on the lattice \cite{Duarte:2016iko}. On the other hand, suppose that $\delta Z_{2}=-\delta Z_{c}$, as would be the case to one loop if we decided to choose our renormalization condition according to $Z_{2}Z_{c}=1$. Then we would have, by \eqref{eq:Zc} and \eqref{eq:sigmaghmu0},
\begin{equation}\label{eq:dZ2dZc0}
\delta Z_{2}=-\delta Z_{c}=-\frac{3\lambda}{4}\left(\frac{2}{\epsilon}+\frac{5}{6}+\ln\frac{\overline{\mu}^{2}}{m^{2}}\right)+O(\mu^{2})\,,
\end{equation}
yielding $\beta_{\lambda}\to\text{constant}\times \lambda^{2}$ in the low-energy limit. This is equivalent to $\lambda(\mu)\propto 1/\ln\mu^{2}\to 0$ as $\mu\to 0$.

While having a running coupling that vanishes logarithmically at zero momentum might not be an issue in and of itself, if we insist that $Z_{2}Z_{c}=1$ in the Taylor scheme, then we run into the conclusion that the gluon propagator remains massless at low energy, when improved by the methods of the Renormalization Group (RG). In order to see this, let us first write down the explicit expression for the RG-improved gluon two-point vertex in the present scheme. Denoting by $\Gamma_{AA}^{(2)\perp}(p;\mu)$ the two-point vertex renormalized by minimal subtraction at the scale $\mu$, one has
\begin{equation}
\Gamma_{AA}^{(2)\perp}(p;\mu)=p^{2}\,\exp\left\{-\int_{\mu}^{p}\frac{d\tilde{\mu}}{\tilde{\mu}}\,\gamma_{A}(\tilde{\mu})\right\}\,.
\end{equation}
As already observed in \cite{Dudal:2005na} based on the relation $m^2\propto g^2\Braket{A^2}$, $Z_{2}Z_{c}=1$ implies $\gamma_{2}=-\gamma_{c}$, so that, by \eqref{eq:bm},
\begin{equation}\label{verband}
\beta_{m^{2}}=\gamma_{c}m^{2}\,.
\end{equation}
The gluon anomalous dimension $\gamma_{A}$ can then be re-written as
\begin{equation}
\gamma_{A}=\frac{\beta_{\lambda}}{\lambda}-2\,\frac{\beta_{m^{2}}}{m^{2}}\,,
\end{equation}
where in deriving the above equality we used the Taylor condition. We then compute
\begin{align}
\int_{\mu}^{p}\frac{d\tilde{\mu}}{\tilde{\mu}}\,\gamma_{A}(\tilde{\mu})&=\int_{\mu}^{p}d\tilde{\mu}\left(\frac{1}{\lambda}\frac{d\lambda}{d\tilde{\mu}}-\frac{2}{m^{2}}\frac{dm^{2}}{d\tilde{\mu}}\right)=\\
\notag&=\ln\left(\frac{\lambda(p)}{\lambda(\mu)}\frac{m^{4}(\mu)}{m^{4}(p)}\right)\,.
\end{align}
Our final expression for the RG-improved gluon two-point vertex therefore reads
\begin{equation}\label{eq:gluonrg}
\Gamma_{AA}^{(2)\perp}(p;\mu)=p^{2}\frac{\lambda(\mu)}{\lambda(p)}\frac{m^{4}(p)}{m^{4}(\mu)}\ .
\end{equation}
We remark that this expression is valid in any scheme in which $\gamma_{2}=-\gamma_{c}$ holds together with the Taylor condition.

Now, since according to \eqref{eq:gammacmu0} (and by dimensional counting)
\begin{equation}
\gamma_{c}(\mu)\propto\lambda\,\frac{\mu^{2}}{m^{2}}\qquad(\mu\to 0)\,,
\end{equation}
using $\beta_{m^{2}}=\gamma_{c}m^{2}$, we find that
\begin{equation}
\mu\frac{dm^{2}}{d\mu}\propto \lambda \mu^{2}\qquad(\mu \to 0)\,.
\end{equation}
Provided that the running coupling vanishes at zero momentum (whether logarithmically or quadratically is irrelevant), the above equation implies that $m^{2}(\mu)$ saturates to a constant in the low-energy limit. Therefore, as $p\to 0$, we find that
\begin{align}\label{eq:glueasymp0}
\Gamma_{AA}^{(2)\perp}(p;\mu)\sim \frac{p^{2}}{\lambda(p)}\qquad(p\to 0)\,.
\end{align}
We now see why having a logarithmically vanishing coupling spoils the infrared behavior of the RG-improved gluon propagator: if $\lambda(p)\sim1/\ln p^{2}$, then $\Gamma_{AA}^{(2)\perp}(p)\sim p^{2}\ln p^{2}$, implying that the propagator diverges at $p=0$.

Ultimately, the reason why we found this divergence is that the specific choice $Z_{2}Z_{c}=1$, which to one loop is equivalent to $\delta Z_{2}=-\delta Z_{c}$, kills the deep-infrared $m^{2}/\mu^{2}$ term in the beta function $\beta_{\lambda}$. Using such a condition, however, is not at all forced on us by any consistency requirement: so long as the divergences in $\delta Z_{2}$ and $\delta Z_{c}$ are taken to be equal with an opposite sign, we are free to choose the finite part of $Z_{2}$ according to the needs of our scheme. Therefore, instead of picking the single value that yields a diverging propagator, we decide to use a slightly different renormalization condition for $\delta Z_{2}$, namely
\begin{equation}\label{eq:dZ2dZc}
\delta Z_{2} = -\delta Z_{c}+\lim_{\mu\to 0}[\delta Z_{c}]_{\text{fin.}}\,,
\end{equation}
where $[\delta Z_{c}]_{\text{fin.}}$ is the finite part of $\delta Z_{c}$. In other words, we start from $\delta Z_{c}$ and subtract from it its zero-energy finite term. Note that in the MOM scheme any $\overline{\mu}$-dependent term must actually be regarded as part of the divergence, since $\overline{\mu}$ is an arbitrary scale introduced by dimensional regularization which needs to be absorbed by renormalization.

Going back to \eqref{eq:dZ2dZc0}, we see that, to leading order, the above condition is equivalent to
\begin{equation}
\delta Z_{2} = -\delta Z_{c}+\frac{5\lambda}{8}\quad\text{or}\quad Z_{2}Z_{c}=1+\frac{5\lambda}{8}\,.
\end{equation}
Immediately, we find that in the scheme defined by \eqref{eq:dZ2dZc}, which henceforth we will refer to as the Dinamically Infrared-Safe (DIS) scheme, the beta function $\beta_{\lambda}$ regains its infrared $m^{2}/\mu^{2}$ term:
\begin{equation}
\beta_{\lambda}=\frac{5\lambda^{2}}{4}\frac{m^{2}}{\mu^{2}}+O(\mu^{2})\,.
\end{equation}
As discussed earlier, this implies that $\lambda$ vanishes quadratically as $\mu\to 0$. Furthermore, since the difference between $\delta Z_{2}$ and $-\delta Z_{c}$ is a $\mu$-independent constant, the relation $\gamma_{2}=-\gamma_{c}$ still holds despite being $Z_{2}Z_{c}\neq 1$. Therefore, the expression in \eqref{eq:gluonrg} for the RG-improved gluon two-point vertex is still valid in the DIS scheme, as well as the asymptotic behavior described by \eqref{eq:glueasymp0}. However, this time $\lambda(p)\sim p^{2}$, so that
\begin{align}
\Gamma_{AA}^{(2)\perp}(p;\mu)\to \text{constant}\qquad (p\to 0)
\end{align}
and the propagator saturates to a constant at zero momentum, as it should be.

In the present scheme, the gluon and ghost anomalous dimensions $\gamma_{A}$ and $\gamma_{c}$ read
\begin{widetext}
\begin{align}\label{eq:gammaAfin}
\gamma_{A}&=-\frac{\lambda}{6t^{3}} \bigg[17 t^3-\frac{163}{2} t^2+12t-t^5 \ln t+(2t-3) (t-2)^2(t+1)^2 \ln (t+1)+\\
\notag&\qquad\qquad\qquad +t^{3/2}\sqrt{t+4} \left(t^3-9 t^2+20t-36\right) \ln\left(\frac{\sqrt{t+4}-\sqrt{t}}{\sqrt{t+4}+\sqrt{t}}\right)\bigg]\,,\label{eq:gammacfin}\\
\gamma_{c}&=-\frac{\lambda}{2t^{2}} \left[2t^{2}+2t-t^3 \ln t+(t-2)(t+1)^{2} \ln (t+1)\right]\,,
\end{align}
\end{widetext}
where $t=\mu^{2}/m^{2}$. In the UV, the ordinary Yang-Mills behavior, namely
\begin{align}
\gamma_{A}\to-\frac{13\lambda}{3}\,,\qquad\gamma_{c}\to -\frac{3\lambda}{2}\qquad(\mu\to \infty)\,,
\end{align}
is recovered. As we will see, the RG-improved gluon and ghost propagators computed from these anomalous dimensions are in very good agreement with the lattice data over a wide range of momenta.

Finally, let us write down an expression analogous to \eqref{eq:gluonrg} for the RG-improved ghost two-point vertex. In general, the two-point vertex $\Gamma_{\overline{c}c}^{(2)}(p;\mu)$ renormalized by minimal subtraction at the scale $\mu$ can be written as
\begin{equation}
\Gamma_{\overline{c}c}^{(2)}(p;\mu)=p^{2}\,\exp\left\{-\int_{\mu}^{p}\frac{d\tilde{\mu}}{\tilde{\mu}}\,\gamma_{c}(\tilde{\mu})\right\}\,.
\end{equation}
Since in our scheme $\gamma_{c}=\beta_{m^{2}}/m^{2}$, we compute
\begin{align}
\int_{\mu}^{p}\frac{d\tilde{\mu}}{\tilde{\mu}}\,\gamma_{c}(\tilde{\mu})&=\int_{\mu}^{p}d\tilde{\mu}\,\frac{1}{m^{2}}\frac{dm^{2}}{d\tilde{\mu}}=\\
\notag&=\ln\left(\frac{m^{2}(p)}{m^{2}(\mu)}\right)\,.
\end{align}
In particular, the ghost two-point vertex can be expressed in terms of the running gluon mass squared $m^{2}(p)$ as
\begin{equation}
\Gamma_{\overline{c}c}^{(2)}(p;\mu)=p^{2}\frac{m^{2}(\mu)}{m^{2}(p)}\,.
\end{equation}
The above equation tells us that the ghost propagator diverges like $1/p^{2}$ at zero momentum, and that the ghost form factor---as a function of momentum---has the same behavior as the running gluon mass squared, being essentially equal to it modulo a constant factor of $m^{-2}(\mu)$.

In the next section, we will redefine the gap equation in order to derive suitable initial conditions for the RG flow of the theory. These will be used to compute the RG-improved gluon and ghost propagators, which we will then compare with the lattice data in a later section.

\subsection{Gap equation revisited and the RG flow}
The gap equation allows us to fix the initial value of the gluon mass $m^{2}$ in the RG flow starting from the initial renormalization scale $\mu$ and coupling $\lambda$. Analogously as for the propagators, we will benefit from the RG invariance of the effective potential \eqref{eq:Vm0} to optimize it. We can first work in the $\MSbar$ scheme to do so. As per construction $\omu\frac{d}{d\omu} V=0$, or,
\begin{equation}\label{LL1}
  \left(\omu \frac{\p}{\p \omu}+\beta(g^2)\frac{\p}{\p g^2}+\beta_{m^2}(g^2)\frac{\p}{\p m^2}\right)V=0,
\end{equation}
we can use this RG equation to resum all leading logs. Therefore, we set
\begin{equation}\label{LL2}
  V_{LL}(m)=\frac{9}{13} \frac{N^2-1}{N}\frac{m^4}{2g^2}\sum_{n=0}^{\infty} v_n u^n,
\end{equation}
with $v_0=1$ and $u=g^2 \ln (m^2/\omu^2)$. Setting $\beta(g^2)=-2g^2\sum_{n=0}^\infty \beta_n (g^2)^{n+1}$, $\beta_{m^2}(g^2)/m^2= g^2\sum_{n=0}^\infty \gamma_n (g^2)^{n}$, we get from \eqref{LL1} at leading order
\begin{equation}\label{LL3}
  (\gamma_0+\beta_0)\sum_{n=0}^{\infty} v_n u^n -(\beta_0u+1)\sum_{n=0}^{\infty} (n+1) v_{n+1}u^n=0.
\end{equation}
This can be rephrased as
\begin{equation}\label{LL4}
(\gamma_0+\beta_0)F(u) - (\beta_0u+1)F'(u)=0
\end{equation}
with $F(u)=\sum_{n=0}^{\infty} v_n u^n$. Using $v_0=1$, the solution reads
\begin{equation}\label{LL5}
F(u)=(1+\beta_0 u)^{1+\gamma_0/\beta_0}
\end{equation}
and the optimized potential becomes
\begin{eqnarray}\label{LL6}
  V_{LL}(m)&=&\\&& \hspace{-1cm}\frac{9}{13} \frac{N^2-1}{N}\frac{m^4(\omu)}{2g^2(\omu)} \left(1+\beta_0g^2(\omu)\ln\frac{m^2(\omu)}{\omu^2}\right)^{1+\gamma_0/\beta_0}.\nonumber
\end{eqnarray}
The above RG methodology was borrowed from \cite{Kastening:1991gv}, see also \cite{Bando:1992np} for further interesting comments about RG log resummations. In Appendix~\ref{app3}, we have taken a closer look at the effective potential.

It can be easily seen that at this order, the last factor amounts to replacing $\frac{m^4(\omu)}{2g^2(\omu)}\to \frac{m^4(m)}{2g^2(m)}$, clearly showing potentially large logs are resummed, and thereby also establishing the explicit $\omu$-independence of the improved effective potential, see also \cite{Bando:1992np}.

Expressing everything in terms of $\lambda$, we obtain as solution of $\frac{\p V_{LL}}{\p m} = 0$ in the $\MSbar$ scheme
\begin{equation}\label{eq:gapsol}
m_{\text{sol.}, \MSbartiny}(\omu)=\omu e^{-\frac{13}{88}-\frac{3}{22\lambda_{\MSbartiny}(\omu)}}\,,
\end{equation}
as $\beta_0 = \frac{11N}{3(16\pi^2)}$, $\gamma_0=-\frac{3N}{2(16\pi^2)}$, see e.g.~\cite{Gracey:2002yt} and \eqref{verband}. For later usage in Appendix~\ref{app3}, we already mention $\beta_1 = \frac{34N^2}{3(16\pi^2)^2}$, $\gamma_1=-\frac{95N^2}{24(16\pi^2)^2}$.

In order to make use of this solution in the DIS scheme, we first need to perform a scheme conversion from $\MSbar$ to DIS, using the appropriate renormalization factors. Going back to \eqref{eq:Taylor} and \eqref{eq:massZs} we see that, at a fixed renormalization scale,
\begin{align}\label{conversion}
m^{2}_{\DIS}&=\frac{Z_{2,\DIS}Z_{c,\DIS}^{2}}{Z_{2, \MSbartiny}Z_{c,\MSbartiny}^{2}}\,m^{2}_{\MSbartiny}\,,\\
\lambda_{\DIS}&=\frac{Z_{A,\DIS}Z_{c,\DIS}^{2}}{Z_{A, \MSbartiny}Z_{c, \MSbartiny}^{2}}\,\lambda_{\MSbartiny}\,,
\end{align}
or, to one loop, since $Z_{2, \MSbartiny}Z_{c, \MSbartiny}=1$,
\begin{align}
m^{2}_{\DIS}&=\left[1+\frac{5\lambda}{8}+(\delta Z_{c,\DIS}-\delta Z_{c, \MSbartiny})\right]\,m^{2}_{\MSbartiny}\,,\label{eq:massMStoDIS}\\
\lambda_{\DIS}&=\Big[1+(\delta Z_{A,\DIS}-\delta Z_{A, \MSbartiny})+\label{eq:lambdaMStoDIS}\\
\notag&\qquad\quad+2(\delta Z_{c,\DIS}-\delta Z_{c, \MSbartiny})\Big]\,\lambda_{\MSbartiny}\,.
\end{align}
Of course, to lowest order, the mass and coupling on which the renormalization factors depend can be computed in any of the two schemes.

\begin{figure}
\includegraphics[width=0.39\textwidth]{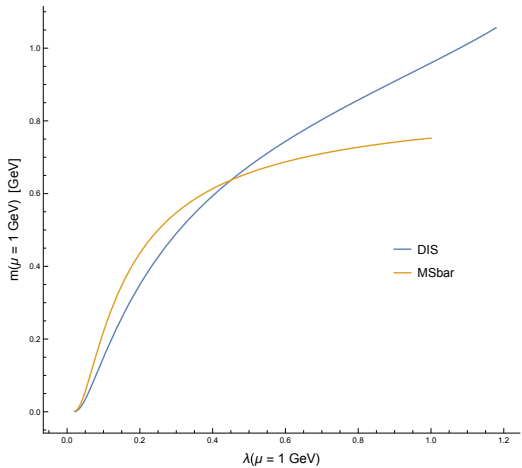}
\caption{Solutions of the gap equation in the DIS and $\MSbar$ schemes at the renormalization scale $\mu_{0}=1$~GeV.}
\label{fig:gapeq}
\end{figure}

In what follows, we will fix the renormalization scale to be equal to $\mu_{0}=1$~GeV---i.e., the scale at which we start the renormalization group flow---and use $\lambda_{\DIS}(\mu_{0})$ as our independent variable. Then we will compute $\lambda_{\MSbartiny}(\mu_{0})$ as a function of $\lambda_{\DIS}(\mu_{0})$ using \eqref{eq:lambdaMStoDIS}, plug the result into \eqref{eq:gapsol} to obtain the solution of the $\MSbar$ resummed gap equation, and finally convert the solution to the DIS scheme using \eqref{eq:massMStoDIS}. Doing so yields the DIS mass $m^{2}_{\DIS}(\mu_{0})$ corresponding to the DIS coupling $\lambda_{\DIS}(\mu_{0})$ by virtue of the $\MSbar$ gap equation. In Fig.~\ref{fig:gapeq} we show the solutions of the gap equation, both in the $\MSbar$ and in the DIS scheme, at the scale $\mu_{0}=1$~GeV. As we can see, the conversion modifies only slightly the relation between the mass and coupling, at least for not so large values of $\lambda$.

\begin{figure}
\includegraphics[width=0.43\textwidth]{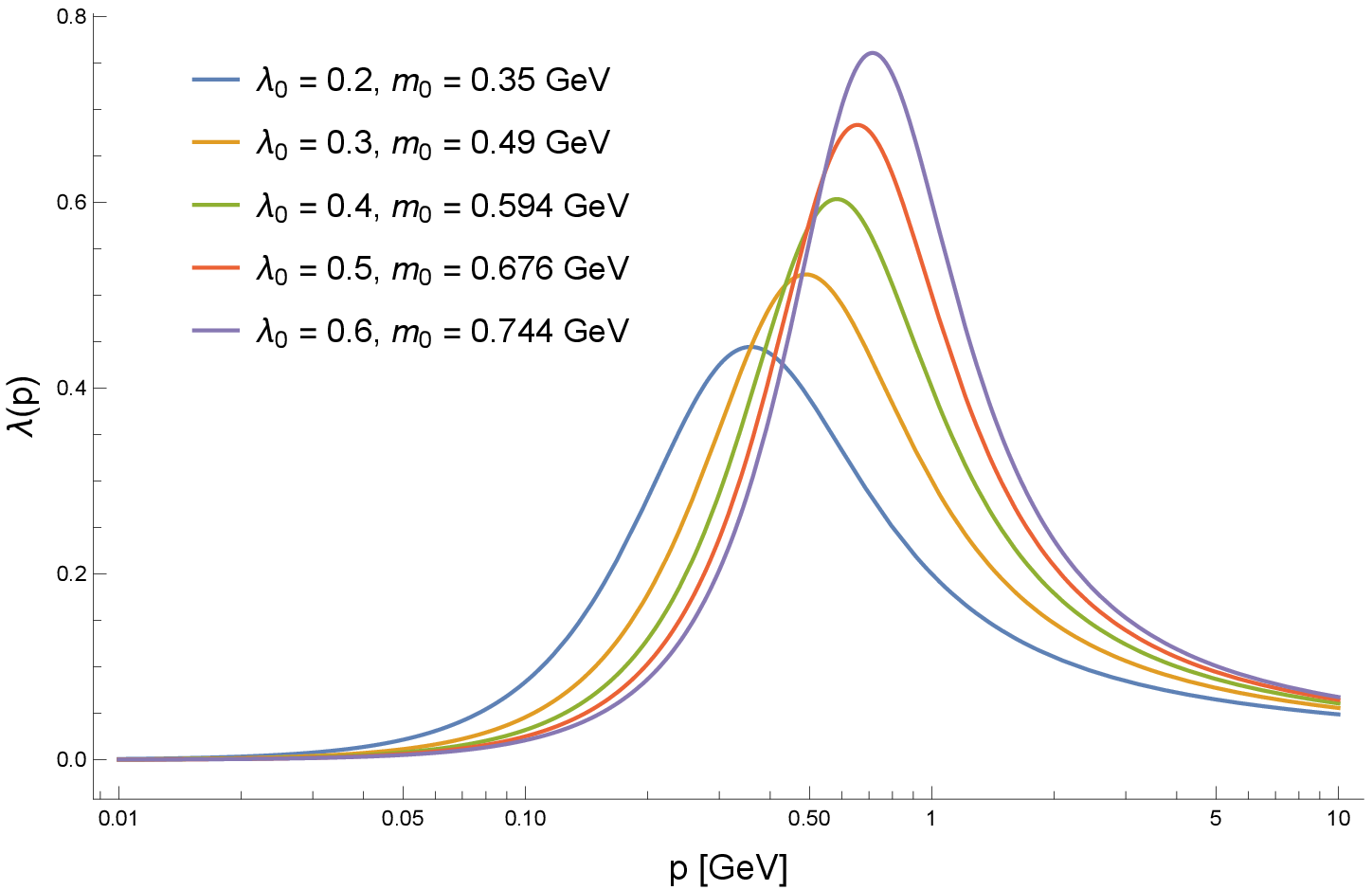}
\vskip 10pt
\includegraphics[width=0.43\textwidth]{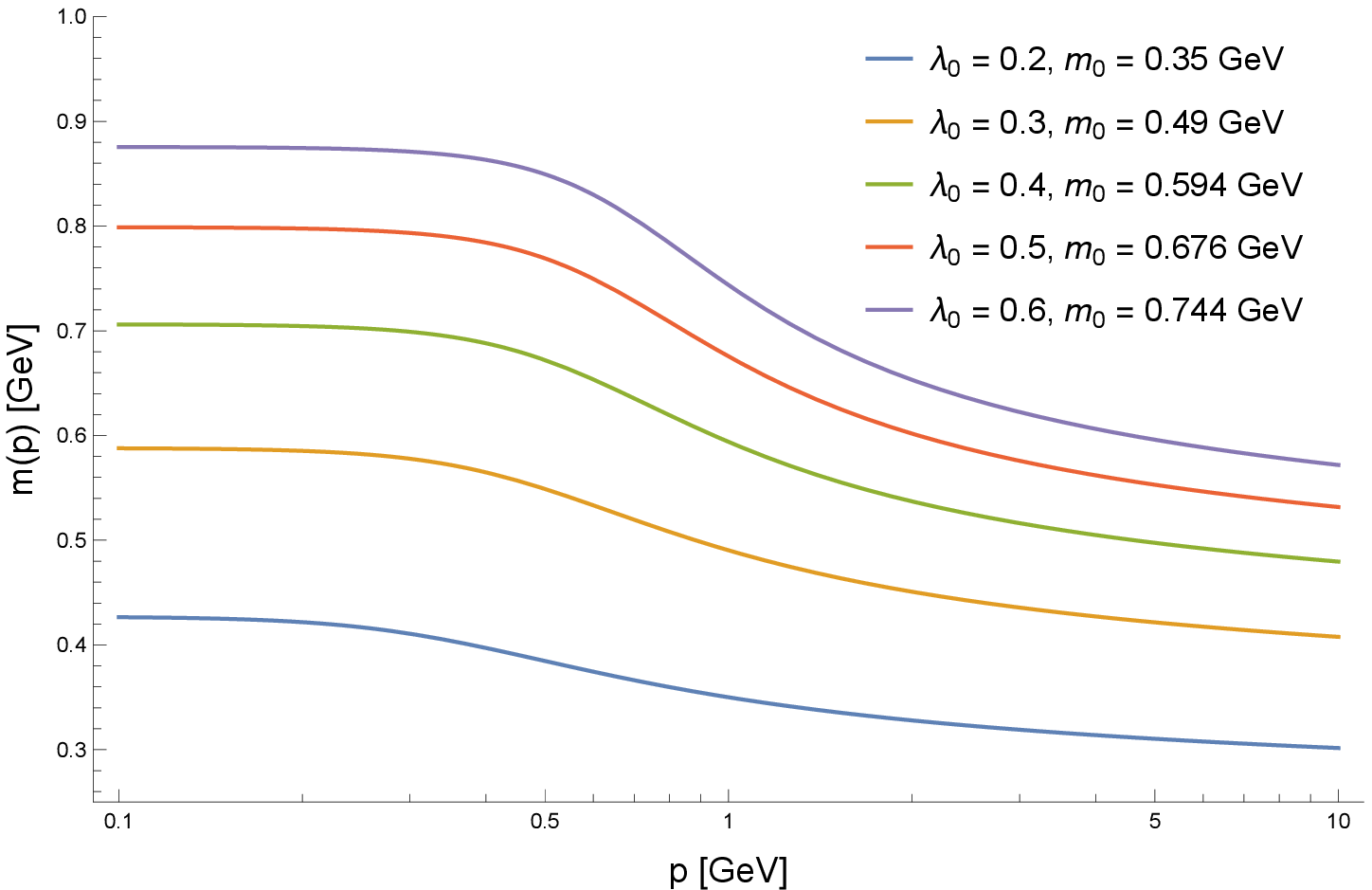}
\caption{DIS running coupling (top) and mass (bottom) for different values of the coupling $\lambda_{0}$ at the initial scale $\mu_{0}=1$~GeV, with the corresponding mass $m_{0}$ computed by solving the gap equation.}
\label{fig:running}
\end{figure}

Dropping the subscript for the DIS quantities, with the DIS $m_{0}=m(\mu_{0})=m_{\text{sol}.}$ computed at the initial scale $\mu_{0}$ as a function of the DIS coupling $\lambda_{0}=\lambda(\mu_{0})$ as detailed above, in Fig.~\ref{fig:running} we display some of the solutions to the beta-function equations in the DIS scheme. As we can see, the running coupling has no infrared Landau pole\footnote{We checked that this behavior persists to very large values of the coupling $\lambda_{0}$ at the initial scale $\mu_{0}=1$~GeV. Indeed, we could not find any value of $\lambda_{0}$ for which the running coupling diverges.} and vanishes like $p^{2}$ as $p\to 0$, as we anticipated in the last section. The absence of Landau poles in the running coupling is a necessary condition for the self-consistency of any perturbative approach to Quantum Chromodynamics, and one which is shared by most of the models which treat the gluons as massive---see e.g.~\cite{Pelaez:2021tpq,Comitini:2020ozt}. Indeed, it is precisely the existence of a gluon mass scale that makes it possible for the running coupling to change its behavior and decrease as the momentum decreases. At large energies, since $\beta_{\lambda}\to -\frac{22\lambda^{2}}{3}$ as $\mu\to\infty$ by \eqref{eq:gammaAfin} and \eqref{eq:gammacfin}, the coupling runs just like in ordinary Yang-Mills theory.

Finally, at zero momentum the running gluon mass saturates to a constant $m(0)$, again in agreement with our analysis of Sec.~IIIC. Since $\beta_{m^{2}}<0$, with $m_{0}=m_{\text{sol.}}$ an increasing function of $\lambda_{0}$, $m(0)$ increases with the initial value of the coupling. It attains the expected order of magnitude when $\lambda_{0}\gtrsim 0.2-0.3$, which corresponds to values of the coupling $\lambda\gtrsim0.4-0.5$ at the peak (equivalently, $\alpha_{s}\gtrsim 1.7-2.1$). In the UV, we find that
\begin{equation}
\beta_{m^{2}}\to-\frac{3\lambda m^{2}}{2}\qquad(\mu\to \infty)\,,
\end{equation}
yielding a mass that decreases like $\lambda^{\frac{9}{44}}(\mu)\sim[\ln(\mu^{2})]^{-\frac{9}{44}}$ at large energies, thereby restoring the massless, asymptotically free, UV limit.

\subsection{Comparison with the lattice data}

\begin{figure}
\includegraphics[width=0.43\textwidth]{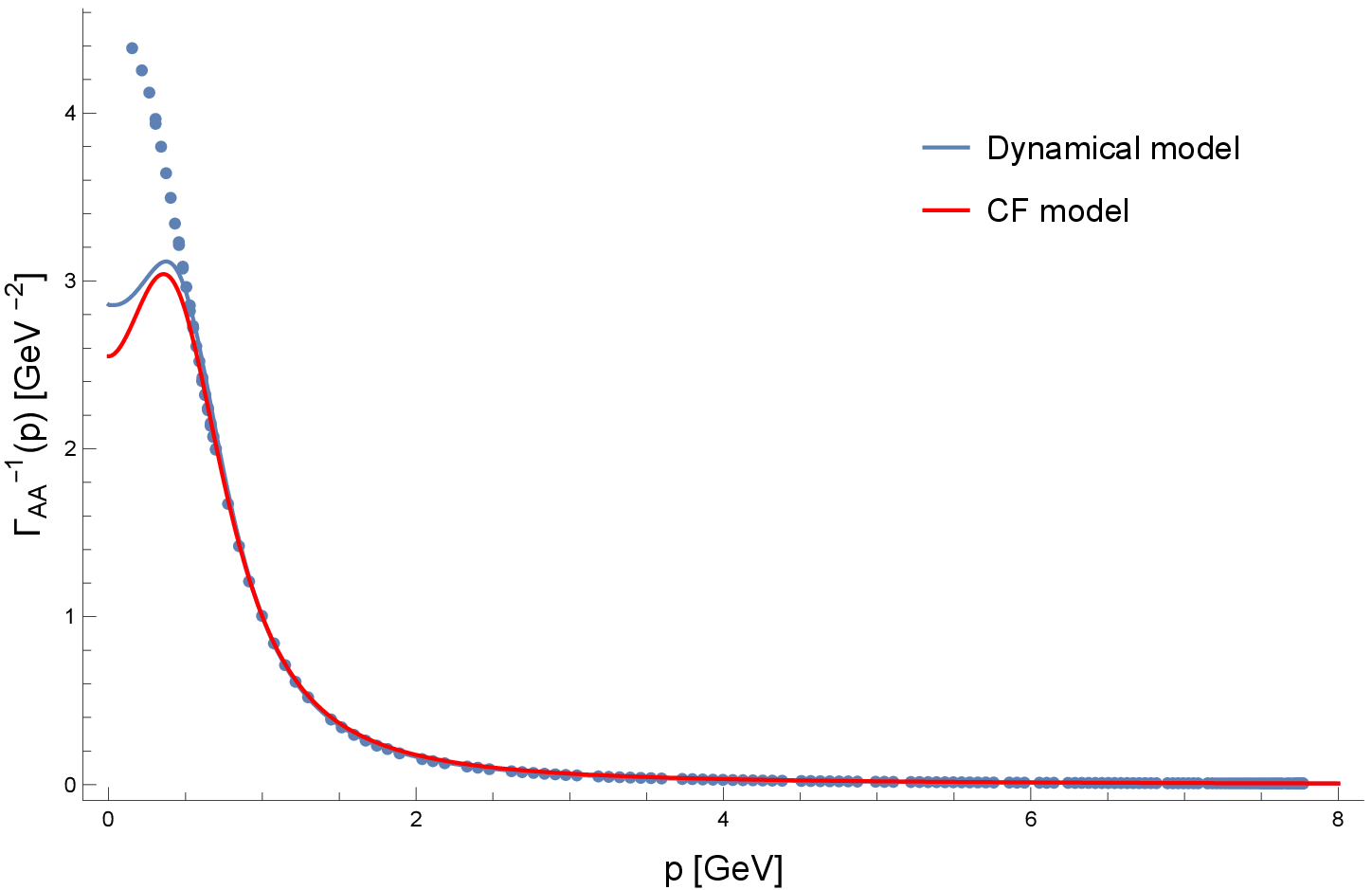}
\includegraphics[width=0.43\textwidth]{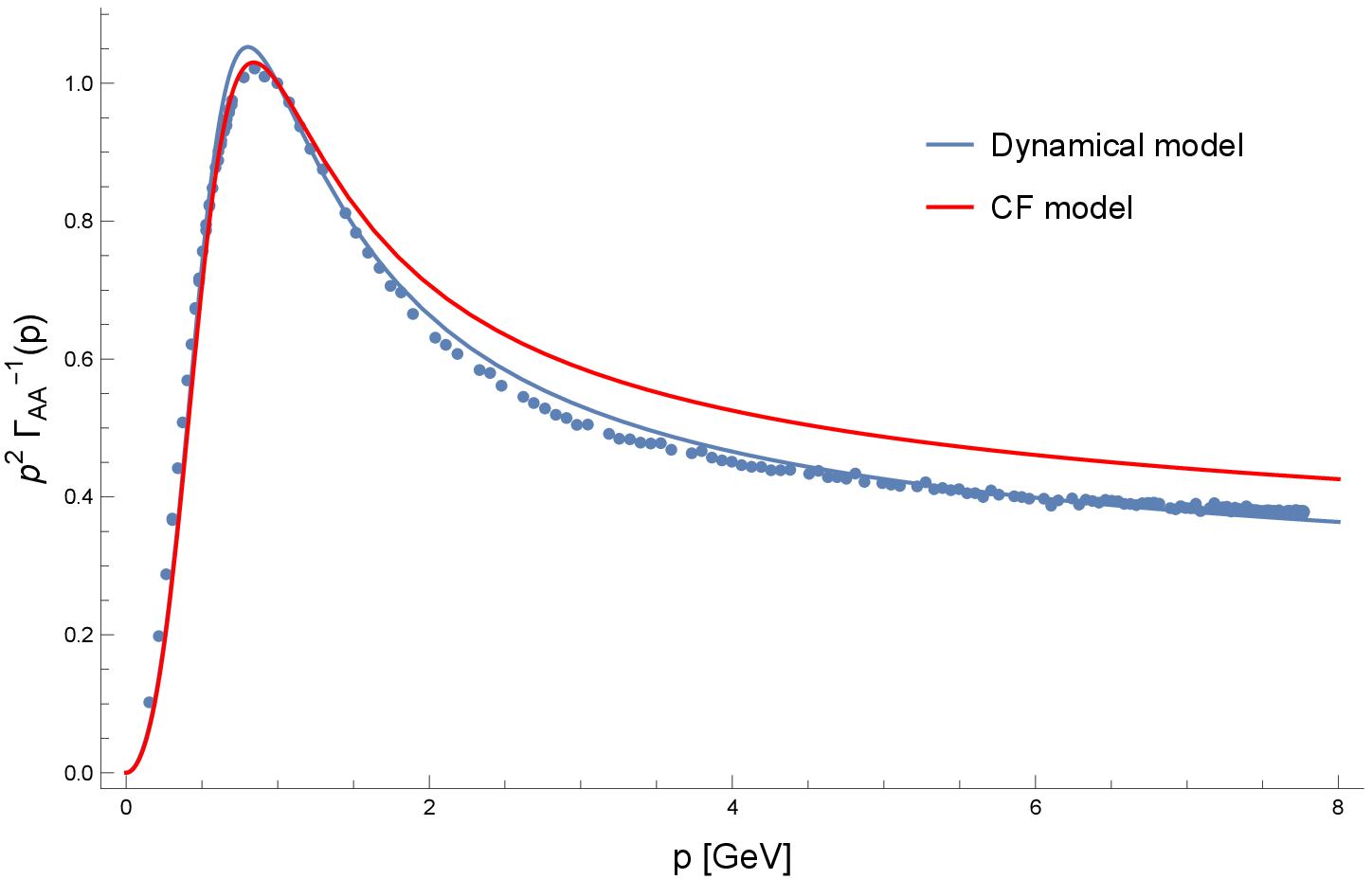}
\includegraphics[width=0.43\textwidth]{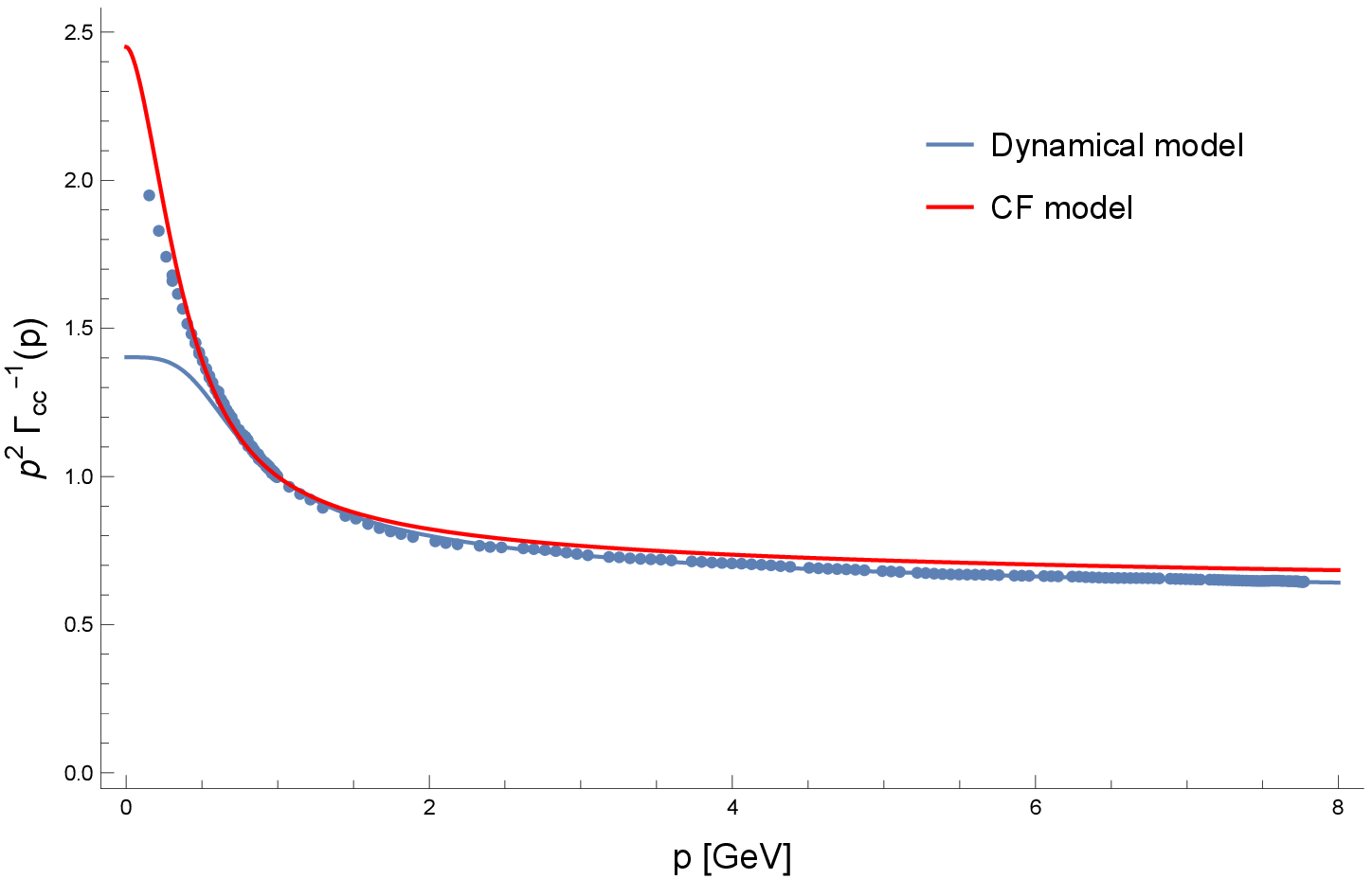}
\caption{RG-improved gluon propagator (top), gluon form factor (middle) and ghost form factor (bottom) in the DIS scheme, renormalized at $\mu_{0}=1$~GeV, together with the lattice data of \cite{Duarte:2016iko} and analogous CF model results (red curves) for comparison, see the text for details. The initial value of the coupling $\lambda_{0}=0.473$ was obtained by fitting the combined lattice data for the gluon and ghost form factors. The initial value of the gluon mass $m_{0}=0.655$~GeV was computed by using the gap equation as described in Sec.~IIID.}
\label{fig:latticecomp}
\end{figure}

We are now in a position to compare the predictions of our model with the lattice data, more precisely the $80^4, \beta=6.0$ gluon and ghost data sets of \cite{Duarte:2016iko}, see also \cite{Dudal:2018cli}. In Fig.~\ref{fig:latticecomp} we show the RG-improved gluon propagator, gluon form factor and ghost form factor renormalized at the scale $\mu_{0}=1$~GeV. In order to obtain the two-point functions, we fitted the initial value of the coupling\footnote{In order to simplify the numerical calculations, we used the $\overline{\text{MS}}$ coupling $\lambda_{\MSbartiny}(\mu_{0})$ at $\mu_{0}$ as the fit parameter, and then obtained the DIS scheme $\lambda_{0}$ using Eqs.~\eqref{eq:gapsol} and \eqref{eq:lambdaMStoDIS}.} by the least-squares method using the combined lattice data of \cite{Duarte:2016iko} for the gluon and ghost form factors, with the gluon mass $m_{0}=m_{\text{sol.}}$ computed from the gap equation as a dependent input, as described in the previous section. The fit yielded a value of $\lambda_{0}=0.473$, which corresponds to $\alpha_{s}(\mu_{0})=1.981$, $m_{0}=0.655$~GeV, and $\lambda_{\MSbartiny}(\mu_{0})=0.316$.

Both the form factors turn out to be in very good agreement with the lattice data at moderate to high energies---up to $8$~GeV---, despite the gluon falling slightly below the lattice in the UV, and slightly above it at intermediate energies. On the other hand, at momenta $p\lesssim0.5$~GeV, the RG-improved functions are suppressed with respect to their lattice analogues. This behavior is not unseen at one loop in massive expansions of Yang-Mills theory \cite{Comitini:2020}, at least as far as the gluon is concerned. Indeed, it is shown by the CF model itself, when the latter is fitted to the lattice data by the same procedure used for our fit\footnote{\label{CFfoot}We should remark that the CF fit shown in Fig.~\ref{fig:latticecomp} is not part of the original studies on the subject, but rather it was made anew starting from the combined gluon and ghost lattice form factors of \cite{Duarte:2016iko} for the purpose of comparison. In more detail, the RG-improved CF functions were computed in the so-called infrared-safe scheme \cite{Tissier:2011ey}, rescaled so that $\Gamma_{AA}^{(2)\perp}(p)/p^{2}=1$ at $p=1$~GeV, like in our DIS scheme. The rescaling factor could as well be determined by leaving it as a free parameter of the fit, in which case the CF propagators would significantly improve---especially in the UV. However, we chose not to do so, in order to make the comparison to our results more immediate.}. Previous studies of the CF model \cite{Gracey:2019xom} suggest that this behavior could improve by going to two loops.

\begin{figure}
\includegraphics[width=0.43\textwidth]{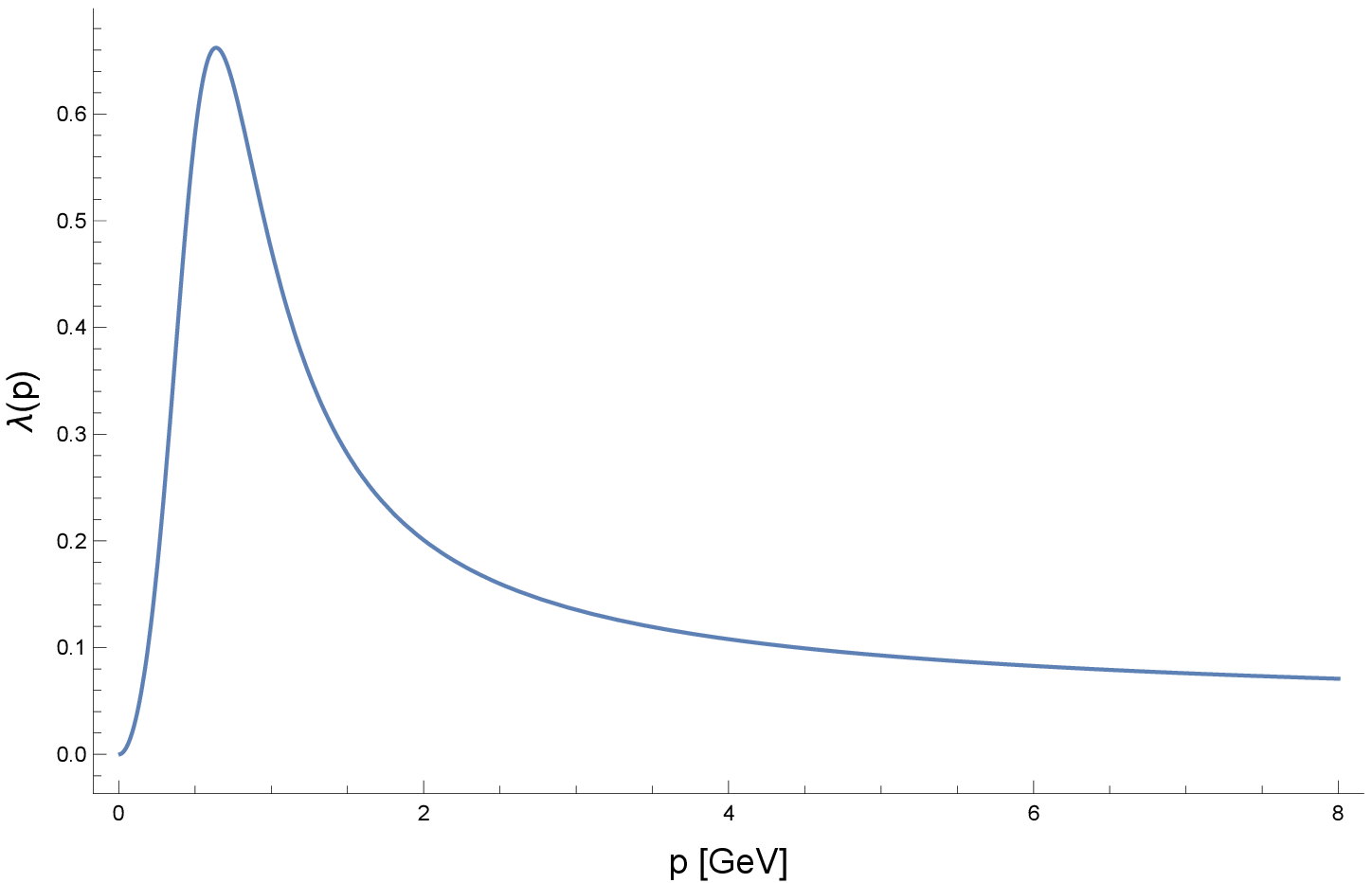}
\vskip 10pt
\includegraphics[width=0.43\textwidth]{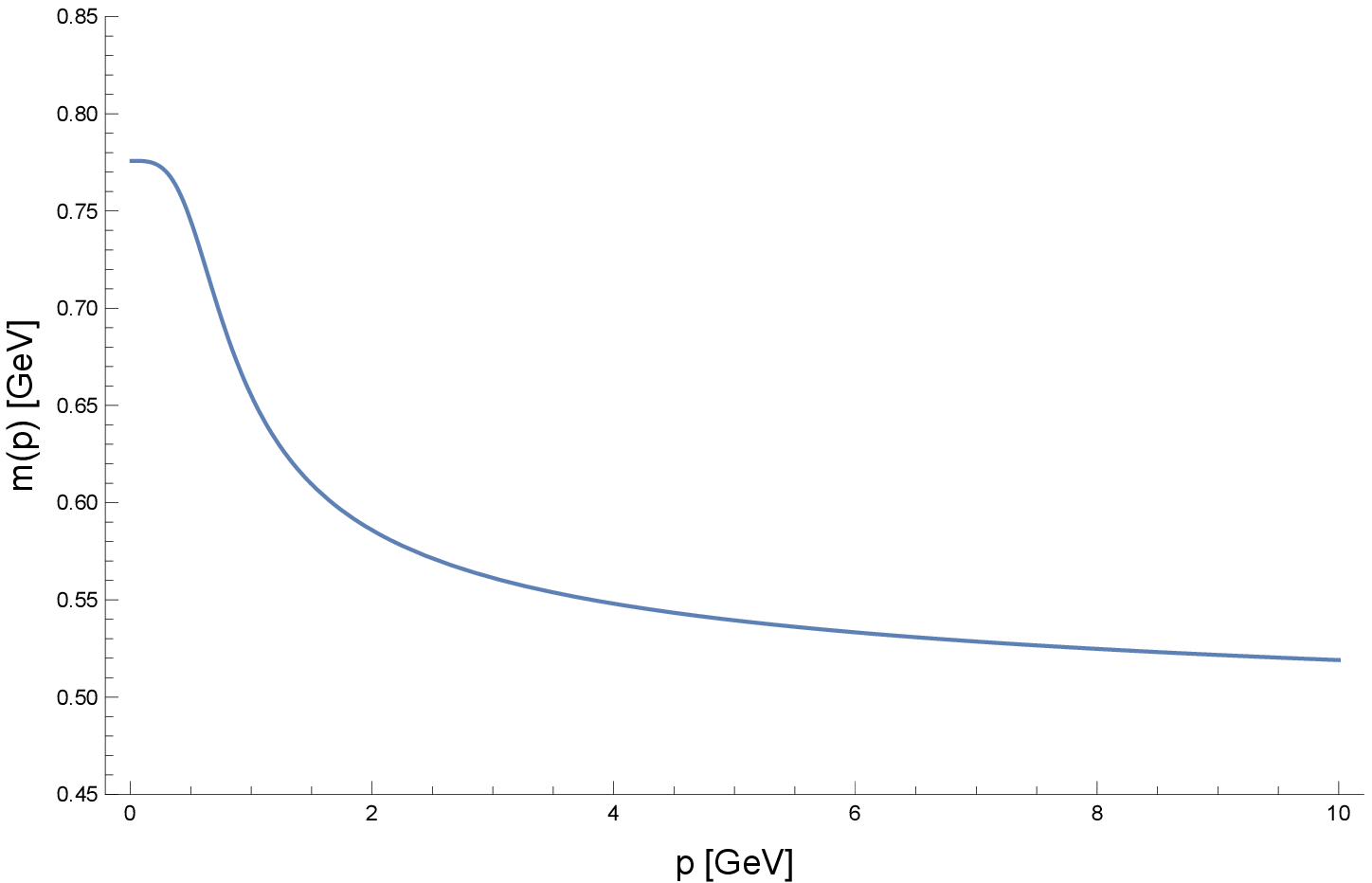}
\caption{Running coupling (top) and mass (bottom) corresponding to the fit in Fig.~\ref{fig:latticecomp}.}
\label{fig:running2}
\end{figure}

The running coupling $\lambda(p)$ and mass $m(p)$ given the initial value $\lambda_{0}=0.473$ and the gap equation are shown in Fig.~\ref{fig:running2}. The maximum of the coupling occurs at $p\approx 0.64$~GeV, where $\lambda\approx0.66$ (i.e. $\alpha_{s}\approx 2.77$). At zero momentum, the running mass saturates to $m(0)\approx 0.78$~GeV.

\begin{figure}
\includegraphics[width=0.43\textwidth]{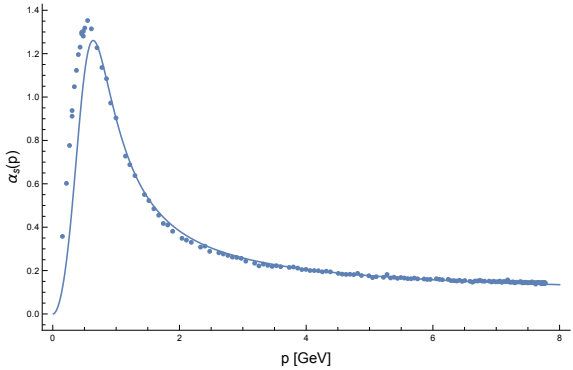}
\caption{Running coupling corresponding to the fit in Fig.~\ref{fig:latticecomp}, rescaled by a factor of $1/2.2$. Lattice data from \cite{Duarte:2016iko}. See the text for details.}
\label{fig:alphalatt}
\end{figure}

In Fig.~\ref{fig:alphalatt} we compare the lattice data for the Taylor coupling to our running coupling $\alpha_{s}(p)$. Since the fitted $\lambda_{0}$ (equivalently, $\alpha_{s}(\mu_{0})$) is a one-loop estimate of the coupling, a rescaling of $\alpha_{s}(p)$ is needed in order to match the lattice. In our case, we had to divide $\alpha_{s}(p)$ by 2.2 in order to align the former both to the lattice UV tail and to the value of $\alpha_{s}$ at the initial renormalization scale $\mu_{0}$. At $\mu_{0}=1$~GeV, the rescaled value of the coupling $\alpha_{s}$ is found to be $0.90$.

\subsection{Testing the stability of the DIS scheme}

In order to test the stability of the DIS scheme, it is useful to extend the results presented in Secs.~IIIC-IIIE to different initial renormalization scales $\mu_{0}$ and to alternative renormalization parameters.

For the purposes of this section, we shall denote with $\Delta Z$ the quantity defined as
\begin{equation}
\Delta Z=Z_{2}Z_{c}-1\ .
\end{equation}
In our previous analysis, we took $\Delta Z=\frac{5\lambda}{8}$, the latter being equal to the one-loop zero-momentum limit of the finite part of $\delta Z_{c}$. While natural in many respects, this choice is far from unique: in Sec.~IIIC we saw that any non-zero $\Delta Z$ of the form $\text{const.}\times \lambda$ yields a gluon propagator which saturates to a finite value as $p\to 0$. It thus makes sense to ask whether choosing a constant other than $5/8$ in $\Delta Z$ would substantially alter our results.

\begin{table}
\def\arraystretch{1.2}
\centering
\begin{tabular}{c||c|c|c}
$\mu_{0}/\Delta Z$&$\lambda/2$&$5\lambda/8$&$3\lambda/4$\\
\hline\hline
$1$~GeV&$(0.516,\,0.673)$&$(0.473,\,0.655)$&$(0.439,\,0.639)$\\
\hline
$2$~GeV&$(0.179,\,0.622)$&$(0.168,\,0.587)$&$(0.160,\,0.560)$\\
\hline
$5$~GeV&$(0.085,\,0.567)$&$(0.082,\,0.533)$&$(0.079,\,0.505)$
\end{tabular}
\caption{Parameters obtained by fitting the lattice data to the DIS scheme RG-improved gluon and ghost form factors, for different initial renormalization scales $\mu_{0}$ and renormalization parameters $\Delta Z$. The cells contain the values $(\lambda_{0},\, m_{0}$~[GeV]$)$ of the fitted initial DIS coupling constant $\lambda_{0}=\lambda(\mu_{0})$ and of the initial DIS gluon mass parameter $m_{0}=m_{0}(\mu_{0})$ computed by solving the gap equation.}
\label{tab:fitparams}
\end{table}

\begin{figure}
\includegraphics[width=0.43\textwidth]{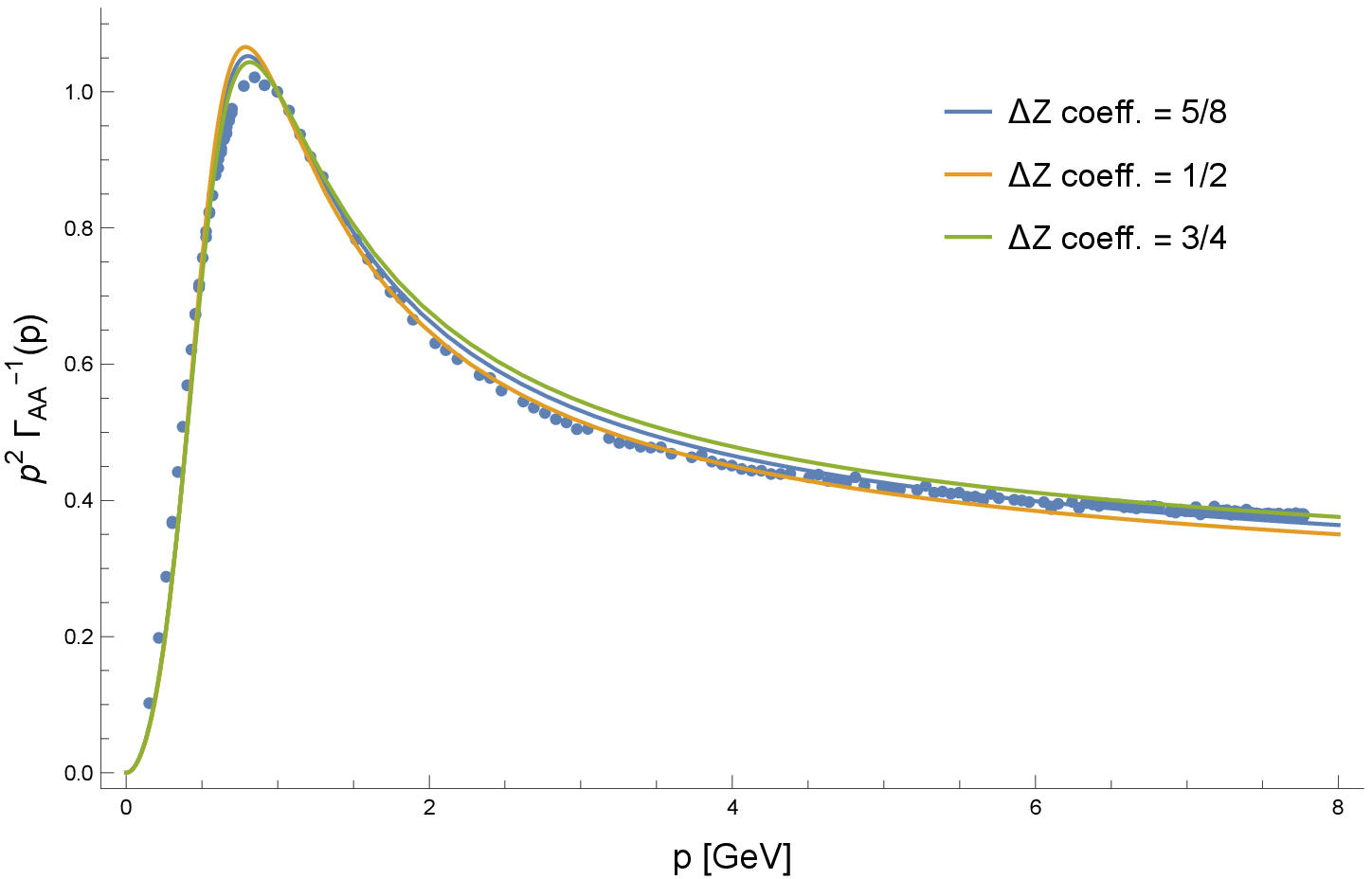}
\includegraphics[width=0.43\textwidth]{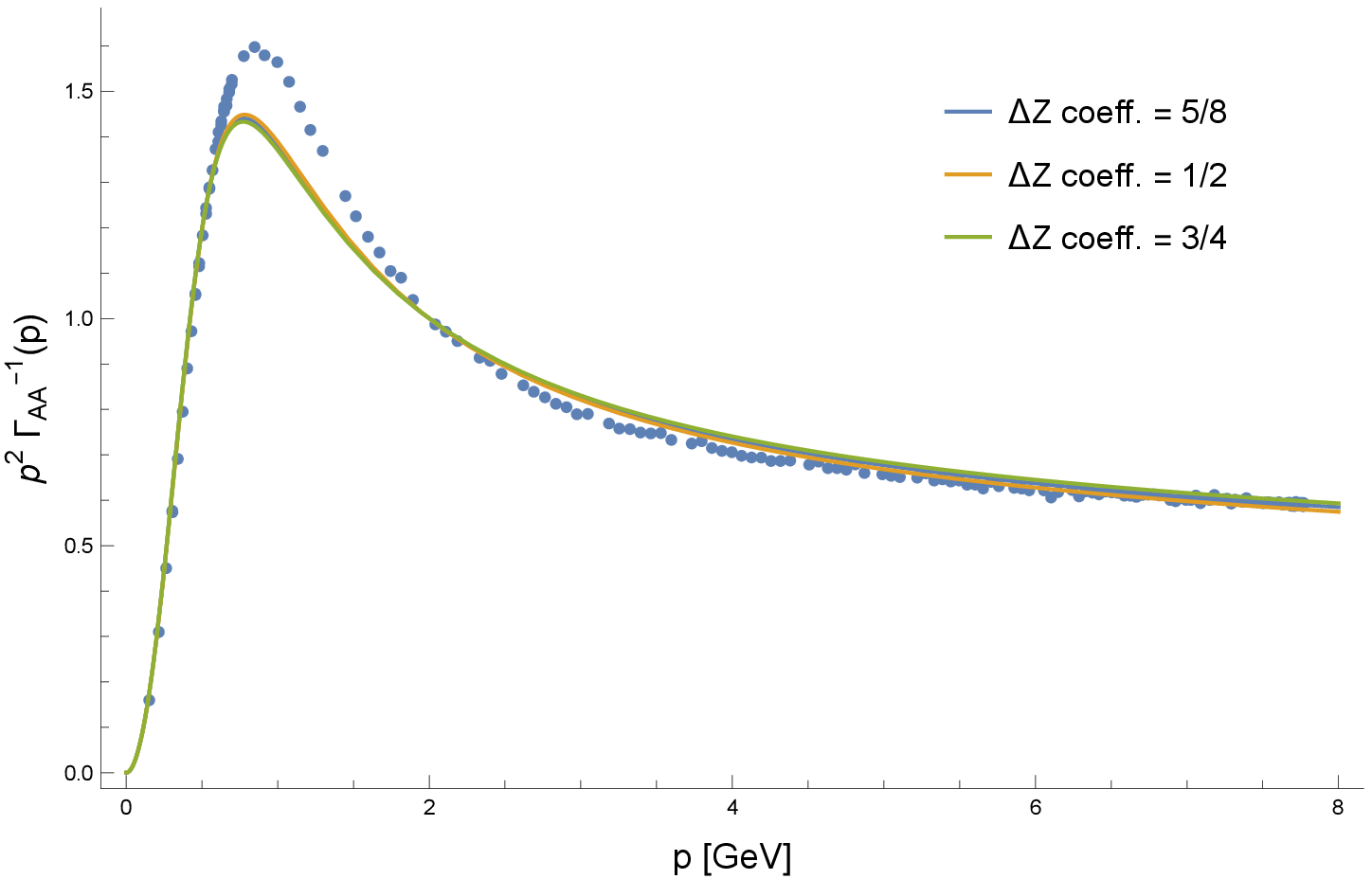}
\includegraphics[width=0.43\textwidth]{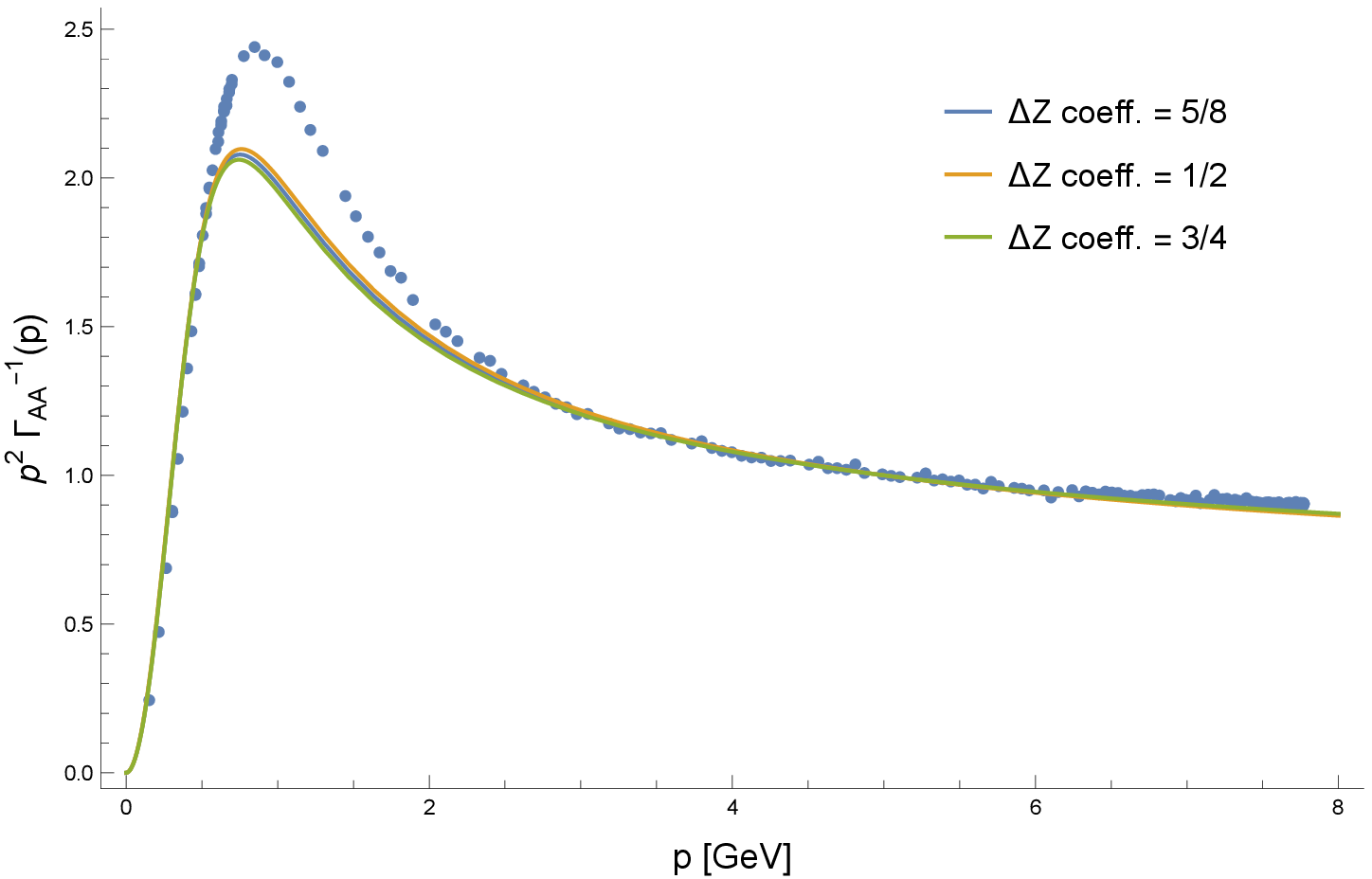}
\caption{RG-improved gluon form factor in the DIS scheme at different initial renormalization scales $\mu_{0}$ and for different $\Delta Z$'s. Top: $\mu_{0}=1$~GeV. Middle: $\mu_{0}=2$~GeV. Bottom: $\mu_{0}=5$~GeV. See the text for details.}
\label{fig:distestsgluon}
\end{figure}

\begin{figure}
\includegraphics[width=0.43\textwidth]{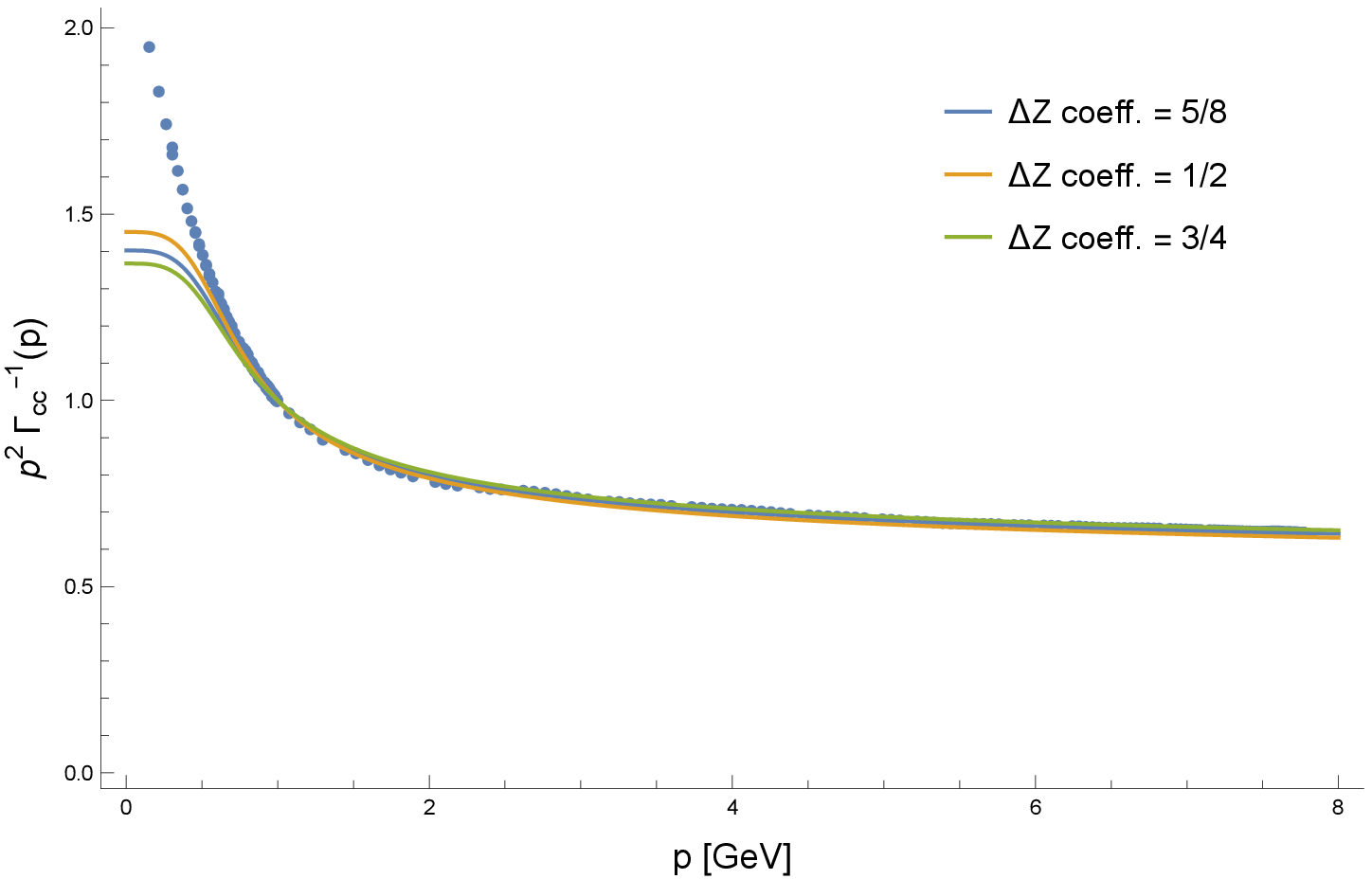}
\includegraphics[width=0.43\textwidth]{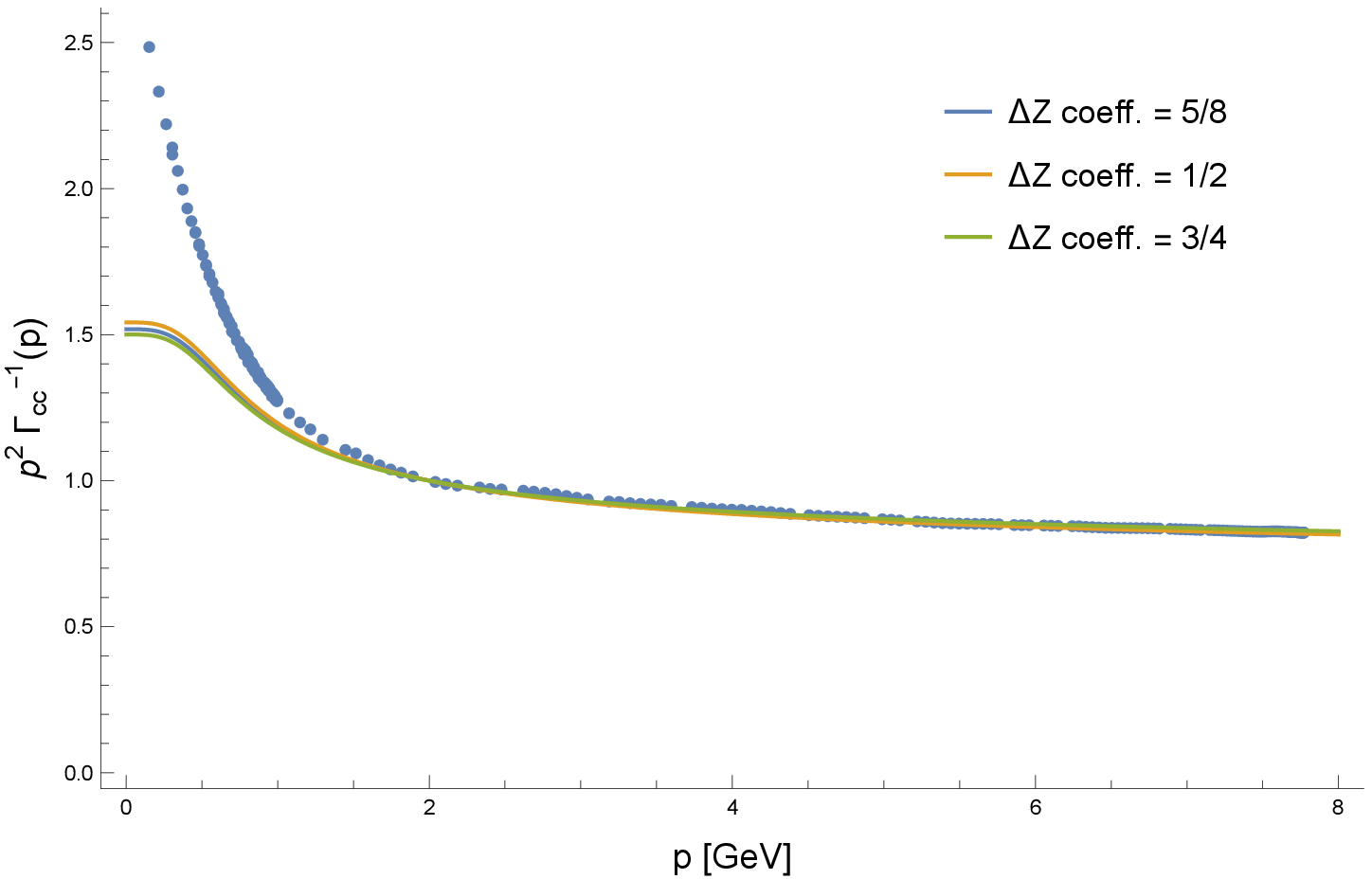}
\includegraphics[width=0.43\textwidth]{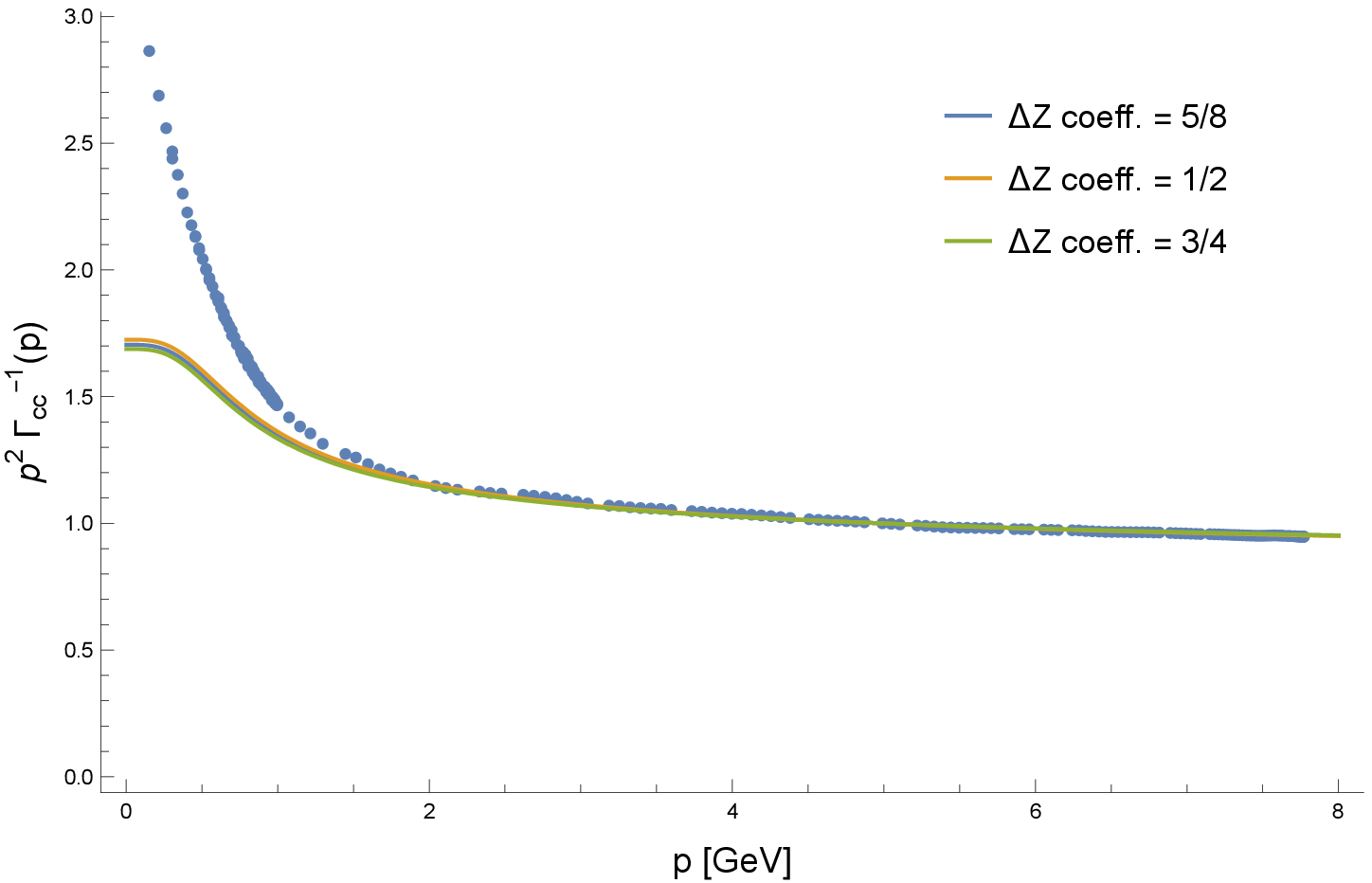}
\caption{RG-improved ghost form factor in the DIS scheme at different initial renormalization scales $\mu_{0}$ and for different $\Delta Z$'s. Plots as in Fig.~\ref{fig:distestsgluon}.}
\label{fig:distestsghost}
\end{figure}

In Figs.~\ref{fig:distestsgluon}~and~\ref{fig:distestsghost} we answer this question by fitting the lattice data while setting $\Delta Z=\frac{5\lambda}{8}^{+20\%}_{-20\%}$ -- that is, $\Delta Z=\frac{\lambda}{2},\frac{3\lambda}{4}$. Additionally, we integrate the RG flow starting from three different initial renormalization scales -- $\mu_{0}=1$~GeV$, 2$~GeV and $5$~GeV -- in order to test the scale-dependence of the scheme. The fit parameters -- reported in Tab.~\ref{tab:fitparams} -- were obtained by adapting the procedure laid out in Sec.~IIID-IIIE to the new renormalization parameters.

It is clear from the fits that increasing or decreasing the value of the coefficient of $\lambda$ in $\Delta Z$ does not have a significant impact on the qualitative behavior of the scheme. Quantitatively, a change in the initial values of the coupling constant $\lambda_{0}$ largely compensates for the difference in the coefficients, although some deviation from our previous results can be observed in the intermediate- to high-energy regime of the gluon sector and in the low-energy regime of the ghost sector at $\mu_{0}=1$~GeV and -- less prominently -- at $\mu_{0}=2$~GeV.

As for the change of the initial renormalization scale, increasing $\mu_{0}$ yields a worse match with the lattice data, especially at intermediate energies in the gluon sector and at low energies in the ghost sector. Nonetheless, we assess that the agreement between the dynamical model/DIS scheme and the lattice can be still considered satisfactory, given that the present results are obtained by a one-loop calculation and by fitting a single free parameter.


\subsection{SU(2)}

\begin{figure}
\includegraphics[width=0.43\textwidth]{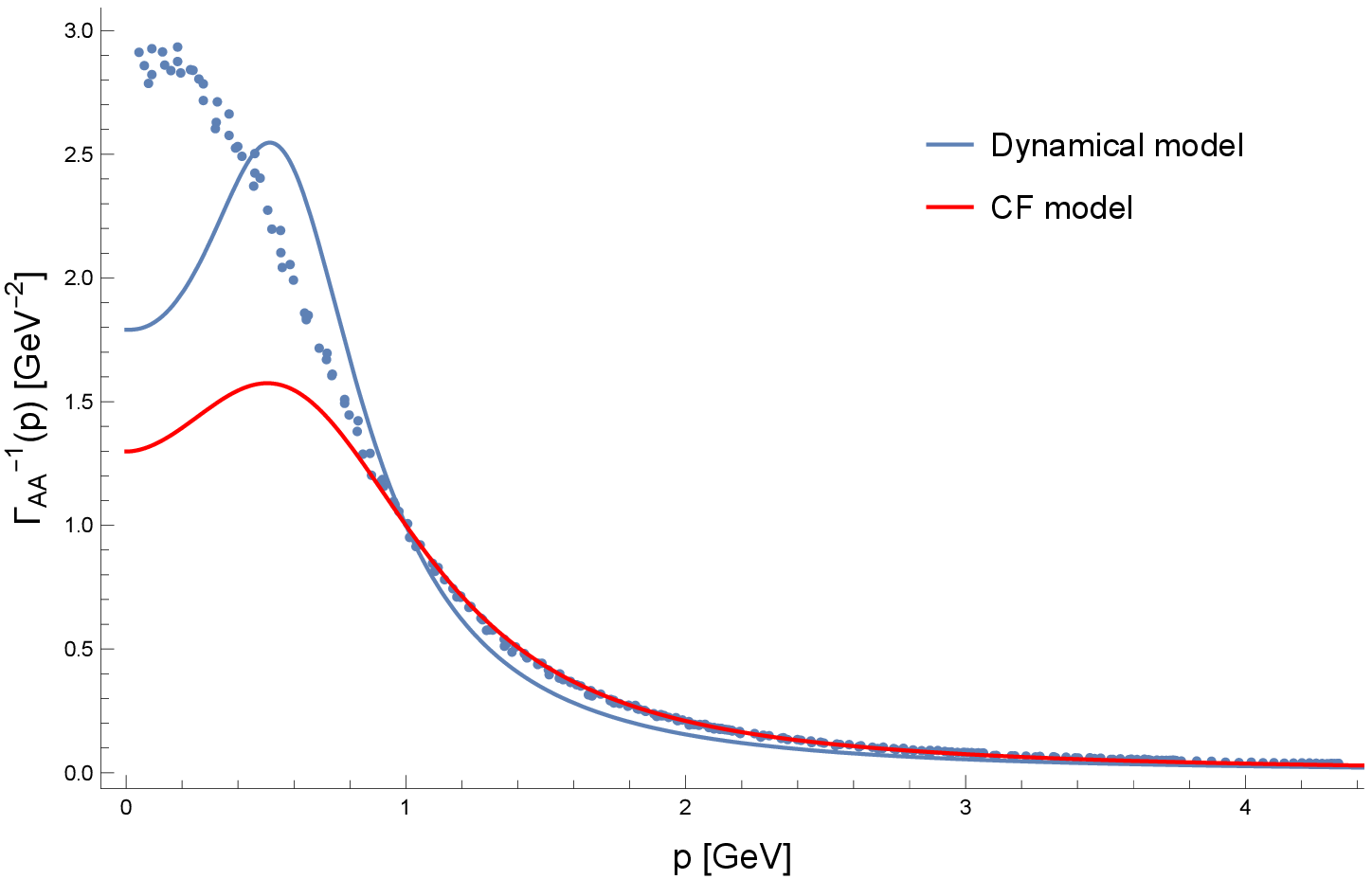}
\includegraphics[width=0.43\textwidth]{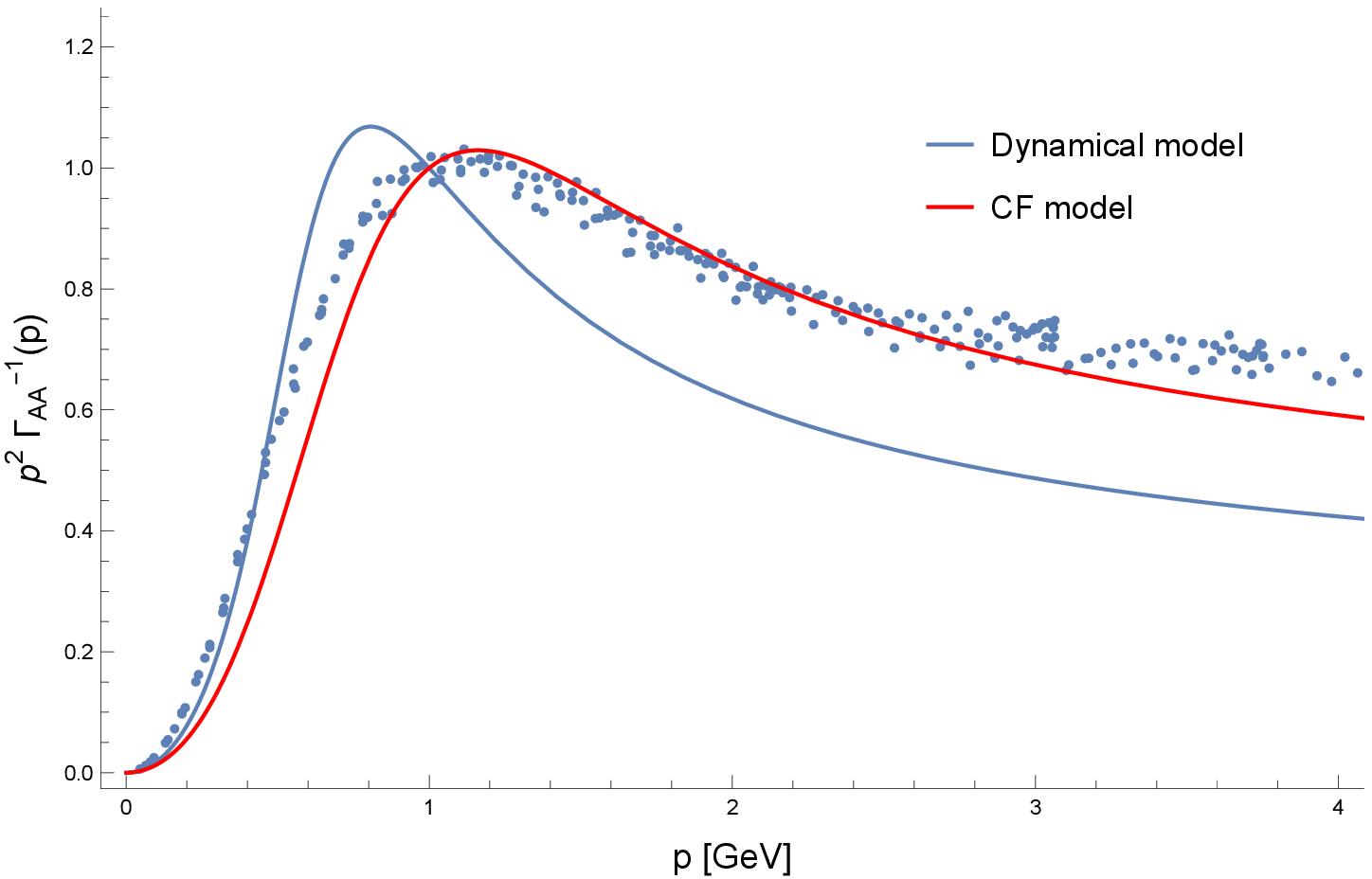}
\includegraphics[width=0.43\textwidth]{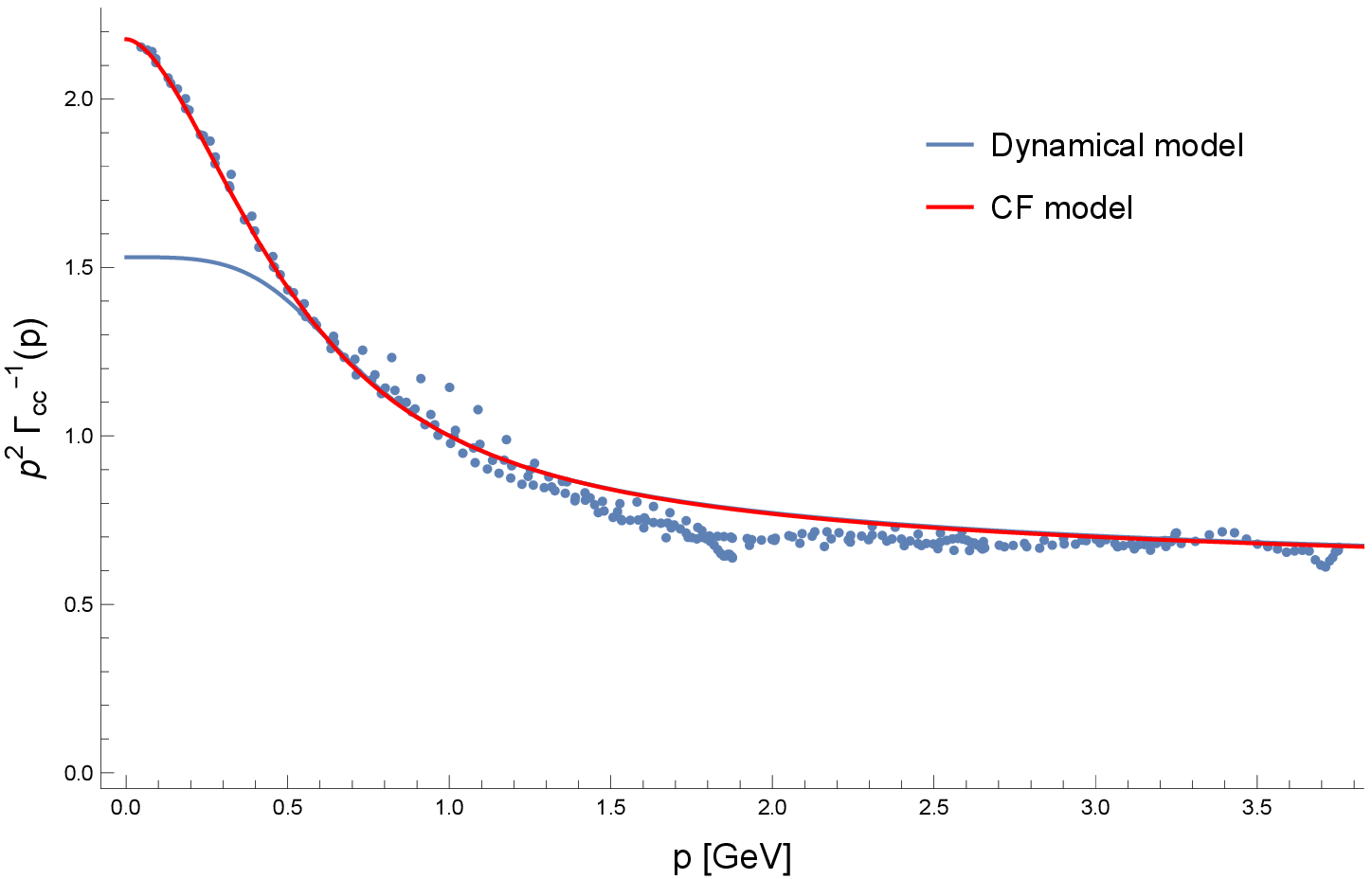}
\caption{RG-improved SU(2) gluon propagator (top), gluon form factor (middle) and ghost form factor (bottom) in the DIS scheme, renormalized at $\mu_{0}=1$~GeV, together with the lattice data of \cite{Cucchieri:2007rg,Cucchieri:2008fc,Cucchieri:2012cb,Cucchieri:2016jwg} and analogous CF model results (red curves) for comparison. The gluon and the ghost form factors were fitted simultaneously, see the text for details.}
\label{fig:su2mixed}
\end{figure}

\begin{figure}
\includegraphics[width=0.43\textwidth]{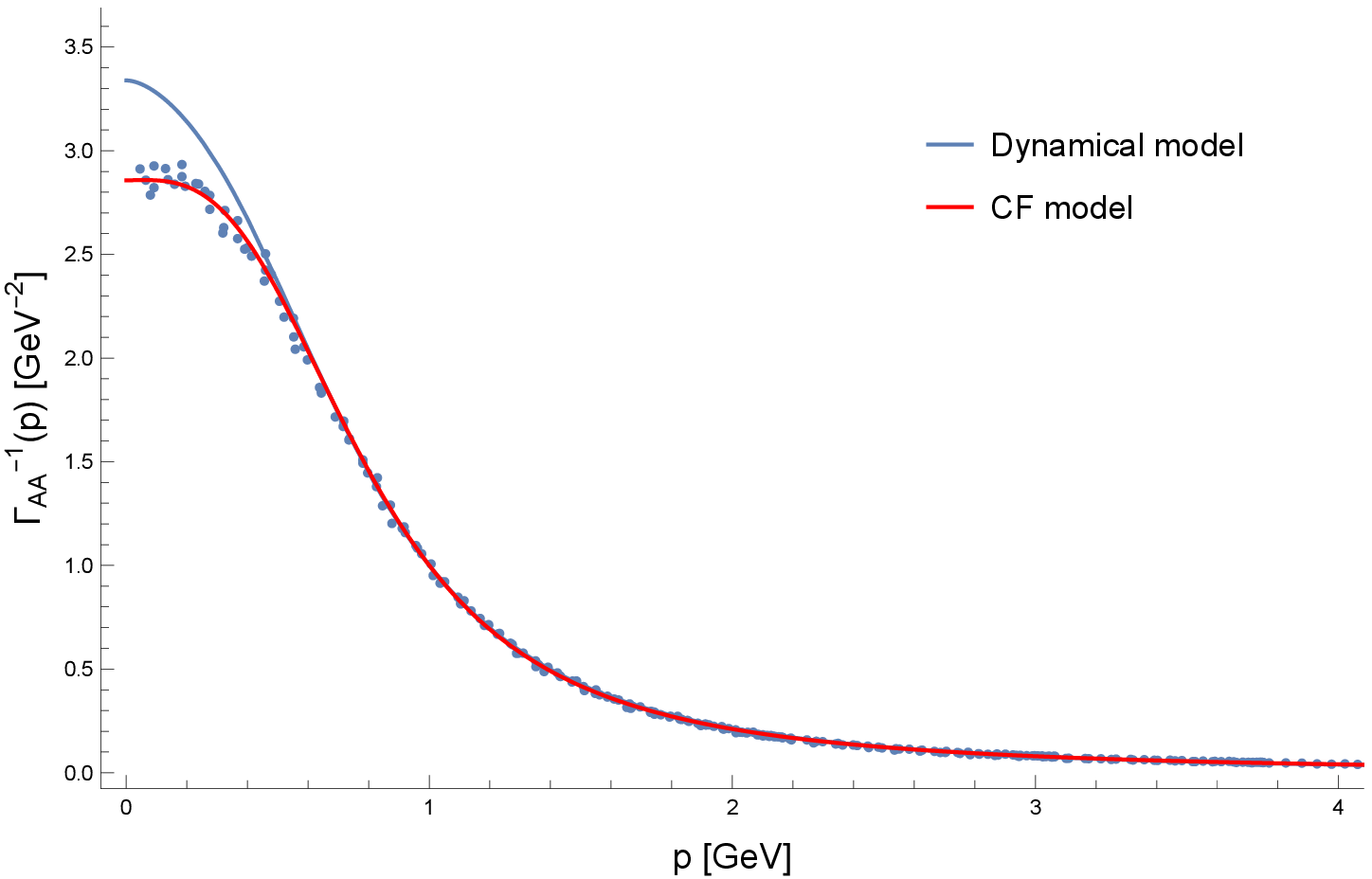}
\includegraphics[width=0.43\textwidth]{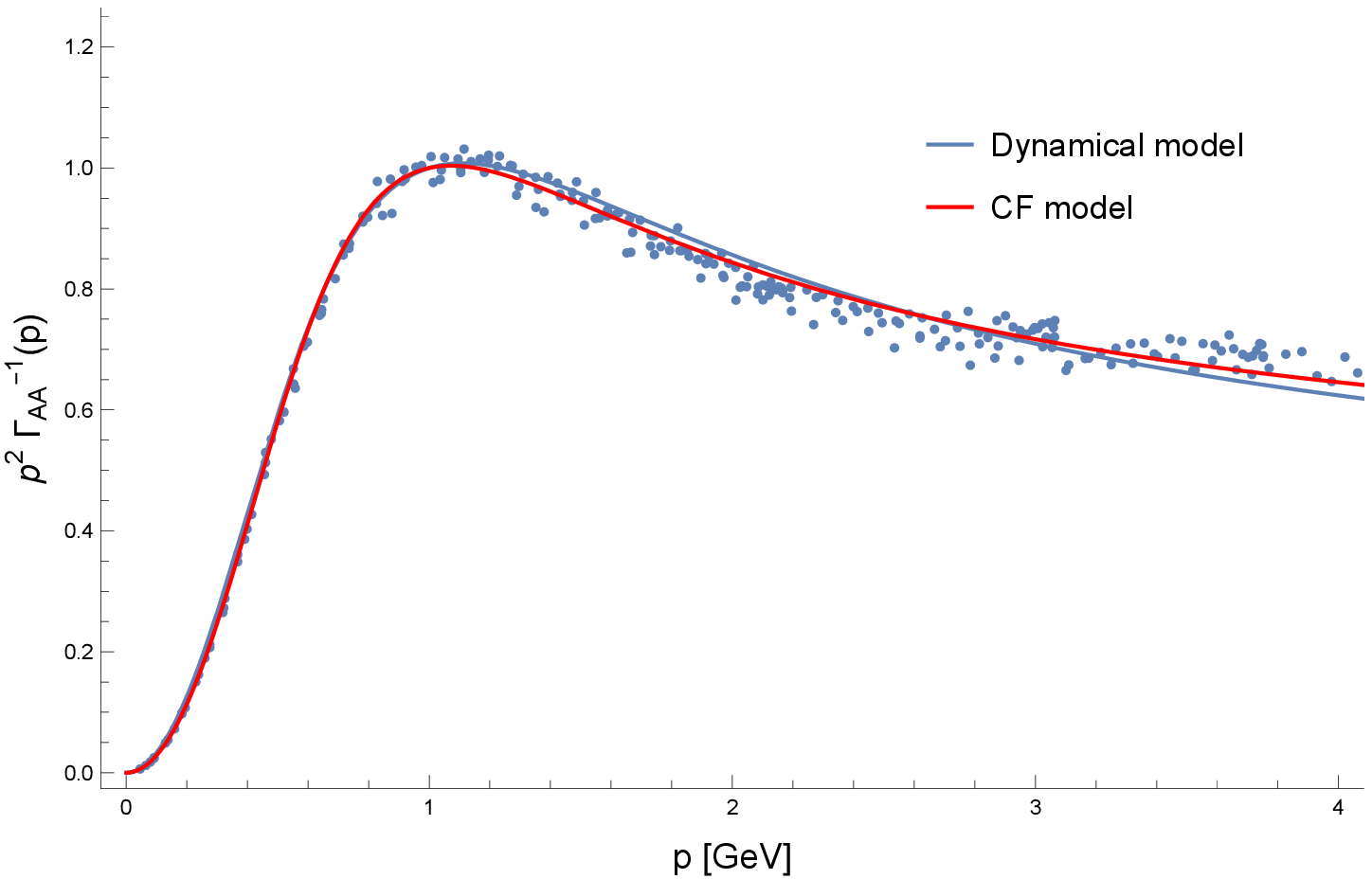}
\includegraphics[width=0.43\textwidth]{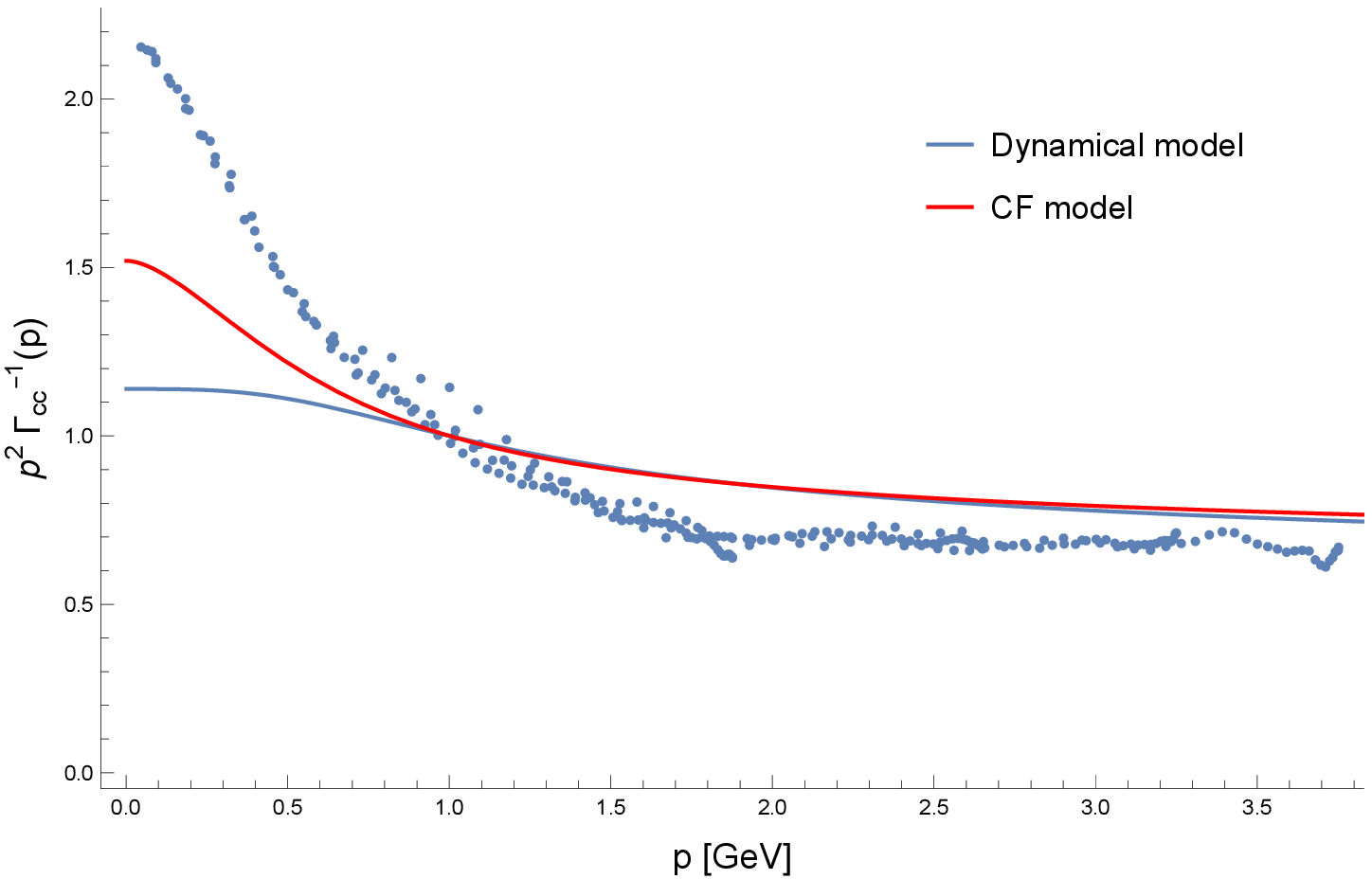}
\caption{RG-improved SU(2) gluon propagator (top), gluon form factor (middle) and ghost form factor (bottom) in the DIS scheme, renormalized at $\mu_{0}=1$~GeV, together with the lattice data of \cite{Cucchieri:2007rg,Cucchieri:2008fc,Cucchieri:2012cb,Cucchieri:2016jwg} and analogous CF model results (red curves) for comparison. Fit of the gluon form factor alone, see the text for details.}
\label{fig:su2gluon}
\end{figure}

To end our overview of the dynamical model, we should mention that some preliminary tests were performed in order to evaluate whether the DIS scheme is also capable of capturing, qualitatively and quantitatively speaking, the SU(2) pure Yang-Mills dynamics. In general, as displayed in Figs.~\ref{fig:su2mixed} and~\ref{fig:su2gluon}, these tests showed a worse agreement with the lattice data of \cite{Cucchieri:2007rg,Cucchieri:2008fc,Cucchieri:2012cb,Cucchieri:2016jwg} in comparison to SU(3). We believe this could be mostly due to a failure of the simple one-loop approximation in the ghost sector, exemplified by the fact that we were not able to obtain a fit of the lattice data for the ghost form factor alone.

In more detail, we performed simple two-parameter fits of the SU(2) gluon and ghost form factors, both separately and simultaneously, at the initial renormalization scale $\mu_{0}=1$~GeV. The simultaneous fit of the two form factors -- shown in Fig.~\ref{fig:su2mixed} -- yielded an unsatisfactory agreement with the lattice data of \cite{Cucchieri:2007rg,Cucchieri:2008fc,Cucchieri:2012cb,Cucchieri:2016jwg}, especially in the gluon sector. A good match of the gluon form factor/propagator, on the other hand, was obtained at the expense of the ghost form factor by fitting the former on its own -- see Fig.~\ref{fig:su2gluon}. Interestingly, in this second case, the fitted value of the coupling constant $\lambda(\mu_{0})=0.35$ was found to be nearly half that obtained by simultaneously fitting the two form factors -- i.e., $\lambda(\mu_{0})=0.59$. As for the fit of the ghost form factor alone, our employed algorithm was not able to reach convergence and provide us with meaningful values of the parameters.

Some insight into these results can be gained by taking the Curci-Ferrari model as a reference. A reiteration of the SU(2) Curci-Ferrari fits \cite{Gracey:2019xom} using the procedure described in Note \ref{CFfoot} -- see Figs.~\ref{fig:su2mixed} and~\ref{fig:su2gluon} -- displays most of the features we just reported for the DIS scheme: fitting the gluon form factor alone does not yield a good agreement with the lattice data for the ghost sector, which only improves when a simultaneous fit of both the form factors is performed; this in turn requires a more-than-doubled value of the coupling constant\footnote{$\lambda(\mu_{0})=0.63$ and $\lambda(\mu_{0})=0.30$ for the simultaneous gluon-ghost and gluon-only CF fits, respectively.}, and leads to a worse match in the gluon sector. We should note that, as far as the dynamical model/DIS scheme is concerned, this kind of behavior is unseen in SU(3), where fitting the lattice data to the gluon and/or the ghost form factors, either simultaneously or separately, yields essentially the same value of the coupling constant\footnote{At $\mu_{0}=1$~GeV, $\lambda(\mu_{0})=0.47$ and $\lambda(\mu_{0})=0.48$ for the one-parameter gluon-only and ghost-only DIS SU(3) fits, respectively.}. Despite these similarities, however, there is one major difference between the two models: just like in SU(3), instead of quickly saturating and deviating from the lattice data as in the DIS scheme, the one-loop RG-improved Curci-Ferrari ghost form factor manages to reproduce the data well down into the deep IR. In fact, at variance with the DIS scheme, fitting the ghost form factor alone is actually possible within the CF model -- at the price of a further increase in the value of the coupling constant\footnote{$\lambda(\mu_{0})=0.92$ for the ghost-only CF fit.}.

We can then conclude that two main factors are at play in making the DIS scheme perform worse in SU(2) than in SU(3): 1. at one loop, a sub-optimal agreement within the ghost sector makes it harder to obtain a good overall fit of the lattice data; 2. this mismatch pushes towards larger values of the coupling constant, were the one-loop approximation is less trustworthy. Since finding a good two-parameter fit of the SU(2) data was not possible within the present approach, we did not try implementing the gap equation and attempt a one-parameter fit. We leave a more in-depth analysis of the SU(2) case to a future study.

\section{Conclusions}
We showed that the non-vanishing of a non-local BRST invariant mass dimension two condensate in pure Yang-Mills theory can be probed in any linear covariant gauge by minimizing an effective potential for the condensate, leading to a dynamical gluon mass which value is fixed in terms of a gap equation. For renormalization purposes of the potential, we had to introduce a novel coupling, but by resorting to a procedure known as the reduction of couplings, this parameter could be expressed as a power series in the usual coupling constant.

The computation of the potential and condensate was carried out in Landau gauge, in which case the non-local condensate reduces to just $\Braket{A^2}$ and computations drastically simplify.

The renormalization group improvement of the gluon and ghost propagators was performed in
a newly introduced renormalization scheme termed the Dynamically-Infrared-Safe (DIS) scheme. In the infrared, the dynamically massive model reproduces the expected, non-perturbative behavior
of pure Yang-Mills theory not only qualitatively---by the saturation of the gluon
propagator at zero momentum---but also quantitatively, as demonstrated by a comparison
with the lattice data. In the UV, where the effects of the gluon condensate are negligible as shown explicitly by renormalization group arguments,
it reduces to ordinary perturbation theory.

Our results thus indicate that the BRST invariant
gluon condensate is a good candidate for explaining by which mechanism
dynamical mass generation occurs in the gluon sector of pure Yang-Mills theory\footnote{Needless to say, other approaches
to dynamical gluon mass generation exist than those mentioned already, see for example \cite{Aguilar:2004sw,Aguilar:2008xm,Fischer:2008uz,Cyrol:2016tym,Huber:2018ned,Horak:2022aqx}.}. Albeit that the eventually dynamically massive model shares many similarities with the Curci-Ferrari model, it is different as highlighted in the text.

A logical next step would be to extend the propagator and renormalization group analysis to other linear covariant gauges, in which case the Nielsen identities \cite{Nielsen:1975fs,Capri:2016gut,Napetschnig:2021ria,Siringo:2022nok} might prove valuable to get a grip on the gauge (in)dependent contributions. Going beyond the Landau gauge will also require to take into proper account the then explicit non-local nature of the condensate, based on \cite{Capri:2017npq}.

\section*{Acknowledgments}
G. C. acknowledges financial support from the Istituto Nazionale di Fisica Nucleare (INFN), ``Strongly Interacting Matter at high density and temperature" (SIM) project, and from the University of Catania, ``Linea di Intervento 2 for HQCDyn". D. D. acknowledges financial support from École Polytechnique (Institut Polytechnique de Paris) and from Centre national de la recherche scientifique (CNRS), next to the warm hospitality at Centre de Physique Théorique (CPhT), which made possible this work. S. P. S. would like to thank the Brazilian agencies Conselho Nacional de Desenvolvimento Científico e Tecnológico (CNPq) and Fundação de Amparo à Pesquisa do Estado do Rio de Janeiro (FAPERJ) for financial support. This study was financed in part by the Coordenação de Aperfeiçoamento de Pessoal de Nível Superior—Brasil (CAPES)—Financial Code 001 (M. N. F.). We thank O.~Oliveira and P.~Silva for providing us with the SU(3) and A.~Cucchieri and T.~Mendes for the SU(2) lattice data. We thank U.~Reinosa for collaboration during the early stages of this work and many useful discussions during the preparation of the manuscript.

\appendix

\section{Perturbative decoupling of the determinant $\pmb{\det\left(\Lambda(\xi)\right)}$ in dimensional regularization}\label{app1}

In order to localize the BRST-invariant gluon field $A_{\mu}^{h}$, in Sec.~IIA we introduced a unity into the partition function in the form
\begin{equation}\label{unityapp0}
1=\mathcal{N}\int \mathcal{D}\xi\,\mathcal{D}\tau\,\mathcal{D}\bar\eta\,\mathcal{D}\eta\,e^{-S_1}\det(\Lambda(\xi))\,.
\end{equation}
The latter is obtained by a change of variables $F\to\xi$ in the functional integral
\begin{align}\label{unityapp1}
1&=\int\mathcal{D}F\ \delta(F)=\qquad(F=\partial\cdot A^{h})\\
\notag&=\int\mathcal{D}\xi\ \det\left(\frac{\delta (\partial\cdot A^{h})}{\delta\xi}\right)\,\delta(\partial\cdot A^{h})=\\
\notag&=\int\mathcal{D}\xi\ \det\left(-\partial\cdot D(A^{h})\Lambda(\xi)\right)\,\delta(\partial\cdot A^{h})\,,
\end{align}
followed by the factorization of the functional determinant and the re-writing of $\det(-\partial\cdot D(A^{h}))$ in terms of a functional integral over a pair of ghost fields $(\eta, \overline{\eta})$, and of $\delta(\partial\cdot A^{h})$ in terms of its $\tau$-Fourier transform. This is analogous to the Faddeev-Popov procedure which is routinely carried out to derive the partition function of pure Yang-Mills theory in the Landau gauge, with the ghosts $\eta$ and $\overline{\eta}$ in place of the standard ghost fields $c$ and $\overline{c}$, and the Fourier field $\tau$ in place of the Nakanishi-Lautrup field $b$.

In the context of gauge fixing, thanks to the gauge-invariance of the partition function, one usually exchanges $A_{\mu}^{h}$ for $A_{\mu}$ in \eqref{unityapp0}-\eqref{unityapp1} by a change of the gluon field integration variables, after which the only $\xi$-dependent term left in the unity is the determinant $\det(\Lambda(\xi))$. It is easy to see, then, that the $\xi$-integral decouples from the rest of the partition function, so that the former can be absorbed into the normalization factor $\mathcal{N}$. On the other hand, our introduction of the unity in Sec.~IIA is carried out at a stage in which the partition function is already gauge-fixed. Therefore, no change of variable $A_{\mu}^{h}\to A_{\mu}$ can be performed, and, in a generic linear covariant gauge, the non-decoupled field $\xi$ must be treated on the same footing as the other dynamical fields of the theory.

Nevertheless, it can still be shown that the determinant $\det(\Lambda(\xi))$ does not perturbatively contribute to the $n$-point functions of the theory, as long as it is defined in dimensional regularization. As a consequence, when doing calculations in perturbation theory using dimensional regularization, the determinant can be suppressed by setting $\det(\Lambda(\xi))=1$.

In order to prove our statement, we first rewrite the determinant in terms of a functional integral over a new pair of ghost fields $(\lambda,\overline{\lambda})$,
\begin{equation}\label{eq:detlambda0}
\det(\Lambda(\xi))=\int\mathcal{D}\overline{\lambda}\mathcal{D}\lambda\ \exp\left\{-\int d^{d}x\ \overline{\lambda}^{a}\Lambda_{ab}(\xi)\lambda^{b}\right\}\,.
\end{equation}
Since perturbatively
\begin{equation}
\Lambda_{ab}(\xi)=\delta_{ab}-\frac{g}{2}\,f_{abc}\,\xi^{c}+\frac{g^{2}}{3!}\,f_{ace}f_{edb}\,\xi^{c}\xi^{d}+\cdots\,,
\end{equation}
We may reexpress \eqref{eq:detlambda0} as
\begin{equation}
\det(\Lambda(\xi))=\int\mathcal{D}\overline{\lambda}\mathcal{D}\lambda\ e^{-(I_{0}+I_{1})}\,,
\end{equation}
where
\begin{align}
I_{0}&=\int d^{d}x\ \overline{\eta}^{a}\eta^{a}\,,\\
I_{1}&=\int d^{d}x\ \overline{\eta}^{a}\Omega_{ab}(\xi)\eta^{b}\,,
\end{align}
having defined
\begin{align}
\Omega_{ab}(\xi)=\Lambda_{ab}(\xi)-\delta_{ab}\,.
\end{align}
The action term $I_{1}$ contains the interaction between $(\lambda,\overline{\lambda})$ and $\xi$. The latter is quadratic in the ghost fields, with its $\xi$-dependence encoded in the function $\Omega_{ab}(\xi)$. $I_{0}$, on the other hand, contains the zero-order ghost propagator, which is easily seen to be $Q^{ab}(p)=\delta^{ab}$ in momentum space, or $Q^{ab}(x)=\delta^{ab}\delta(x)$ in coordinate space.

\begin{figure}
\includegraphics[width=0.18\textwidth]{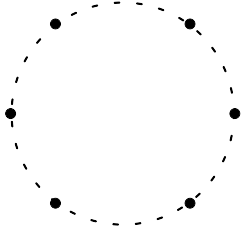}
\caption{Loop contributing to the ghost average in \eqref{eq:avgn0conn} (example for the case $n=6$). The dashed line is the $(\lambda,\overline{\lambda})$ zero-order propagator.}
\label{fig:lambdaloop}
\end{figure}

Now, consider the vacuum expectation value $\langle\mathcal{O}\rangle$ of an operator $\mathcal{O}$ which does not depend on the newly-introduced fields $(\lambda,\overline{\lambda})$. This can be computed as
\begin{align}\label{eq:lambdaovev}
\langle\mathcal{O}\rangle=\frac{\langle\mathcal{O}e^{-I_{1}}\rangle_{0}}{\langle e^{-I_{1}}\rangle_{0}}&=\langle\mathcal{O}e^{-I_{1}}\rangle_{0,\text{conn.}}=\\
\notag&=\sum_{n=0}^{+\infty}\frac{(-1)^{n}}{n!}\langle\mathcal{O}I_{1}^{n}\rangle_{0,\text{conn.}}\,,
\end{align}
where the subscript 0 denotes that the average is to be taken with respect to the action $I_{0}$ plus any other $(\lambda,\overline{\lambda})$-independent term originally present in the full action of the theory. $\langle\mathcal{O}I_{1}^{n}\rangle_{0,\text{conn.}}$ explicitly reads
\begin{align}\label{eq:avgn0conn}
&\langle\mathcal{O}I_{1}^{n}\rangle_{0,\text{conn.}}=\int \prod_{i=1}^{n} d^{d}x_{i}\ \left\langle\mathcal{O}\prod_{j=1}^{n}\Omega_{a_{j}b_{j}}(\xi(x_{j}))\right\rangle_{00,\text{conn.}}\times\\
\notag&\quad\times\left\langle\overline{\lambda}^{a_{1}}(x_{1})\lambda^{b_{1}}(x_{1})\cdots \overline{\lambda}^{a_{n}}(x_{n})\lambda^{b_{n}}(x_{n})\right\rangle_{\text{gh.},\text{conn.}}\,,
\end{align}
where the subscript 00 denotes that the first average is to be taken with respect to the full, $(\lambda,\overline{\lambda})$-independent action, whereas the subscript ``gh.'' denotes that the second average is to be taken with respect to the zero-order ghost action $I_{0}$. Diagrammatically, for each $n\geq 1$, the ghost average receives contributions from a single ghost loop, depicted in Fig.~\ref{fig:lambdaloop}. In coordinate space, suppressing the color structure, the diagram reads
\begin{equation}
(-1)(n-1)!\,\delta(x_{1}-x_{2})\cdots\delta(x_{n-1}-x_{n})\delta(x_{n}-x_{1})\,,
\end{equation}
or, equivalently,
\begin{equation}
(-1)(n-1)!\,\delta(0)\int d^{d}x\ \prod_{i=1}^{n}\delta(x_{i}-x)\,,
\end{equation}
where $\delta(0)$ is a Dirac delta in coordinate space,
\begin{equation}\label{eq:deltagh0}
\delta(0)=\int\frac{d^{d}q}{(2\pi)^{d}}\ 1\,.
\end{equation}
Therefore, for $n\geq 1$,
\begin{align}
\langle\mathcal{O}I_{1}^{n}\rangle_{0,\text{conn.}}&=(-1)(n-1)!\,\delta(0)\times\\
\notag&\times\int d^{d}x\ \left\langle\mathcal{O}\text{Tr}\left\{\Omega^{n}(\xi(x))\right\}\right\rangle_{00,\text{conn.}}\,,
\end{align}
where $\Omega^{n}(\xi)$ is the matrix product of $n$ factors of $\Omega(\xi)$ and the trace is taken over the color indices.

In dimensional regularization, the integral in \eqref{eq:deltagh0} vanishes \cite{Collins:1984xc}. It follows that $\langle\mathcal{O}I_{1}^{n}\rangle_{0,\text{conn.}}=0$ for every $n\geq 1$, so that, going back to \eqref{eq:lambdaovev},
\begin{equation}\label{eq:appafin}
\langle\mathcal{O}\rangle=\langle\mathcal{O}\rangle_{0}=\langle\mathcal{O}\rangle_{00}\,,
\end{equation}
where to obtain $\langle\mathcal{O}\rangle_{00}$ we have integrated out the free ghost action $I_{0}$ from $\langle\mathcal{O}\rangle_{0}$. What \eqref{eq:appafin} means is that the perturbative corrections to the vacuum expectation value $\langle\mathcal{O}\rangle$ due to the determinant $\det(\Lambda(\xi))$ vanish in dimensional regularization. Therefore, the vacuum expectation value of any operator $\mathcal{O}$ in the full theory can be computed by setting $\det(\Lambda(\xi))=1$ in its dimensionally regularized partition function.

One may have noticed that our proof---aside from dimensional regularization---relies exclusively on the fact that $\Lambda(\xi)$ is equal to the unit matrix to lowest order in perturbation theory. The question arises, then, whether the proof is general enough to apply to the determinant of any such matrix. The answer is that, in general, it does not. Indeed, setting $\delta(0)=0$ in dimensional regularization is allowed if and only if the calculations can be carried out without spoiling the symmetries of the theory.

While Lorentz invariance is clearly preserved by the action in \eqref{eq:detlambda0}, showing that  the latter does not violate the BRST invariance of the full action of the theory requires us to extend the symmetry to the ghost fields $\lambda$ and $\overline{\lambda}$. Indeed, a straightforward calculation starting from \eqref{brst2} and the definition of $\Lambda_{ab}(\xi)$ in \eqref{eq:lambdadef} yields
\begin{equation}
s\Lambda_{ab}(\xi)=\Lambda_{ac}(\xi)\,\Psi^{c}_{b}(c,\xi)\,,
\end{equation}
with
\begin{equation}
\Psi^{a}_{b}(c,\xi)=-\frac{\partial(s\xi^{a})}{\partial \xi^{b}}\,,
\end{equation}
so that the ghosts must have non-vanishing BRST transformations if \eqref{eq:detlambda0} is to be invariant. Since the BRST transformation does not act on the anti-ghost index of $\Lambda_{ab}(\xi)$, it is reasonable to define
\begin{align}\label{eq:ext2brst}
s\lambda^{a}=-\Psi^{a}_{b}(c,\xi)\,\lambda^{b}\,,\qquad s\overline{\lambda}^{a}=0\,,
\end{align}
where $s\lambda^{a}$ is chosen so that $s(\Lambda\lambda)=0$. \eqref{eq:detlambda0}---and the full action of the theory together with it---is invariant with respect to these extended BRST transformations. The nilpotency of the extended BRST operator is then easily proved by observing that $s^{2}\Lambda_{ab}(\xi)=0$---which holds thanks to the nilpotency of $s$ on the fields $\xi$ and $c$---implies that
\begin{equation}
0=s^{2}\Lambda_{ab}=\Lambda_{ac}\left(\Psi^{c}_{d}\Psi^{d}_{b}+s\Psi_{b}^{c}\right)\,,
\end{equation}
that is, $s\Psi=-\Psi^{2}$. When plugged into \eqref{eq:ext2brst}, the latter ensures that $s^{2}\lambda^{a}=s^{2}\overline{\lambda}^{a}=0$.

\section{Unity at work}\label{app2}

As shown in the main text, a one-loop computation shows that the $\sigma$-action with $\sigma \to \sigma_0 + \delta\sigma$ leads to
\begin{equation}
\Pi^{\text{1-loop}} = p^2 + m^2 + \Pi_{CF}^{\text{1-loop}}(p^2,m^2) + \Pi_{\text{extra}}^{\text{1-loop}}
\end{equation}
where the mass term will be cancelled by the self energy contributions originating from the extra terms in the action, when using the gap equation $\frac{\p V}{\p \sigma} =0$ (cf.~\eqref{eq:final}).  With the ``extra'' part, we mean the diagrams that are generated by the extra vertices arising from the $\delta{\sigma}$-part of the $\sigma$-Lagrangian, meaning the $\delta{\sigma} A^2$- and $A^4$-vertices, see \eqref{eq:AAsigma}, \eqref{eq:AAAA}.

This will actually be a more generic feature of the dynamical model. More precisely, the tree level mass will always cancel against certain contributions coming from the extra diagrams.

Let us now try to show this by making use of the ``power of unity''. We will work with a simplified notation, to make things clear. But the general argument readily applies to the case under study.

Let us consider a theory, Theory 1, with a partition function $\int [\ddd A] e^{-S(A)}$, but we could also add a unity to get, Theory 2 with a partition function
\begin{equation}
\int \mathcal{D} A e^{-\frac{1}{\hbar}S(A)} \times \int \mathcal{D} \sigma e^{-\frac{1}{2\hbar} \int \ddd x^4 \left( \sigma - \frac{g A^2}{2}\right)^2}
\end{equation}
We understand that this is still the same theory as the (exact) Gaussian integral over $\sigma$ yields a unity\footnote{We did not write global normalization factors.}. This means that (connected) correlation functions remain unchanged, $\expval{A \cdots A}_1 = \expval{A \cdots A}_2$ \footnote{With ``$A$'' a shorthand for all original fields.}. We have temporarily introduced the (loop counting) factor $\hbar$. Clearly, by identifying order per order in $\hbar$, the equivalence between $\expval{A \cdots A}_1$ and $\expval{A \cdots A}_2$ will also hold order per order in perturbation theory, seen as formal power series in $\hbar$.

Let us illustrate it explicitly: the extra vertices introduced by the unity, \eqref{eq:AAsigma} and \eqref{eq:AAAA}, are visually represented by
\begin{center}
	\begin{tikzpicture}
	\begin{feynman}
	\vertex (a) at (-1,0) {\(\sigma \)};
	\vertex (b) at (0,0) ;
	\vertex (c) at (0.75,0.75) {\(A\)};
	\vertex (d) at (0.75,-0.75) {\(A\)};
		
	\diagram {
		(a) -- [plain] (b),
		(b) -- [gluon] (c),
		(b) -- [gluon] (d),
	};

 	\vertex [right=0.3em of b] {\(g\)};

	\end{feynman}
	\end{tikzpicture}
	\hspace{2cm}
	\begin{tikzpicture}
	\begin{feynman}
	\vertex (a) at (0,0) ;
	\vertex (b) at (-0.75,0.75) {\(A\)};
	\vertex (c) at (0.75,0.75) {\(A\)};
	\vertex (d) at (0.75,-0.75) {\(A\)};
	\vertex (e) at (-0.75,-0.75) {\(A\)};
		
	\diagram {
		(a) -- [gluon] (b),
		(a) -- [gluon] (c),
		(a) -- [gluon] (d),
		(a) -- [gluon] (e),
	};

	\vertex [right=0.3em of a] {\( -3g^2 \)};
	
	\end{feynman}
	\end{tikzpicture}
\end{center}
We get at one-loop a contribution proportional to
\begin{eqnarray}\label{eq:UnityDiagrams}
\expval{AA}_{extra}^{conn} &=& 
\begingroup%
  \makeatletter%
  \providecommand\color[2][]{%
    \errmessage{(Inkscape) Color is used for the text in Inkscape, but the package 'color.sty' is not loaded}%
    \renewcommand\color[2][]{}%
  }%
  \providecommand\transparent[1]{%
    \errmessage{(Inkscape) Transparency is used (non-zero) for the text in Inkscape, but the package 'transparent.sty' is not loaded}%
    \renewcommand\transparent[1]{}%
  }%
  \providecommand\rotatebox[2]{#2}%
  \ifx\svgwidth\undefined%
    \setlength{\unitlength}{427.428576bp}%
    \ifx\svgscale\undefined%
      \relax%
    \else%
      \setlength{\unitlength}{\unitlength * \real{\svgscale}}%
    \fi%
  \else%
    \setlength{\unitlength}{\svgwidth}%
  \fi%
  \global\let\svgwidth\undefined%
  \global\let\svgscale\undefined%
  \makeatother%
\begin{tikzpicture}
\begin{feynman}
\vertex (a) at (0,0) ;
\vertex (b) at (1,0) ;
\vertex (c) at (2,0) ;
\vertex (d) at (1,0.75) ;
\vertex (e) at (1,1.5) ;

\diagram {
	(a) -- [gluon] (b),
	(b) -- [gluon] (c),
	(b) -- [plain] (d),
	(d) -- [gluon, half left] (e),
	(d) -- [gluon, half right] (e),
};

\end{feynman}
\end{tikzpicture}
\endgroup%
 + 
\begingroup%
  \makeatletter%
  \providecommand\color[2][]{%
    \errmessage{(Inkscape) Color is used for the text in Inkscape, but the package 'color.sty' is not loaded}%
    \renewcommand\color[2][]{}%
  }%
  \providecommand\transparent[1]{%
    \errmessage{(Inkscape) Transparency is used (non-zero) for the text in Inkscape, but the package 'transparent.sty' is not loaded}%
    \renewcommand\transparent[1]{}%
  }%
  \providecommand\rotatebox[2]{#2}%
  \ifx\svgwidth\undefined%
    \setlength{\unitlength}{427.428576bp}%
    \ifx\svgscale\undefined%
      \relax%
    \else%
      \setlength{\unitlength}{\unitlength * \real{\svgscale}}%
    \fi%
  \else%
    \setlength{\unitlength}{\svgwidth}%
  \fi%
  \global\let\svgwidth\undefined%
  \global\let\svgscale\undefined%
  \makeatother%
\begin{tikzpicture}
\begin{feynman}
\vertex (a) at (0,0) ;
\vertex (b) at (1,0) ;
\vertex (c) at (2,0) ;
\vertex (d) at (3,0) ;

\diagram {
	(a) -- [gluon] (b),
	(b) -- [plain, half left] (c),
	(b) -- [gluon] (c),
	(c) -- [gluon] (d),
};

\end{feynman}
\end{tikzpicture}
\endgroup%
\nonumber\\ &+& 
\begingroup%
  \makeatletter%
  \providecommand\color[2][]{%
    \errmessage{(Inkscape) Color is used for the text in Inkscape, but the package 'color.sty' is not loaded}%
    \renewcommand\color[2][]{}%
  }%
  \providecommand\transparent[1]{%
    \errmessage{(Inkscape) Transparency is used (non-zero) for the text in Inkscape, but the package 'transparent.sty' is not loaded}%
    \renewcommand\transparent[1]{}%
  }%
  \providecommand\rotatebox[2]{#2}%
  \ifx\svgwidth\undefined%
    \setlength{\unitlength}{427.428576bp}%
    \ifx\svgscale\undefined%
      \relax%
    \else%
      \setlength{\unitlength}{\unitlength * \real{\svgscale}}%
    \fi%
  \else%
    \setlength{\unitlength}{\svgwidth}%
  \fi%
  \global\let\svgwidth\undefined%
  \global\let\svgscale\undefined%
  \makeatother%
\begin{tikzpicture}
\begin{feynman}
\vertex (a) at (0,0) ;
\vertex (b) at (1,0) ;
\vertex (c) at (2,0) ;
\vertex (d) at (1,1.5) ;

\diagram {
	(a) -- [gluon] (b),
	(b) -- [gluon] (c),
	(b) -- [gluon, half left] (d),
	(b) -- [gluon, half right] (d),
};

\end{feynman}
\end{tikzpicture}
\endgroup%
 = 0
\end{eqnarray}
using the trivial (constant) propagator $\expval{\sigma \sigma} = 1$ and the correct symmetry factors, we end up
\begin{equation}
\expval{AA}_{\text{extra}}^{1PI} =  g^2 \int \expval{AA} + \frac{1}{2} g^2 \int \expval{AA} - \frac{3}{2} g^2 \int \expval{AA}
\end{equation}
which is indeed equal to zero.

Next, we consider the possibility that $\sigma$ develops a vacuum expectation value, namely
\begin{equation}
\sigma \to \expval{\sigma} + \delta{\sigma} \equiv \sigma + \delta\sigma
\end{equation}
with per definition $\expval{\delta{\sigma}} = 0$. In this case, the partition function will read
\begin{eqnarray}\label{eq:Sunity}
	&&\int \mathcal{D}A e^{-S[A]- \int \ddd^4x g \sigma A^2} \int \mathcal{D}\delta\sigma e^{- \frac{1}{2}  \int \ddd^4 x \left( \delta{\sigma} - \frac{g A^2}{2}\right)^2}\nonumber\\
&&\times e^{- \int \ddd^4 x \left( \frac{\sigma^2}{2} \right) }
\end{eqnarray}
The first exponential will still yield a unity, the second consists of a constant and will play no role in correlation functions (but it does in the quantum effective action). The term linear in $\delta\sigma$ we dropped, as this will cancel order per order by removing all $\delta\sigma$-tadpole graphs, which is equivalent to $\expval{\delta\sigma}\equiv0$, or, extremization of the effective action w.r.t.~$\sigma$, that is, the gap equation, see e.g.~\cite{Peskin:1995ev}. Finally, there is also an effective mass term for the $A$-field that we included in the original action, that is, with a ``dynamically massive'' $A$-field.

At one-loop, we would now get
\begin{eqnarray}\label{1example}
\expval{AA}_{extra}^{conn} &=&  +  \nonumber\\&+& \nonumber\\ 
&=& + 
\end{eqnarray}
These two diagrams are completely similar to those in the last line of \eqref{eq:g2a}, when resummed into the inverse propagator they will annihilate the tree level mass term upon using the gap equation $\frac{\p V}{\p \sigma}=0$, as it follows from direct computation also here. The first diagram in the first line is nonzero, but it will cancel against other (non-written) tadpole contributions. Notice that we did not explicitly use the underlying unity at this point.

We can reconsider the 3 diagrams in the first line of \eqref{1example} also from the unity viewpoint though. In that case, the 3 diagrams cancel against each other, just as before. However, this means we have removed one, but not all, tadpole contributions, meaning that there will be a net tadpole correction to $\expval{AA}_{extra}^{1PI}$, which actually corresponds to minus the first diagram, upon amputating the external $A$-legs of course. But due to the gap equation, this is nothing else than the tree level mass, up to the sign.

It is clear this observation will continue to hold through at any order $n$: we can always keep the necessary tadpole (sub)graphs to get all diagrams that make up the unity order per order (the sum of which diagrams will then lead to a zero), the remaining contribution will be the (usually never written) ``counterterm'' that eliminates the tadpoles, up to the sign. As at one loop, this will exactly kill the tree level mass due to  the gap equation. Overall, we will thus be left by the same diagrams of the Curci-Ferrari model, up to the tree level mass. Notice though that the mass running in the loops will still need an appropriate re-expansion up to the considered order, as the actual mass is defined from $m^2\equiv-\mu^\epsilon\sigma/\zeta(g^2)$, see \eqref{echtemassa}. This will lead to further differences with the Curci-Ferrari case.

 \section{A few more words about the effective potential}\label{app3}
 \begin{figure}[h]
\includegraphics[width=0.43\textwidth]{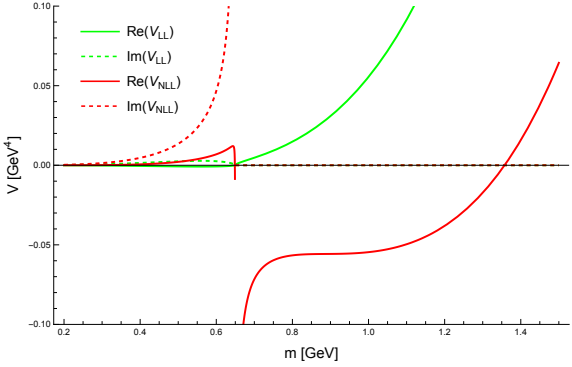}
\caption{Leading $V_{LL}(m)$ and next-to-leading log $V_{NLL}(m)$ renormalization group improved effective potential  at $\omu_0=1~\text{GeV}$ with choice of parameter $\lambda_{\MSbartiny}(\mu_{0})=0.316$ (see Section III.E).}
\label{fig:RGV}
\end{figure}

 \begin{figure}[h]
\includegraphics[width=0.43\textwidth]{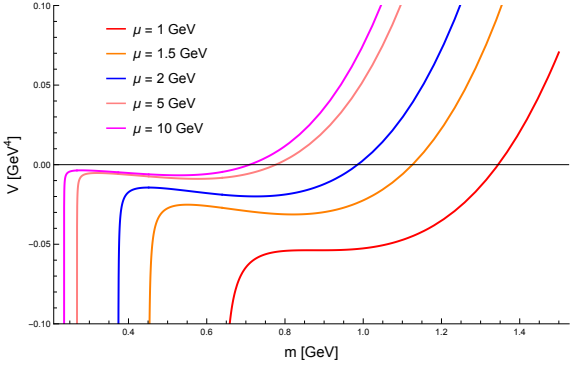}
\caption{Next-to-leading log $V_{NLL}(m)$ renormalization group improved effective potential for various values of $\omu$.}
\label{fig:RGV2}
\end{figure}
In \eqref{LL2}, we constructed an RG improved potential up to leading log (LL) order. The solution \eqref{eq:gapsol} corresponds to a genuine stationary point as $\frac{\p V_{LL}}{\p m}=0$. Upon closer inspection we notice it actually corresponds to a minimum of $\text{Re}(V)$, and a maximum of $\text{Im}(V)$. We kept however using \eqref{eq:gapsol} as a solution to the gap equation in the remainder of the paper\footnote{There is actually also the solution at $m^2=0$. However, this is not to be trusted. To keep the expansion under control, we should then assume $\mu\sim 0$, implying $g^2\sim \infty$ in the $\MSbar$ scheme, thence invalidating the expansion. The relevant expansion parameter at $\omu_0=1~\text{GeV}$, $\lambda_{\MSbartiny}(\mu_{0})=0.316$, is sufficiently small to trust the used expansion more or less, with moreover $m_{\text{sol.}}/\omu_0$ close to $1$.}. Before discussing the validity of this approximation, we note that we chose to minimize $V_{LL}(m(\omu))$ w.r.t.~to the variable $m(\omu)$. Up to the considered order, as noted below \eqref{LL6}, one can reinterpret it in terms of $V(m(\omu=m))$ and thence minimize w.r.t.~$m(m)$ rather than $m(\omu)$. We did not do so, as our eventual interest lies in the RG flow of the propagators in the DIS scheme, with initial conditions set at $\omu_0$. This would require, after having minimized w.r.t.~$m(m)$, to flow this $\MSbar$ solution from $\omu=m$ to $\omu=\omu_0$ using the $\MSbar$ anomalous dimensions, followed by converting it to the desired initial value in the DIS scheme via the conversion \eqref{conversion}. As we do not expect the $\MSbar$ scheme to be trustworthy at lower scales, we intentionally first set $\omu=\omu_0$ and then directly extract an estimate for the $\MSbar$ initial conditions at the chosen, high enough, scale $\omu_0$, followed by the conversion to DIS.

In passing, we notice here that for the LL minimum solution \eqref{eq:gapsol}, $\omu \frac{d}{d\omu}m_{\text{sol.},\MSbartiny}=0$ up to the considered order. Although it might look strange to find a RG invariant solution of the LL gap equation, this will be always the case in the LL case, which after all is based on the zeroth order (``classical'') potential, dressed with logs. This common feature can be easily checked using the related formalisms of \cite{Kastening:1991gv,Bando:1992np} as for single scale theory, the LL potential will always have the structure of \eqref{LL6} with a solution similar to \eqref{eq:gapsol}. This no-running of the solution is thus an artefact of the LL approximation.

This being said, we can now have a look at what happens when we include the  next-to-leading-log (NLL) corrections. As we know the first term of the NLL series, we can use this to our advantage. Indeed, generalizing \eqref{LL2}, we may write
\begin{eqnarray}\label{LL2bis}
  V_{NLL}(m)&=&\frac{9}{13} \frac{N^2-1}{N}\frac{m^4}{2g^2}\left(\sum_{n=0}^{\infty} v_n u^n+g^2\sum_{n=0}^{\infty} w_n u^n\right),\nonumber\\
  &=& \frac{9}{13} \frac{N^2-1}{N}\frac{m^4}{2g^2}\left(F(u) + g^2 G(u)\right)
\end{eqnarray}
where $w_0=-\frac{113 N}{288\pi^2}$, as it follows from \eqref{eq:Vm0}. The function $F(u)$ was already determined in \eqref{LL5}, and upon imposing \eqref{LL1} to next-to-leading order, we get
\begin{eqnarray}\label{LL3bis}
  &&(\beta_1+\gamma_1)F(u) + \gamma_0 G(u)\nonumber\\ &&- (1+\beta_0u)G'(u)+\left(\frac{\gamma_0}{2}-\beta_1u\right)F'(u)=0
\end{eqnarray}
which solves uniquely to
\begin{eqnarray}\label{LL4bis}
G(u)&=&\frac{1}{288 \beta_0^2\pi ^2 } (1+\beta_0 u)^{\gamma_0/\beta0}\times\\
&&\hspace{-1.5cm}\left(-113N\beta_0^2-288\beta_0\beta_1\gamma_0\pi^2u+288\beta_0\gamma_1\pi^2u+\ln(1+\beta_0u)\times\right.\nonumber\\&&\left.\hspace{-1.5cm}\pi^2(288\beta_0\beta_1 +144\beta_0^2\gamma_0+288\beta_1\gamma_0+288\beta_1\gamma_0^2+144\beta_0\gamma_0^2)\right)\nonumber
\end{eqnarray}
As a check on this result, re-expanding \eqref{LL2bis} up to order $g^2$ gives the correct terms upon comparison with the two-loop $\MSbar$ effective potential as it was computed first in \cite{Verschelde:2001ia}, later on verified in \cite{Gracey:2004bk}, up to the constant term of the two-loop piece of course, which is the first of the next-to-next-to-leading log order terms.

Focussing on the SU(3) case and setting as before $\lambda_{\MSbartiny}(\mu_{0})=0.316$ at $\omu=\mu_{0}=1~\text{GeV}$, we come to Fig.~\ref{fig:RGV}, where both the LL and NLL potential are shown. We see that going to next-to-leading log order in the RG improved expansion does shift the location of the minimum to around 0.8~GeV, but it does land in the region where the improved NLL potential is real-valued. For the record, we refrain from using this minimum in Section III.V, as we only solved the two-point functions RG flow at leading order. Actually, the potential is rather flat in the region of interest, so let us see what happens in terms of a variable $\omu$, which will also allow to verify if the expected UV RG asymptotics is reached. We first convert the fit value $\lambda_{\MSbartiny}(\mu_{0})=0.316$ into the corresponding $\lms$-value through inversion of the two-loop expression
\begin{equation}\label{LL100}
  g^2(\omu)=\frac{1}{\beta_0\ln\frac{\omu^2}{\lms^2}}\left(1-\frac{\beta_1}{\beta_0}\frac{\ln\ln\frac{\omu^2}{\lms^2}}{\beta_0\ln\frac{\omu^2}{\lms^2}}\right)
\end{equation}
at $\omu=\mu_0$, yielding $\lms=0.630~\text{GeV}$. Feeding this and \eqref{LL100} back into $V_{NLL}(m)$ allows one to show the potential for various values of $\omu$, as shown in Fig.~\ref{fig:RGV2}. We observe a clear, real-valued minimum which lowers for growing $\omu$.
 \begin{figure}[t]
\includegraphics[width=0.43\textwidth]{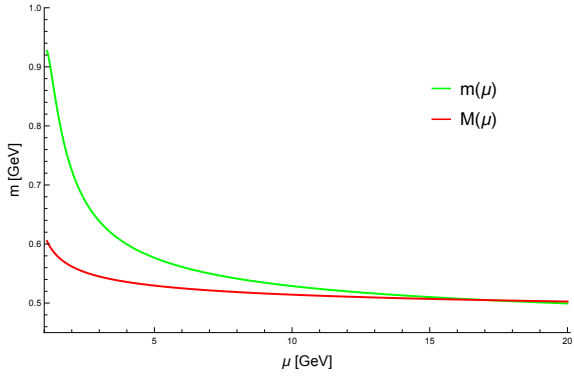}
\caption{Running minimum of $V_{NLL}(m)$ (green) compared to the expected UV asymptotics (red), see \eqref{fitm} .}
\label{fig:RGV3}
\end{figure}
At last, solving for the minimum leads to Fig.~\ref{fig:RGV3}, shown together with a fitted (over the interval $\mu=[5,10]~\text{GeV}$) \begin{equation}M(\omu)=0.571\left(\ln\frac{\omu}{\lms}\right)^{-9/88},\label{fitm}\end{equation} consistent with the expected UV RG behaviour in terms of $\partial \ln m/\partial \ln \omu=(\gamma_0/2) g^2+ \ldots$. For too low values of $\omu$, i.e.~too close to the $\MSbar$ Landau pole at $\omu=\lms$, we should not trust the results anymore. The Landau pole is eventually also what pushes the potentials in FIG.~\ref{fig:RGV2} to $-\infty$ and eventually into the complex region for too small $m$, making the quantity $\beta_0 g^2(\omu)\ln\frac{m^2}{\omu^2}$ too large. In any case, log RG resummations capture information from all orders, whilst the gap equation imposes a further constraint between terms of different orders. It would be interesting to investigate more sophisticated RG improvements of the effective potential, so that for example the RG flow of the solution is strictly consistent with the anomalous dimension of the mass at the chosen order, for any value of the RG scale, but this falls beyond the scope of the current paper. This might also further complicate the numerics (stability) of the fitting procedure. We plan to come back to this in future work.

\bibliography{cfbib}

\begin{thebibliography}{59}%
\makeatletter
\providecommand \@ifxundefined [1]{%
 \@ifx{#1\undefined}
}%
\providecommand \@ifnum [1]{%
 \ifnum #1\expandafter \@firstoftwo
 \else \expandafter \@secondoftwo
 \fi
}%
\providecommand \@ifx [1]{%
 \ifx #1\expandafter \@firstoftwo
 \else \expandafter \@secondoftwo
 \fi
}%
\providecommand \natexlab [1]{#1}%
\providecommand \enquote  [1]{``#1''}%
\providecommand \bibnamefont  [1]{#1}%
\providecommand \bibfnamefont [1]{#1}%
\providecommand \citenamefont [1]{#1}%
\providecommand \href@noop [0]{\@secondoftwo}%
\providecommand \href [0]{\begingroup \@sanitize@url \@href}%
\providecommand \@href[1]{\@@startlink{#1}\@@href}%
\providecommand \@@href[1]{\endgroup#1\@@endlink}%
\providecommand \@sanitize@url [0]{\catcode `\\12\catcode `\$12\catcode
  `\&12\catcode `\#12\catcode `\^12\catcode `\_12\catcode `\%12\relax}%
\providecommand \@@startlink[1]{}%
\providecommand \@@endlink[0]{}%
\providecommand \url  [0]{\begingroup\@sanitize@url \@url }%
\providecommand \@url [1]{\endgroup\@href {#1}{\urlprefix }}%
\providecommand \urlprefix  [0]{URL }%
\providecommand \Eprint [0]{\href }%
\providecommand \doibase [0]{https://doi.org/}%
\providecommand \selectlanguage [0]{\@gobble}%
\providecommand \bibinfo  [0]{\@secondoftwo}%
\providecommand \bibfield  [0]{\@secondoftwo}%
\providecommand \translation [1]{[#1]}%
\providecommand \BibitemOpen [0]{}%
\providecommand \bibitemStop [0]{}%
\providecommand \bibitemNoStop [0]{.\EOS\space}%
\providecommand \EOS [0]{\spacefactor3000\relax}%
\providecommand \BibitemShut  [1]{\csname bibitem#1\endcsname}%
\let\auto@bib@innerbib\@empty
\bibitem [{\citenamefont {Curci}\ and\ \citenamefont
  {Ferrari}(1976{\natexlab{a}})}]{Curci:1976bt}%
  \BibitemOpen
  \bibfield  {author} {\bibinfo {author} {\bibfnamefont {G.}~\bibnamefont
  {Curci}}\ and\ \bibinfo {author} {\bibfnamefont {R.}~\bibnamefont
  {Ferrari}},\ }\href {https://doi.org/10.1007/BF02729999} {\bibfield
  {journal} {\bibinfo  {journal} {Nuovo Cim.}\ }\textbf {\bibinfo {volume}
  {A32}},\ \bibinfo {pages} {151} (\bibinfo {year}
  {1976}{\natexlab{a}})}\BibitemShut {NoStop}%
\bibitem [{\citenamefont {Curci}\ and\ \citenamefont
  {Ferrari}(1976{\natexlab{b}})}]{Curci:1976kh}%
  \BibitemOpen
  \bibfield  {author} {\bibinfo {author} {\bibfnamefont {G.}~\bibnamefont
  {Curci}}\ and\ \bibinfo {author} {\bibfnamefont {R.}~\bibnamefont
  {Ferrari}},\ }\href {https://doi.org/10.1007/BF02730056} {\bibfield
  {journal} {\bibinfo  {journal} {Nuovo Cim. A}\ }\textbf {\bibinfo {volume}
  {35}},\ \bibinfo {pages} {1} (\bibinfo {year} {1976}{\natexlab{b}})},\
  \bibinfo {note} {[Erratum: Nuovo Cim.A 47, 555 (1978)]}\BibitemShut {NoStop}%
\bibitem [{\citenamefont {Gracey}(2003)}]{Gracey:2002yt}%
  \BibitemOpen
  \bibfield  {author} {\bibinfo {author} {\bibfnamefont {J.~A.}\ \bibnamefont
  {Gracey}},\ }\href {https://doi.org/10.1016/S0370-2693(02)03077-0} {\bibfield
   {journal} {\bibinfo  {journal} {Phys. Lett.}\ }\textbf {\bibinfo {volume}
  {B552}},\ \bibinfo {pages} {101} (\bibinfo {year} {2003})},\ \Eprint
  {https://arxiv.org/abs/hep-th/0211144} {arXiv:hep-th/0211144 [hep-th]}
  \BibitemShut {NoStop}%
\bibitem [{\citenamefont {Tissier}\ and\ \citenamefont
  {Wschebor}(2010)}]{Tissier:2010ts}%
  \BibitemOpen
  \bibfield  {author} {\bibinfo {author} {\bibfnamefont {M.}~\bibnamefont
  {Tissier}}\ and\ \bibinfo {author} {\bibfnamefont {N.}~\bibnamefont
  {Wschebor}},\ }\href {https://doi.org/10.1103/PhysRevD.82.101701} {\bibfield
  {journal} {\bibinfo  {journal} {Phys. Rev.}\ }\textbf {\bibinfo {volume}
  {D82}},\ \bibinfo {pages} {101701} (\bibinfo {year} {2010})},\ \Eprint
  {https://arxiv.org/abs/1004.1607} {arXiv:1004.1607 [hep-ph]} \BibitemShut
  {NoStop}%
\bibitem [{\citenamefont {Tissier}\ and\ \citenamefont
  {Wschebor}(2011)}]{Tissier:2011ey}%
  \BibitemOpen
  \bibfield  {author} {\bibinfo {author} {\bibfnamefont {M.}~\bibnamefont
  {Tissier}}\ and\ \bibinfo {author} {\bibfnamefont {N.}~\bibnamefont
  {Wschebor}},\ }\href {https://doi.org/10.1103/PhysRevD.84.045018} {\bibfield
  {journal} {\bibinfo  {journal} {Phys. Rev.}\ }\textbf {\bibinfo {volume}
  {D84}},\ \bibinfo {pages} {045018} (\bibinfo {year} {2011})},\ \Eprint
  {https://arxiv.org/abs/1105.2475} {arXiv:1105.2475 [hep-th]} \BibitemShut
  {NoStop}%
\bibitem [{\citenamefont {Pelaez}\ \emph {et~al.}(2013)\citenamefont {Pelaez},
  \citenamefont {Tissier},\ and\ \citenamefont {Wschebor}}]{Pelaez:2013cpa}%
  \BibitemOpen
  \bibfield  {author} {\bibinfo {author} {\bibfnamefont {M.}~\bibnamefont
  {Pelaez}}, \bibinfo {author} {\bibfnamefont {M.}~\bibnamefont {Tissier}},\
  and\ \bibinfo {author} {\bibfnamefont {N.}~\bibnamefont {Wschebor}},\ }\href
  {https://doi.org/10.1103/PhysRevD.88.125003} {\bibfield  {journal} {\bibinfo
  {journal} {Phys. Rev. D}\ }\textbf {\bibinfo {volume} {88}},\ \bibinfo
  {pages} {125003} (\bibinfo {year} {2013})},\ \Eprint
  {https://arxiv.org/abs/1310.2594} {arXiv:1310.2594 [hep-th]} \BibitemShut
  {NoStop}%
\bibitem [{\citenamefont {Reinosa}\ \emph
  {et~al.}(2017{\natexlab{a}})\citenamefont {Reinosa}, \citenamefont {Serreau},
  \citenamefont {Tissier},\ and\ \citenamefont {Wschebor}}]{Reinosa:2017qtf}%
  \BibitemOpen
  \bibfield  {author} {\bibinfo {author} {\bibfnamefont {U.}~\bibnamefont
  {Reinosa}}, \bibinfo {author} {\bibfnamefont {J.}~\bibnamefont {Serreau}},
  \bibinfo {author} {\bibfnamefont {M.}~\bibnamefont {Tissier}},\ and\ \bibinfo
  {author} {\bibfnamefont {N.}~\bibnamefont {Wschebor}},\ }\href
  {https://doi.org/10.1103/PhysRevD.96.014005} {\bibfield  {journal} {\bibinfo
  {journal} {Phys. Rev.}\ }\textbf {\bibinfo {volume} {D96}},\ \bibinfo {pages}
  {014005} (\bibinfo {year} {2017}{\natexlab{a}})},\ \Eprint
  {https://arxiv.org/abs/1703.04041} {arXiv:1703.04041 [hep-th]} \BibitemShut
  {NoStop}%
\bibitem [{\citenamefont {Gracey}\ \emph {et~al.}(2019)\citenamefont {Gracey},
  \citenamefont {Pel\'aez}, \citenamefont {Reinosa},\ and\ \citenamefont
  {Tissier}}]{Gracey:2019xom}%
  \BibitemOpen
  \bibfield  {author} {\bibinfo {author} {\bibfnamefont {J.~A.}\ \bibnamefont
  {Gracey}}, \bibinfo {author} {\bibfnamefont {M.}~\bibnamefont {Pel\'aez}},
  \bibinfo {author} {\bibfnamefont {U.}~\bibnamefont {Reinosa}},\ and\ \bibinfo
  {author} {\bibfnamefont {M.}~\bibnamefont {Tissier}},\ }\href
  {https://doi.org/10.1103/PhysRevD.100.034023} {\bibfield  {journal} {\bibinfo
   {journal} {Phys. Rev. D}\ }\textbf {\bibinfo {volume} {100}},\ \bibinfo
  {pages} {034023} (\bibinfo {year} {2019})},\ \Eprint
  {https://arxiv.org/abs/1905.07262} {arXiv:1905.07262 [hep-th]} \BibitemShut
  {NoStop}%
\bibitem [{\citenamefont {Dall'Olio}\ and\ \citenamefont
  {Weber}(2022)}]{DallOlio:2020xpu}%
  \BibitemOpen
  \bibfield  {author} {\bibinfo {author} {\bibfnamefont {P.}~\bibnamefont
  {Dall'Olio}}\ and\ \bibinfo {author} {\bibfnamefont {A.}~\bibnamefont
  {Weber}},\ }\href {https://doi.org/10.1016/j.aop.2022.168801} {\bibfield
  {journal} {\bibinfo  {journal} {Annals Phys.}\ }\textbf {\bibinfo {volume}
  {439}},\ \bibinfo {pages} {168801} (\bibinfo {year} {2022})},\ \Eprint
  {https://arxiv.org/abs/2012.02389} {arXiv:2012.02389 [hep-th]} \BibitemShut
  {NoStop}%
\bibitem [{\citenamefont {Pel\'aez}\ \emph {et~al.}(2021)\citenamefont
  {Pel\'aez}, \citenamefont {Reinosa}, \citenamefont {Serreau}, \citenamefont
  {Tissier},\ and\ \citenamefont {Wschebor}}]{Pelaez:2021tpq}%
  \BibitemOpen
  \bibfield  {author} {\bibinfo {author} {\bibfnamefont {M.}~\bibnamefont
  {Pel\'aez}}, \bibinfo {author} {\bibfnamefont {U.}~\bibnamefont {Reinosa}},
  \bibinfo {author} {\bibfnamefont {J.}~\bibnamefont {Serreau}}, \bibinfo
  {author} {\bibfnamefont {M.}~\bibnamefont {Tissier}},\ and\ \bibinfo {author}
  {\bibfnamefont {N.}~\bibnamefont {Wschebor}},\ }\href
  {https://doi.org/10.1088/1361-6633/ac36b8} {\bibfield  {journal} {\bibinfo
  {journal} {Rept. Prog. Phys.}\ }\textbf {\bibinfo {volume} {84}},\ \bibinfo
  {pages} {124202} (\bibinfo {year} {2021})},\ \Eprint
  {https://arxiv.org/abs/2106.04526} {arXiv:2106.04526 [hep-th]} \BibitemShut
  {NoStop}%
\bibitem [{\citenamefont {Barrios}\ \emph {et~al.}(2021)\citenamefont
  {Barrios}, \citenamefont {Gracey}, \citenamefont {and},\ and\ \citenamefont
  {Reinosa}}]{Barrios:2021cks}%
  \BibitemOpen
  \bibfield  {author} {\bibinfo {author} {\bibfnamefont {N.}~\bibnamefont
  {Barrios}}, \bibinfo {author} {\bibfnamefont {J.~A.}\ \bibnamefont {Gracey}},
  \bibinfo {author} {\bibfnamefont {M.~P.}\ \bibnamefont {and}},\ and\ \bibinfo
  {author} {\bibfnamefont {U.}~\bibnamefont {Reinosa}},\ }\href
  {https://doi.org/10.1103/PhysRevD.104.094019} {\bibfield  {journal} {\bibinfo
   {journal} {Phys. Rev. D}\ }\textbf {\bibinfo {volume} {104}},\ \bibinfo
  {pages} {094019} (\bibinfo {year} {2021})},\ \Eprint
  {https://arxiv.org/abs/2103.16218} {arXiv:2103.16218 [hep-th]} \BibitemShut
  {NoStop}%
\bibitem [{\citenamefont {Reinosa}\ \emph {et~al.}(2015)\citenamefont
  {Reinosa}, \citenamefont {Serreau}, \citenamefont {Tissier},\ and\
  \citenamefont {Wschebor}}]{Reinosa:2014zta}%
  \BibitemOpen
  \bibfield  {author} {\bibinfo {author} {\bibfnamefont {U.}~\bibnamefont
  {Reinosa}}, \bibinfo {author} {\bibfnamefont {J.}~\bibnamefont {Serreau}},
  \bibinfo {author} {\bibfnamefont {M.}~\bibnamefont {Tissier}},\ and\ \bibinfo
  {author} {\bibfnamefont {N.}~\bibnamefont {Wschebor}},\ }\href
  {https://doi.org/10.1103/PhysRevD.91.045035} {\bibfield  {journal} {\bibinfo
  {journal} {Phys. Rev.}\ }\textbf {\bibinfo {volume} {D91}},\ \bibinfo {pages}
  {045035} (\bibinfo {year} {2015})},\ \Eprint
  {https://arxiv.org/abs/1412.5672} {arXiv:1412.5672 [hep-th]} \BibitemShut
  {NoStop}%
\bibitem [{\citenamefont {Reinosa}\ \emph
  {et~al.}(2017{\natexlab{b}})\citenamefont {Reinosa}, \citenamefont {Serreau},
  \citenamefont {Tissier},\ and\ \citenamefont
  {Tresmontant}}]{Reinosa:2016iml}%
  \BibitemOpen
  \bibfield  {author} {\bibinfo {author} {\bibfnamefont {U.}~\bibnamefont
  {Reinosa}}, \bibinfo {author} {\bibfnamefont {J.}~\bibnamefont {Serreau}},
  \bibinfo {author} {\bibfnamefont {M.}~\bibnamefont {Tissier}},\ and\ \bibinfo
  {author} {\bibfnamefont {A.}~\bibnamefont {Tresmontant}},\ }\href
  {https://doi.org/10.1103/PhysRevD.95.045014} {\bibfield  {journal} {\bibinfo
  {journal} {Phys. Rev.}\ }\textbf {\bibinfo {volume} {D95}},\ \bibinfo {pages}
  {045014} (\bibinfo {year} {2017}{\natexlab{b}})},\ \Eprint
  {https://arxiv.org/abs/1606.08012} {arXiv:1606.08012 [hep-th]} \BibitemShut
  {NoStop}%
\bibitem [{\citenamefont {Maelger}\ \emph {et~al.}(2018)\citenamefont
  {Maelger}, \citenamefont {Reinosa},\ and\ \citenamefont
  {Serreau}}]{Maelger:2018vow}%
  \BibitemOpen
  \bibfield  {author} {\bibinfo {author} {\bibfnamefont {J.}~\bibnamefont
  {Maelger}}, \bibinfo {author} {\bibfnamefont {U.}~\bibnamefont {Reinosa}},\
  and\ \bibinfo {author} {\bibfnamefont {J.}~\bibnamefont {Serreau}},\ }\href
  {https://doi.org/10.1103/PhysRevD.98.094020} {\bibfield  {journal} {\bibinfo
  {journal} {Phys. Rev.}\ }\textbf {\bibinfo {volume} {D98}},\ \bibinfo {pages}
  {094020} (\bibinfo {year} {2018})},\ \Eprint
  {https://arxiv.org/abs/1805.10015} {arXiv:1805.10015 [hep-th]} \BibitemShut
  {NoStop}%
\bibitem [{\citenamefont {Serreau}\ and\ \citenamefont
  {Tissier}(2012)}]{Serreau:2012cg}%
  \BibitemOpen
  \bibfield  {author} {\bibinfo {author} {\bibfnamefont {J.}~\bibnamefont
  {Serreau}}\ and\ \bibinfo {author} {\bibfnamefont {M.}~\bibnamefont
  {Tissier}},\ }\href {https://doi.org/10.1016/j.physletb.2012.04.041}
  {\bibfield  {journal} {\bibinfo  {journal} {Phys. Lett. B}\ }\textbf
  {\bibinfo {volume} {712}},\ \bibinfo {pages} {97} (\bibinfo {year} {2012})},\
  \Eprint {https://arxiv.org/abs/1202.3432} {arXiv:1202.3432 [hep-th]}
  \BibitemShut {NoStop}%
\bibitem [{\citenamefont {Tissier}(2018)}]{Tissier:2017fqf}%
  \BibitemOpen
  \bibfield  {author} {\bibinfo {author} {\bibfnamefont {M.}~\bibnamefont
  {Tissier}},\ }\href {https://doi.org/10.1016/j.physletb.2018.07.043}
  {\bibfield  {journal} {\bibinfo  {journal} {Phys. Lett. B}\ }\textbf
  {\bibinfo {volume} {784}},\ \bibinfo {pages} {146} (\bibinfo {year}
  {2018})},\ \Eprint {https://arxiv.org/abs/1711.08694} {arXiv:1711.08694
  [hep-th]} \BibitemShut {NoStop}%
\bibitem [{\citenamefont {Reinosa}\ \emph {et~al.}(2021)\citenamefont
  {Reinosa}, \citenamefont {Serreau}, \citenamefont {Terin},\ and\
  \citenamefont {Tissier}}]{Reinosa:2020skx}%
  \BibitemOpen
  \bibfield  {author} {\bibinfo {author} {\bibfnamefont {U.}~\bibnamefont
  {Reinosa}}, \bibinfo {author} {\bibfnamefont {J.}~\bibnamefont {Serreau}},
  \bibinfo {author} {\bibfnamefont {R.~C.}\ \bibnamefont {Terin}},\ and\
  \bibinfo {author} {\bibfnamefont {M.}~\bibnamefont {Tissier}},\ }\href
  {https://doi.org/10.21468/SciPostPhys.10.2.035} {\bibfield  {journal}
  {\bibinfo  {journal} {SciPost Phys.}\ }\textbf {\bibinfo {volume} {10}},\
  \bibinfo {pages} {035} (\bibinfo {year} {2021})},\ \Eprint
  {https://arxiv.org/abs/2004.12413} {arXiv:2004.12413 [hep-th]} \BibitemShut
  {NoStop}%
\bibitem [{\citenamefont {Verschelde}\ \emph {et~al.}(2001)\citenamefont
  {Verschelde}, \citenamefont {Knecht}, \citenamefont {Van~Acoleyen},\ and\
  \citenamefont {Vanderkelen}}]{Verschelde:2001ia}%
  \BibitemOpen
  \bibfield  {author} {\bibinfo {author} {\bibfnamefont {H.}~\bibnamefont
  {Verschelde}}, \bibinfo {author} {\bibfnamefont {K.}~\bibnamefont {Knecht}},
  \bibinfo {author} {\bibfnamefont {K.}~\bibnamefont {Van~Acoleyen}},\ and\
  \bibinfo {author} {\bibfnamefont {M.}~\bibnamefont {Vanderkelen}},\ }\href
  {https://doi.org/10.1016/S0370-2693(01)00929-7} {\bibfield  {journal}
  {\bibinfo  {journal} {Phys. Lett.}\ }\textbf {\bibinfo {volume} {B516}},\
  \bibinfo {pages} {307} (\bibinfo {year} {2001})},\ \Eprint
  {https://arxiv.org/abs/hep-th/0105018} {arXiv:hep-th/0105018 [hep-th]}
  \BibitemShut {NoStop}%
\bibitem [{\citenamefont {Zimmermann}(1985)}]{Zimmermann:1984sx}%
  \BibitemOpen
  \bibfield  {author} {\bibinfo {author} {\bibfnamefont {W.}~\bibnamefont
  {Zimmermann}},\ }\href {https://doi.org/10.1007/BF01206187} {\bibfield
  {journal} {\bibinfo  {journal} {Commun. Math. Phys.}\ }\textbf {\bibinfo
  {volume} {97}},\ \bibinfo {pages} {211} (\bibinfo {year} {1985})}\BibitemShut
  {NoStop}%
\bibitem [{\citenamefont {Heinemeyer}\ \emph {et~al.}(2019)\citenamefont
  {Heinemeyer}, \citenamefont {Mondrag{\'o}n}, \citenamefont {Tracas},\ and\
  \citenamefont {Zoupanos}}]{Heinemeyer:2019vbc}%
  \BibitemOpen
  \bibfield  {author} {\bibinfo {author} {\bibfnamefont {S.}~\bibnamefont
  {Heinemeyer}}, \bibinfo {author} {\bibfnamefont {M.}~\bibnamefont
  {Mondrag{\'o}n}}, \bibinfo {author} {\bibfnamefont {N.}~\bibnamefont
  {Tracas}},\ and\ \bibinfo {author} {\bibfnamefont {G.}~\bibnamefont
  {Zoupanos}}\ }(\bibinfo {year} {2019})\ \Eprint
  {https://arxiv.org/abs/1904.00410} {arXiv:1904.00410 [hep-ph]} \BibitemShut
  {NoStop}%
\bibitem [{\citenamefont {Dell'Antonio}\ and\ \citenamefont
  {Zwanziger}(1991)}]{DellAntonio:1991mms}%
  \BibitemOpen
  \bibfield  {author} {\bibinfo {author} {\bibfnamefont {G.}~\bibnamefont
  {Dell'Antonio}}\ and\ \bibinfo {author} {\bibfnamefont {D.}~\bibnamefont
  {Zwanziger}},\ }\href {https://doi.org/10.1007/BF02099494} {\bibfield
  {journal} {\bibinfo  {journal} {Commun. Math. Phys.}\ }\textbf {\bibinfo
  {volume} {138}},\ \bibinfo {pages} {291} (\bibinfo {year}
  {1991})}\BibitemShut {NoStop}%
\bibitem [{\citenamefont {Lavelle}\ and\ \citenamefont
  {McMullan}(1997)}]{Lavelle:1995ty}%
  \BibitemOpen
  \bibfield  {author} {\bibinfo {author} {\bibfnamefont {M.}~\bibnamefont
  {Lavelle}}\ and\ \bibinfo {author} {\bibfnamefont {D.}~\bibnamefont
  {McMullan}},\ }\href {https://doi.org/10.1016/S0370-1573(96)00019-1}
  {\bibfield  {journal} {\bibinfo  {journal} {Phys. Rept.}\ }\textbf {\bibinfo
  {volume} {279}},\ \bibinfo {pages} {1} (\bibinfo {year} {1997})},\ \Eprint
  {https://arxiv.org/abs/hep-ph/9509344} {arXiv:hep-ph/9509344 [hep-ph]}
  \BibitemShut {NoStop}%
\bibitem [{\citenamefont {Capri}\ \emph {et~al.}(2015)\citenamefont {Capri},
  \citenamefont {Dudal}, \citenamefont {Fiorentini}, \citenamefont {Guimaraes},
  \citenamefont {Justo}, \citenamefont {Pereira}, \citenamefont {Mintz},
  \citenamefont {Palhares}, \citenamefont {Sobreiro},\ and\ \citenamefont
  {Sorella}}]{Capri:2015ixa}%
  \BibitemOpen
  \bibfield  {author} {\bibinfo {author} {\bibfnamefont {M.~A.~L.}\
  \bibnamefont {Capri}}, \bibinfo {author} {\bibfnamefont {D.}~\bibnamefont
  {Dudal}}, \bibinfo {author} {\bibfnamefont {D.}~\bibnamefont {Fiorentini}},
  \bibinfo {author} {\bibfnamefont {M.~S.}\ \bibnamefont {Guimaraes}}, \bibinfo
  {author} {\bibfnamefont {I.~F.}\ \bibnamefont {Justo}}, \bibinfo {author}
  {\bibfnamefont {A.~D.}\ \bibnamefont {Pereira}}, \bibinfo {author}
  {\bibfnamefont {B.~W.}\ \bibnamefont {Mintz}}, \bibinfo {author}
  {\bibfnamefont {L.~F.}\ \bibnamefont {Palhares}}, \bibinfo {author}
  {\bibfnamefont {R.~F.}\ \bibnamefont {Sobreiro}},\ and\ \bibinfo {author}
  {\bibfnamefont {S.~P.}\ \bibnamefont {Sorella}},\ }\href
  {https://doi.org/10.1103/PhysRevD.92.045039} {\bibfield  {journal} {\bibinfo
  {journal} {Phys. Rev.}\ }\textbf {\bibinfo {volume} {D92}},\ \bibinfo {pages}
  {045039} (\bibinfo {year} {2015})},\ \Eprint
  {https://arxiv.org/abs/1506.06995} {arXiv:1506.06995 [hep-th]} \BibitemShut
  {NoStop}%
\bibitem [{\citenamefont {Verschelde}(1995)}]{Verschelde:1995jj}%
  \BibitemOpen
  \bibfield  {author} {\bibinfo {author} {\bibfnamefont {H.}~\bibnamefont
  {Verschelde}},\ }\href {https://doi.org/10.1016/0370-2693(95)00338-L}
  {\bibfield  {journal} {\bibinfo  {journal} {Phys. Lett.}\ }\textbf {\bibinfo
  {volume} {B351}},\ \bibinfo {pages} {242} (\bibinfo {year}
  {1995})}\BibitemShut {NoStop}%
\bibitem [{\citenamefont {Knecht}\ and\ \citenamefont
  {Verschelde}(2001)}]{Knecht:2001cc}%
  \BibitemOpen
  \bibfield  {author} {\bibinfo {author} {\bibfnamefont {K.}~\bibnamefont
  {Knecht}}\ and\ \bibinfo {author} {\bibfnamefont {H.}~\bibnamefont
  {Verschelde}},\ }\href {https://doi.org/10.1103/PhysRevD.64.085006}
  {\bibfield  {journal} {\bibinfo  {journal} {Phys. Rev. D}\ }\textbf {\bibinfo
  {volume} {64}},\ \bibinfo {pages} {085006} (\bibinfo {year} {2001})},\
  \Eprint {https://arxiv.org/abs/hep-th/0104007} {arXiv:hep-th/0104007}
  \BibitemShut {NoStop}%
\bibitem [{\citenamefont {Dudal}\ \emph {et~al.}(2003)\citenamefont {Dudal},
  \citenamefont {Verschelde},\ and\ \citenamefont {Sorella}}]{Dudal:2002pq}%
  \BibitemOpen
  \bibfield  {author} {\bibinfo {author} {\bibfnamefont {D.}~\bibnamefont
  {Dudal}}, \bibinfo {author} {\bibfnamefont {H.}~\bibnamefont {Verschelde}},\
  and\ \bibinfo {author} {\bibfnamefont {S.~P.}\ \bibnamefont {Sorella}},\
  }\href {https://doi.org/10.1016/S0370-2693(03)00043-1} {\bibfield  {journal}
  {\bibinfo  {journal} {Phys. Lett.}\ }\textbf {\bibinfo {volume} {B555}},\
  \bibinfo {pages} {126} (\bibinfo {year} {2003})},\ \Eprint
  {https://arxiv.org/abs/hep-th/0212182} {arXiv:hep-th/0212182 [hep-th]}
  \BibitemShut {NoStop}%
\bibitem [{\citenamefont {Jackiw}(1974)}]{Jackiw:1974cv}%
  \BibitemOpen
  \bibfield  {author} {\bibinfo {author} {\bibfnamefont {R.}~\bibnamefont
  {Jackiw}},\ }\href {https://doi.org/10.1103/PhysRevD.9.1686} {\bibfield
  {journal} {\bibinfo  {journal} {Phys. Rev.}\ }\textbf {\bibinfo {volume}
  {D9}},\ \bibinfo {pages} {1686} (\bibinfo {year} {1974})}\BibitemShut
  {NoStop}%
\bibitem [{\citenamefont {Peskin}\ and\ \citenamefont
  {Schroeder}(1995)}]{Peskin:1995ev}%
  \BibitemOpen
  \bibfield  {author} {\bibinfo {author} {\bibfnamefont {M.~E.}\ \bibnamefont
  {Peskin}}\ and\ \bibinfo {author} {\bibfnamefont {D.~V.}\ \bibnamefont
  {Schroeder}},\ }\href {http://www.slac.stanford.edu/~mpeskin/QFT.html} {\emph
  {\bibinfo {title} {{An Introduction to quantum field theory}}}}\ (\bibinfo
  {publisher} {Addison-Wesley},\ \bibinfo {address} {Reading, USA},\ \bibinfo
  {year} {1995})\BibitemShut {NoStop}%
\bibitem [{\citenamefont {Capri}\ \emph
  {et~al.}(2018{\natexlab{a}})\citenamefont {Capri}, \citenamefont {Dudal},
  \citenamefont {Guimaraes}, \citenamefont {Pereira}, \citenamefont {Mintz},
  \citenamefont {Palhares},\ and\ \citenamefont {Sorella}}]{Capri:2018ijg}%
  \BibitemOpen
  \bibfield  {author} {\bibinfo {author} {\bibfnamefont {M.~A.~L.}\
  \bibnamefont {Capri}}, \bibinfo {author} {\bibfnamefont {D.}~\bibnamefont
  {Dudal}}, \bibinfo {author} {\bibfnamefont {M.~S.}\ \bibnamefont
  {Guimaraes}}, \bibinfo {author} {\bibfnamefont {A.~D.}\ \bibnamefont
  {Pereira}}, \bibinfo {author} {\bibfnamefont {B.~W.}\ \bibnamefont {Mintz}},
  \bibinfo {author} {\bibfnamefont {L.~F.}\ \bibnamefont {Palhares}},\ and\
  \bibinfo {author} {\bibfnamefont {S.~P.}\ \bibnamefont {Sorella}},\ }\href
  {https://doi.org/10.1016/j.physletb.2018.03.058} {\bibfield  {journal}
  {\bibinfo  {journal} {Phys. Lett.}\ }\textbf {\bibinfo {volume} {B781}},\
  \bibinfo {pages} {48} (\bibinfo {year} {2018}{\natexlab{a}})},\ \Eprint
  {https://arxiv.org/abs/1802.04582} {arXiv:1802.04582 [hep-th]} \BibitemShut
  {NoStop}%
\bibitem [{\citenamefont {Verschelde}\ \emph {et~al.}(1997)\citenamefont
  {Verschelde}, \citenamefont {Schelstraete},\ and\ \citenamefont
  {Vanderkelen}}]{Verschelde:1997jx}%
  \BibitemOpen
  \bibfield  {author} {\bibinfo {author} {\bibfnamefont {H.}~\bibnamefont
  {Verschelde}}, \bibinfo {author} {\bibfnamefont {S.}~\bibnamefont
  {Schelstraete}},\ and\ \bibinfo {author} {\bibfnamefont {M.}~\bibnamefont
  {Vanderkelen}},\ }\href {https://doi.org/10.1007/s002880050540} {\bibfield
  {journal} {\bibinfo  {journal} {Z. Phys. C}\ }\textbf {\bibinfo {volume}
  {76}},\ \bibinfo {pages} {161} (\bibinfo {year} {1997})}\BibitemShut
  {NoStop}%
\bibitem [{\citenamefont {Browne}\ \emph {et~al.}(2006)\citenamefont {Browne},
  \citenamefont {Dudal}, \citenamefont {Gracey}, \citenamefont {Lemes},
  \citenamefont {Sarandy}, \citenamefont {Sobreiro}, \citenamefont {Sorella},\
  and\ \citenamefont {Verschelde}}]{Browne:2006uy}%
  \BibitemOpen
  \bibfield  {author} {\bibinfo {author} {\bibfnamefont {R.~E.}\ \bibnamefont
  {Browne}}, \bibinfo {author} {\bibfnamefont {D.}~\bibnamefont {Dudal}},
  \bibinfo {author} {\bibfnamefont {J.~A.}\ \bibnamefont {Gracey}}, \bibinfo
  {author} {\bibfnamefont {V.~E.~R.}\ \bibnamefont {Lemes}}, \bibinfo {author}
  {\bibfnamefont {M.~S.}\ \bibnamefont {Sarandy}}, \bibinfo {author}
  {\bibfnamefont {R.~F.}\ \bibnamefont {Sobreiro}}, \bibinfo {author}
  {\bibfnamefont {S.~P.}\ \bibnamefont {Sorella}},\ and\ \bibinfo {author}
  {\bibfnamefont {H.}~\bibnamefont {Verschelde}},\ }\href
  {https://doi.org/10.1088/0305-4470/39/25/S06} {\bibfield  {journal} {\bibinfo
   {journal} {J. Phys. A}\ }\textbf {\bibinfo {volume} {39}},\ \bibinfo {pages}
  {7889} (\bibinfo {year} {2006})},\ \Eprint
  {https://arxiv.org/abs/hep-th/0602291} {arXiv:hep-th/0602291} \BibitemShut
  {NoStop}%
\bibitem [{\citenamefont {Gribov}(1978)}]{Gribov:1977wm}%
  \BibitemOpen
  \bibfield  {author} {\bibinfo {author} {\bibfnamefont {V.~N.}\ \bibnamefont
  {Gribov}},\ }\href {https://doi.org/10.1016/0550-3213(78)90175-X} {\bibfield
  {journal} {\bibinfo  {journal} {Nucl. Phys. B}\ }\textbf {\bibinfo {volume}
  {139}},\ \bibinfo {pages} {1} (\bibinfo {year} {1978})}\BibitemShut {NoStop}%
\bibitem [{\citenamefont {Kroff}\ and\ \citenamefont
  {Reinosa}(2018)}]{Kroff:2018ncl}%
  \BibitemOpen
  \bibfield  {author} {\bibinfo {author} {\bibfnamefont {D.}~\bibnamefont
  {Kroff}}\ and\ \bibinfo {author} {\bibfnamefont {U.}~\bibnamefont
  {Reinosa}},\ }\href {https://doi.org/10.1103/PhysRevD.98.034029} {\bibfield
  {journal} {\bibinfo  {journal} {Phys. Rev.}\ }\textbf {\bibinfo {volume}
  {D98}},\ \bibinfo {pages} {034029} (\bibinfo {year} {2018})},\ \Eprint
  {https://arxiv.org/abs/1803.10188} {arXiv:1803.10188 [hep-th]} \BibitemShut
  {NoStop}%
\bibitem [{\citenamefont {Cucchieri}\ \emph {et~al.}(2016)\citenamefont
  {Cucchieri}, \citenamefont {Dudal}, \citenamefont {Mendes},\ and\
  \citenamefont {Vandersickel}}]{Cucchieri:2016jwg}%
  \BibitemOpen
  \bibfield  {author} {\bibinfo {author} {\bibfnamefont {A.}~\bibnamefont
  {Cucchieri}}, \bibinfo {author} {\bibfnamefont {D.}~\bibnamefont {Dudal}},
  \bibinfo {author} {\bibfnamefont {T.}~\bibnamefont {Mendes}},\ and\ \bibinfo
  {author} {\bibfnamefont {N.}~\bibnamefont {Vandersickel}},\ }\href
  {https://doi.org/10.1103/PhysRevD.93.094513} {\bibfield  {journal} {\bibinfo
  {journal} {Phys. Rev.}\ }\textbf {\bibinfo {volume} {D93}},\ \bibinfo {pages}
  {094513} (\bibinfo {year} {2016})},\ \Eprint
  {https://arxiv.org/abs/1602.01646} {arXiv:1602.01646 [hep-lat]} \BibitemShut
  {NoStop}%
\bibitem [{\citenamefont {Siringo}(2015)}]{Siringo:2015gia}%
  \BibitemOpen
  \bibfield  {author} {\bibinfo {author} {\bibfnamefont {F.}~\bibnamefont
  {Siringo}},\ }\href {https://doi.org/10.1103/PhysRevD.92.074034} {\bibfield
  {journal} {\bibinfo  {journal} {Phys. Rev. D}\ }\textbf {\bibinfo {volume}
  {92}},\ \bibinfo {pages} {074034} (\bibinfo {year} {2015})},\ \Eprint
  {https://arxiv.org/abs/1507.00122} {arXiv:1507.00122 [hep-ph]} \BibitemShut
  {NoStop}%
\bibitem [{\citenamefont {Siringo}\ and\ \citenamefont
  {Comitini}(2018)}]{Siringo:2018uho}%
  \BibitemOpen
  \bibfield  {author} {\bibinfo {author} {\bibfnamefont {F.}~\bibnamefont
  {Siringo}}\ and\ \bibinfo {author} {\bibfnamefont {G.}~\bibnamefont
  {Comitini}},\ }\href {https://doi.org/10.1103/PhysRevD.98.034023} {\bibfield
  {journal} {\bibinfo  {journal} {Phys. Rev. D}\ }\textbf {\bibinfo {volume}
  {98}},\ \bibinfo {pages} {034023} (\bibinfo {year} {2018})},\ \Eprint
  {https://arxiv.org/abs/1806.08397} {arXiv:1806.08397 [hep-ph]} \BibitemShut
  {NoStop}%
\bibitem [{\citenamefont {Comitini}\ and\ \citenamefont
  {Siringo}(2020{\natexlab{a}})}]{Comitini:2020ozt}%
  \BibitemOpen
  \bibfield  {author} {\bibinfo {author} {\bibfnamefont {G.}~\bibnamefont
  {Comitini}}\ and\ \bibinfo {author} {\bibfnamefont {F.}~\bibnamefont
  {Siringo}},\ }\href {https://doi.org/10.1103/PhysRevD.102.094002} {\bibfield
  {journal} {\bibinfo  {journal} {Phys. Rev. D}\ }\textbf {\bibinfo {volume}
  {102}},\ \bibinfo {pages} {094002} (\bibinfo {year} {2020}{\natexlab{a}})},\
  \Eprint {https://arxiv.org/abs/2007.04231} {arXiv:2007.04231 [hep-ph]}
  \BibitemShut {NoStop}%
\bibitem [{\citenamefont {Duarte}\ \emph {et~al.}(2016)\citenamefont {Duarte},
  \citenamefont {Oliveira},\ and\ \citenamefont {Silva}}]{Duarte:2016iko}%
  \BibitemOpen
  \bibfield  {author} {\bibinfo {author} {\bibfnamefont {A.~G.}\ \bibnamefont
  {Duarte}}, \bibinfo {author} {\bibfnamefont {O.}~\bibnamefont {Oliveira}},\
  and\ \bibinfo {author} {\bibfnamefont {P.~J.}\ \bibnamefont {Silva}},\ }\href
  {https://doi.org/10.1103/PhysRevD.94.014502} {\bibfield  {journal} {\bibinfo
  {journal} {Phys. Rev.}\ }\textbf {\bibinfo {volume} {D94}},\ \bibinfo {pages}
  {014502} (\bibinfo {year} {2016})},\ \Eprint
  {https://arxiv.org/abs/1605.00594} {arXiv:1605.00594 [hep-lat]} \BibitemShut
  {NoStop}%
\bibitem [{\citenamefont {Dudal}\ \emph {et~al.}(2005)\citenamefont {Dudal},
  \citenamefont {Sobreiro}, \citenamefont {Sorella},\ and\ \citenamefont
  {Verschelde}}]{Dudal:2005na}%
  \BibitemOpen
  \bibfield  {author} {\bibinfo {author} {\bibfnamefont {D.}~\bibnamefont
  {Dudal}}, \bibinfo {author} {\bibfnamefont {R.~F.}\ \bibnamefont {Sobreiro}},
  \bibinfo {author} {\bibfnamefont {S.~P.}\ \bibnamefont {Sorella}},\ and\
  \bibinfo {author} {\bibfnamefont {H.}~\bibnamefont {Verschelde}},\ }\href
  {https://doi.org/10.1103/PhysRevD.72.014016} {\bibfield  {journal} {\bibinfo
  {journal} {Phys. Rev. D}\ }\textbf {\bibinfo {volume} {72}},\ \bibinfo
  {pages} {014016} (\bibinfo {year} {2005})},\ \Eprint
  {https://arxiv.org/abs/hep-th/0502183} {arXiv:hep-th/0502183} \BibitemShut
  {NoStop}%
\bibitem [{\citenamefont {Kastening}(1992)}]{Kastening:1991gv}%
  \BibitemOpen
  \bibfield  {author} {\bibinfo {author} {\bibfnamefont {B.~M.}\ \bibnamefont
  {Kastening}},\ }\href {https://doi.org/10.1016/0370-2693(92)90021-U}
  {\bibfield  {journal} {\bibinfo  {journal} {Phys. Lett. B}\ }\textbf
  {\bibinfo {volume} {283}},\ \bibinfo {pages} {287} (\bibinfo {year}
  {1992})}\BibitemShut {NoStop}%
\bibitem [{\citenamefont {Bando}\ \emph {et~al.}(1993)\citenamefont {Bando},
  \citenamefont {Kugo}, \citenamefont {Maekawa},\ and\ \citenamefont
  {Nakano}}]{Bando:1992np}%
  \BibitemOpen
  \bibfield  {author} {\bibinfo {author} {\bibfnamefont {M.}~\bibnamefont
  {Bando}}, \bibinfo {author} {\bibfnamefont {T.}~\bibnamefont {Kugo}},
  \bibinfo {author} {\bibfnamefont {N.}~\bibnamefont {Maekawa}},\ and\ \bibinfo
  {author} {\bibfnamefont {H.}~\bibnamefont {Nakano}},\ }\href
  {https://doi.org/10.1016/0370-2693(93)90725-W} {\bibfield  {journal}
  {\bibinfo  {journal} {Phys. Lett. B}\ }\textbf {\bibinfo {volume} {301}},\
  \bibinfo {pages} {83} (\bibinfo {year} {1993})},\ \Eprint
  {https://arxiv.org/abs/hep-ph/9210228} {arXiv:hep-ph/9210228} \BibitemShut
  {NoStop}%
\bibitem [{\citenamefont {Dudal}\ \emph {et~al.}(2018)\citenamefont {Dudal},
  \citenamefont {Oliveira},\ and\ \citenamefont {Silva}}]{Dudal:2018cli}%
  \BibitemOpen
  \bibfield  {author} {\bibinfo {author} {\bibfnamefont {D.}~\bibnamefont
  {Dudal}}, \bibinfo {author} {\bibfnamefont {O.}~\bibnamefont {Oliveira}},\
  and\ \bibinfo {author} {\bibfnamefont {P.~J.}\ \bibnamefont {Silva}},\ }\href
  {https://doi.org/10.1016/j.aop.2018.08.019} {\bibfield  {journal} {\bibinfo
  {journal} {Annals Phys.}\ }\textbf {\bibinfo {volume} {397}},\ \bibinfo
  {pages} {351} (\bibinfo {year} {2018})},\ \Eprint
  {https://arxiv.org/abs/1803.02281} {arXiv:1803.02281 [hep-lat]} \BibitemShut
  {NoStop}%
\bibitem [{\citenamefont {Comitini}\ and\ \citenamefont
  {Siringo}(2020{\natexlab{b}})}]{Comitini:2020}%
  \BibitemOpen
  \bibfield  {author} {\bibinfo {author} {\bibfnamefont {G.}~\bibnamefont
  {Comitini}}\ and\ \bibinfo {author} {\bibfnamefont {F.}~\bibnamefont
  {Siringo}},\ }\href@noop {} {\bibfield  {journal} {\bibinfo  {journal} {Phys.
  Rev.}\ }\textbf {\bibinfo {volume} {D102}},\ \bibinfo {pages} {094002}
  (\bibinfo {year} {2020}{\natexlab{b}})}\BibitemShut {NoStop}%
\bibitem [{\citenamefont {Cucchieri}\ and\ \citenamefont
  {Mendes}(2008{\natexlab{a}})}]{Cucchieri:2007rg}%
  \BibitemOpen
  \bibfield  {author} {\bibinfo {author} {\bibfnamefont {A.}~\bibnamefont
  {Cucchieri}}\ and\ \bibinfo {author} {\bibfnamefont {T.}~\bibnamefont
  {Mendes}},\ }\href {https://doi.org/10.1103/PhysRevLett.100.241601}
  {\bibfield  {journal} {\bibinfo  {journal} {Phys. Rev. Lett.}\ }\textbf
  {\bibinfo {volume} {100}},\ \bibinfo {pages} {241601} (\bibinfo {year}
  {2008}{\natexlab{a}})},\ \Eprint {https://arxiv.org/abs/0712.3517}
  {arXiv:0712.3517 [hep-lat]} \BibitemShut {NoStop}%
\bibitem [{\citenamefont {Cucchieri}\ and\ \citenamefont
  {Mendes}(2008{\natexlab{b}})}]{Cucchieri:2008fc}%
  \BibitemOpen
  \bibfield  {author} {\bibinfo {author} {\bibfnamefont {A.}~\bibnamefont
  {Cucchieri}}\ and\ \bibinfo {author} {\bibfnamefont {T.}~\bibnamefont
  {Mendes}},\ }\href {https://doi.org/10.1103/PhysRevD.78.094503} {\bibfield
  {journal} {\bibinfo  {journal} {Phys. Rev. D}\ }\textbf {\bibinfo {volume}
  {78}},\ \bibinfo {pages} {094503} (\bibinfo {year} {2008}{\natexlab{b}})},\
  \Eprint {https://arxiv.org/abs/0804.2371} {arXiv:0804.2371 [hep-lat]}
  \BibitemShut {NoStop}%
\bibitem [{\citenamefont {Cucchieri}\ \emph {et~al.}(2012)\citenamefont
  {Cucchieri}, \citenamefont {Dudal},\ and\ \citenamefont
  {Vandersickel}}]{Cucchieri:2012cb}%
  \BibitemOpen
  \bibfield  {author} {\bibinfo {author} {\bibfnamefont {A.}~\bibnamefont
  {Cucchieri}}, \bibinfo {author} {\bibfnamefont {D.}~\bibnamefont {Dudal}},\
  and\ \bibinfo {author} {\bibfnamefont {N.}~\bibnamefont {Vandersickel}},\
  }\href {https://doi.org/10.1103/PhysRevD.85.085025} {\bibfield  {journal}
  {\bibinfo  {journal} {Phys. Rev.}\ }\textbf {\bibinfo {volume} {D85}},\
  \bibinfo {pages} {085025} (\bibinfo {year} {2012})},\ \Eprint
  {https://arxiv.org/abs/1202.1912} {arXiv:1202.1912 [hep-th]} \BibitemShut
  {NoStop}%
\bibitem [{\citenamefont {Aguilar}\ and\ \citenamefont
  {Natale}(2004)}]{Aguilar:2004sw}%
  \BibitemOpen
  \bibfield  {author} {\bibinfo {author} {\bibfnamefont {A.~C.}\ \bibnamefont
  {Aguilar}}\ and\ \bibinfo {author} {\bibfnamefont {A.~A.}\ \bibnamefont
  {Natale}},\ }\href {https://doi.org/10.1088/1126-6708/2004/08/057} {\bibfield
   {journal} {\bibinfo  {journal} {JHEP}\ }\textbf {\bibinfo {volume} {08}},\
  \bibinfo {pages} {057}},\ \Eprint {https://arxiv.org/abs/hep-ph/0408254}
  {arXiv:hep-ph/0408254} \BibitemShut {NoStop}%
\bibitem [{\citenamefont {Aguilar}\ \emph {et~al.}(2008)\citenamefont
  {Aguilar}, \citenamefont {Binosi},\ and\ \citenamefont
  {Papavassiliou}}]{Aguilar:2008xm}%
  \BibitemOpen
  \bibfield  {author} {\bibinfo {author} {\bibfnamefont {A.~C.}\ \bibnamefont
  {Aguilar}}, \bibinfo {author} {\bibfnamefont {D.}~\bibnamefont {Binosi}},\
  and\ \bibinfo {author} {\bibfnamefont {J.}~\bibnamefont {Papavassiliou}},\
  }\href {https://doi.org/10.1103/PhysRevD.78.025010} {\bibfield  {journal}
  {\bibinfo  {journal} {Phys. Rev. D}\ }\textbf {\bibinfo {volume} {78}},\
  \bibinfo {pages} {025010} (\bibinfo {year} {2008})},\ \Eprint
  {https://arxiv.org/abs/0802.1870} {arXiv:0802.1870 [hep-ph]} \BibitemShut
  {NoStop}%
\bibitem [{\citenamefont {Fischer}\ \emph {et~al.}(2009)\citenamefont
  {Fischer}, \citenamefont {Maas},\ and\ \citenamefont
  {Pawlowski}}]{Fischer:2008uz}%
  \BibitemOpen
  \bibfield  {author} {\bibinfo {author} {\bibfnamefont {C.~S.}\ \bibnamefont
  {Fischer}}, \bibinfo {author} {\bibfnamefont {A.}~\bibnamefont {Maas}},\ and\
  \bibinfo {author} {\bibfnamefont {J.~M.}\ \bibnamefont {Pawlowski}},\ }\href
  {https://doi.org/10.1016/j.aop.2009.07.009} {\bibfield  {journal} {\bibinfo
  {journal} {Annals Phys.}\ }\textbf {\bibinfo {volume} {324}},\ \bibinfo
  {pages} {2408} (\bibinfo {year} {2009})},\ \Eprint
  {https://arxiv.org/abs/0810.1987} {arXiv:0810.1987 [hep-ph]} \BibitemShut
  {NoStop}%
\bibitem [{\citenamefont {Cyrol}\ \emph {et~al.}(2016)\citenamefont {Cyrol},
  \citenamefont {Fister}, \citenamefont {Mitter}, \citenamefont {Pawlowski},\
  and\ \citenamefont {Strodthoff}}]{Cyrol:2016tym}%
  \BibitemOpen
  \bibfield  {author} {\bibinfo {author} {\bibfnamefont {A.~K.}\ \bibnamefont
  {Cyrol}}, \bibinfo {author} {\bibfnamefont {L.}~\bibnamefont {Fister}},
  \bibinfo {author} {\bibfnamefont {M.}~\bibnamefont {Mitter}}, \bibinfo
  {author} {\bibfnamefont {J.~M.}\ \bibnamefont {Pawlowski}},\ and\ \bibinfo
  {author} {\bibfnamefont {N.}~\bibnamefont {Strodthoff}},\ }\href
  {https://doi.org/10.1103/PhysRevD.94.054005} {\bibfield  {journal} {\bibinfo
  {journal} {Phys. Rev. D}\ }\textbf {\bibinfo {volume} {94}},\ \bibinfo
  {pages} {054005} (\bibinfo {year} {2016})},\ \Eprint
  {https://arxiv.org/abs/1605.01856} {arXiv:1605.01856 [hep-ph]} \BibitemShut
  {NoStop}%
\bibitem [{\citenamefont {Huber}(2020)}]{Huber:2018ned}%
  \BibitemOpen
  \bibfield  {author} {\bibinfo {author} {\bibfnamefont {M.~Q.}\ \bibnamefont
  {Huber}},\ }\href {https://doi.org/10.1016/j.physrep.2020.04.004} {\bibfield
  {journal} {\bibinfo  {journal} {Phys. Rept.}\ }\textbf {\bibinfo {volume}
  {879}},\ \bibinfo {pages} {1} (\bibinfo {year} {2020})},\ \Eprint
  {https://arxiv.org/abs/1808.05227} {arXiv:1808.05227 [hep-ph]} \BibitemShut
  {NoStop}%
\bibitem [{\citenamefont {Horak}\ \emph {et~al.}(2022)\citenamefont {Horak},
  \citenamefont {Ihssen}, \citenamefont {Papavassiliou}, \citenamefont
  {Pawlowski}, \citenamefont {Weber},\ and\ \citenamefont
  {Wetterich}}]{Horak:2022aqx}%
  \BibitemOpen
  \bibfield  {author} {\bibinfo {author} {\bibfnamefont {J.}~\bibnamefont
  {Horak}}, \bibinfo {author} {\bibfnamefont {F.}~\bibnamefont {Ihssen}},
  \bibinfo {author} {\bibfnamefont {J.}~\bibnamefont {Papavassiliou}}, \bibinfo
  {author} {\bibfnamefont {J.~M.}\ \bibnamefont {Pawlowski}}, \bibinfo {author}
  {\bibfnamefont {A.}~\bibnamefont {Weber}},\ and\ \bibinfo {author}
  {\bibfnamefont {C.}~\bibnamefont {Wetterich}},\ }\href
  {https://doi.org/10.21468/SciPostPhys.13.2.042} {\bibfield  {journal}
  {\bibinfo  {journal} {SciPost Phys.}\ }\textbf {\bibinfo {volume} {13}},\
  \bibinfo {pages} {042} (\bibinfo {year} {2022})},\ \Eprint
  {https://arxiv.org/abs/2201.09747} {arXiv:2201.09747 [hep-ph]} \BibitemShut
  {NoStop}%
\bibitem [{\citenamefont {Nielsen}(1975)}]{Nielsen:1975fs}%
  \BibitemOpen
  \bibfield  {author} {\bibinfo {author} {\bibfnamefont {N.~K.}\ \bibnamefont
  {Nielsen}},\ }\href {https://doi.org/10.1016/0550-3213(75)90301-6} {\bibfield
   {journal} {\bibinfo  {journal} {Nucl. Phys. B}\ }\textbf {\bibinfo {volume}
  {101}},\ \bibinfo {pages} {173} (\bibinfo {year} {1975})}\BibitemShut
  {NoStop}%
\bibitem [{\citenamefont {Capri}\ \emph {et~al.}(2017)\citenamefont {Capri},
  \citenamefont {Dudal}, \citenamefont {Pereira}, \citenamefont {Fiorentini},
  \citenamefont {Guimaraes}, \citenamefont {Mintz}, \citenamefont {Palhares},\
  and\ \citenamefont {Sorella}}]{Capri:2016gut}%
  \BibitemOpen
  \bibfield  {author} {\bibinfo {author} {\bibfnamefont {M.~A.~L.}\
  \bibnamefont {Capri}}, \bibinfo {author} {\bibfnamefont {D.}~\bibnamefont
  {Dudal}}, \bibinfo {author} {\bibfnamefont {A.~D.}\ \bibnamefont {Pereira}},
  \bibinfo {author} {\bibfnamefont {D.}~\bibnamefont {Fiorentini}}, \bibinfo
  {author} {\bibfnamefont {M.~S.}\ \bibnamefont {Guimaraes}}, \bibinfo {author}
  {\bibfnamefont {B.~W.}\ \bibnamefont {Mintz}}, \bibinfo {author}
  {\bibfnamefont {L.~F.}\ \bibnamefont {Palhares}},\ and\ \bibinfo {author}
  {\bibfnamefont {S.~P.}\ \bibnamefont {Sorella}},\ }\href
  {https://doi.org/10.1103/PhysRevD.95.045011} {\bibfield  {journal} {\bibinfo
  {journal} {Phys. Rev. D}\ }\textbf {\bibinfo {volume} {95}},\ \bibinfo
  {pages} {045011} (\bibinfo {year} {2017})},\ \Eprint
  {https://arxiv.org/abs/1611.10077} {arXiv:1611.10077 [hep-th]} \BibitemShut
  {NoStop}%
\bibitem [{\citenamefont {Napetschnig}\ \emph {et~al.}(2021)\citenamefont
  {Napetschnig}, \citenamefont {Alkofer}, \citenamefont {Huber},\ and\
  \citenamefont {Pawlowski}}]{Napetschnig:2021ria}%
  \BibitemOpen
  \bibfield  {author} {\bibinfo {author} {\bibfnamefont {M.}~\bibnamefont
  {Napetschnig}}, \bibinfo {author} {\bibfnamefont {R.}~\bibnamefont
  {Alkofer}}, \bibinfo {author} {\bibfnamefont {M.~Q.}\ \bibnamefont {Huber}},\
  and\ \bibinfo {author} {\bibfnamefont {J.~M.}\ \bibnamefont {Pawlowski}},\
  }\href {https://doi.org/10.1103/PhysRevD.104.054003} {\bibfield  {journal}
  {\bibinfo  {journal} {Phys. Rev. D}\ }\textbf {\bibinfo {volume} {104}},\
  \bibinfo {pages} {054003} (\bibinfo {year} {2021})},\ \Eprint
  {https://arxiv.org/abs/2106.12559} {arXiv:2106.12559 [hep-ph]} \BibitemShut
  {NoStop}%
\bibitem [{\citenamefont {Siringo}\ and\ \citenamefont
  {Comitini}(2022)}]{Siringo:2022nok}%
  \BibitemOpen
  \bibfield  {author} {\bibinfo {author} {\bibfnamefont {F.}~\bibnamefont
  {Siringo}}\ and\ \bibinfo {author} {\bibfnamefont {G.}~\bibnamefont
  {Comitini}},\ }\href {https://doi.org/10.1103/PhysRevD.106.076014} {\bibfield
   {journal} {\bibinfo  {journal} {Phys. Rev. D}\ }\textbf {\bibinfo {volume}
  {106}},\ \bibinfo {pages} {076014} (\bibinfo {year} {2022})},\ \Eprint
  {https://arxiv.org/abs/2208.03534} {arXiv:2208.03534 [hep-th]} \BibitemShut
  {NoStop}%
\bibitem [{\citenamefont {Capri}\ \emph
  {et~al.}(2018{\natexlab{b}})\citenamefont {Capri}, \citenamefont {van
  Egmond}, \citenamefont {Peruzzo}, \citenamefont {Guimaraes}, \citenamefont
  {Holanda}, \citenamefont {Sorella}, \citenamefont {Terin},\ and\
  \citenamefont {Toledo}}]{Capri:2017npq}%
  \BibitemOpen
  \bibfield  {author} {\bibinfo {author} {\bibfnamefont {M.~A.~L.}\
  \bibnamefont {Capri}}, \bibinfo {author} {\bibfnamefont {D.~M.}\ \bibnamefont
  {van Egmond}}, \bibinfo {author} {\bibfnamefont {G.}~\bibnamefont {Peruzzo}},
  \bibinfo {author} {\bibfnamefont {M.~S.}\ \bibnamefont {Guimaraes}}, \bibinfo
  {author} {\bibfnamefont {O.}~\bibnamefont {Holanda}}, \bibinfo {author}
  {\bibfnamefont {S.~P.}\ \bibnamefont {Sorella}}, \bibinfo {author}
  {\bibfnamefont {R.~C.}\ \bibnamefont {Terin}},\ and\ \bibinfo {author}
  {\bibfnamefont {H.~C.}\ \bibnamefont {Toledo}},\ }\href
  {https://doi.org/10.1016/j.aop.2018.01.009} {\bibfield  {journal} {\bibinfo
  {journal} {Annals Phys.}\ }\textbf {\bibinfo {volume} {390}},\ \bibinfo
  {pages} {214} (\bibinfo {year} {2018}{\natexlab{b}})},\ \Eprint
  {https://arxiv.org/abs/1712.04073} {arXiv:1712.04073 [hep-th]} \BibitemShut
  {NoStop}%
\bibitem [{\citenamefont {Collins}(1986)}]{Collins:1984xc}%
  \BibitemOpen
  \bibfield  {author} {\bibinfo {author} {\bibfnamefont {J.~C.}\ \bibnamefont
  {Collins}},\ }\href {https://doi.org/10.1017/CBO9780511622656} {\emph
  {\bibinfo {title} {{Renormalization}: {An Introduction to Renormalization,
  The Renormalization Group, and the Operator Product Expansion}}}},\ \bibinfo
  {series} {Cambridge Monographs on Mathematical Physics}, Vol.~\bibinfo
  {volume} {26}\ (\bibinfo  {publisher} {Cambridge University Press},\ \bibinfo
  {address} {Cambridge},\ \bibinfo {year} {1986})\BibitemShut {NoStop}%
\bibitem [{\citenamefont {Gracey}(2005)}]{Gracey:2004bk}%
  \BibitemOpen
  \bibfield  {author} {\bibinfo {author} {\bibfnamefont {J.~A.}\ \bibnamefont
  {Gracey}},\ }\href {https://doi.org/10.1140/epjc/s2004-02082-1} {\bibfield
  {journal} {\bibinfo  {journal} {Eur. Phys. J. C}\ }\textbf {\bibinfo {volume}
  {39}},\ \bibinfo {pages} {61} (\bibinfo {year} {2005})},\ \Eprint
  {https://arxiv.org/abs/hep-ph/0411169} {arXiv:hep-ph/0411169} \BibitemShut
  {NoStop}%
\end{thebibliography}%
\bibliographystyle{apsrev4-2}
\end{document}